\documentclass[12pt]{article}
\usepackage{amsmath}
\usepackage{amsthm}
\usepackage{mathrsfs} 
\usepackage{color}
\usepackage{amssymb}
\usepackage{amsfonts}
\usepackage{graphicx}
\usepackage[a4paper,text={150mm,234mm},centering,headsep=8mm,footskip=15mm]{geometry}
\usepackage[font=small,labelfont=bf]{caption}
\usepackage[numbers,sort&compress]{natbib}
\usepackage{enumerate}
\usepackage{setspace}
\definecolor{darkgreen}{rgb}{0.1,0.4,0.22}
\definecolor{someorange}{rgb}{0.555,0.290,0.204}
\definecolor{darkblue}{rgb}{0.1,0.22,0.5}
\usepackage[colorlinks=true
,urlcolor=darkgreen
,anchorcolor=blue
,citecolor=darkblue
,filecolor=blue
,linkcolor=someorange
,menucolor=blue
,linktocpage=true
,pdfproducer=medialab
]{hyperref}
\usepackage{framed}

\def\d{\mathrm{d}}

\begin{document}

\renewcommand{\theequation}{\thesection.\arabic{equation}} 
\numberwithin{equation}{section} 

\begin{titlepage}
\begin{center}
\rightline{\small DESY-16-021}
\vskip 1cm

{\Large \bf Scalar Potential from Higher Derivative $\mathcal{N}=1$ Superspace}
\vskip 1.2cm

{\bf  David Ciupke}

\vskip 0.8cm

{\em DESY, Hamburg, Germany}
\vskip 0.3cm

{\tt david.ciupke@desy.de}
\end{center}

\vskip 1cm

\begin{center} {\bf ABSTRACT } 
\end{center}

\noindent

The supersymmetric completion of higher-derivative operators often requires introducing corrections to the scalar potential. In this paper we study these corrections systematically in the context of theories with $\mathcal{N}=1$ global and local supersymmetry in $D=4$ focusing on ungauged chiral multiplets. In globally supersymmetric theories the most general off-shell effective scalar potential can be captured by a dependence of the K\"ahler potential on additional chiral superfields. For supergravity we find a much richer structure of possible corrections. In this context we classify the leading order and next-to-leading order superspace derivative operators and determine the component forms of a subclass thereof. Moreover, we present an algorithm that simplifies the computation of the respective on-shell action. As particular applications we study the structure of the supersymmetric vacua for these theories and comment on the form of the corrections to shift-symmetric no-scale models. These results are relevant for the computation of effective actions for string compactifications and, in turn, for moduli stabilization and string inflation.
\bigskip

\today

\end{titlepage}

\newpage

\tableofcontents
\newpage
\section{Introduction}
Effective field theory (EFT) constitutes a useful and broadly applicable framework. Of special interest are effective descriptions of theories with global or local supersymmetry. While effective field theories generically include higher-derivative operators, for supersymmetric effective theories the completion of the higher-derivatives often requires introducing corrections to the scalar potential. 
To emphasize the general importance of these higher-derivative operators and the additional corrections to the scalar potential, consider an effective $\mathcal{N}=1$ supergravity in four dimensions which is valid up to some cut-off scale $\Lambda \leq M_p$. For example in theories obtained from superstring-compactifications this cut-off scale might be the string scale $M_s$ or the Kaluza-Klein scale $M_{KK}$. In particular and as we will see later on, there exist infinitely many $\Lambda$-suppressed operators in the scalar potential which are present both in the effective higher-derivative supergravity as well as the global limit thereof. Therefore, these operators are dominant over the Planck-suppressed operators in the ordinary scalar potential of $\mathcal{N}=1$ supergravity and, thus, play a role in several scenarios. In particular, they affect the vacuum structure and are relevant to moduli stabilization, but are also important in the context of inflation. Briefly said, the goal of this work is to understand the connection between higher-derivative operators for global and local supersymmetry and their respective corrections to the scalar potential systematically.

To begin with it is necessary to gain an understanding of higher-derivative operators for effective theories with rigid supersymmetry. These will be the subject of the first part of this paper. Here we consider theories with ungauged chiral multiplets only. See also \cite{Cecotti:1986jy, Buchbinder:1994iw, Buchbinder:1995ideas, Antoniadis:2007xc, Khoury:2010gb, Farakos:2013zsa, Nitta:2014fca, Dudas:2015vka} for earlier investigations of higher-derivative operators for globally supersymmetric theories. Here we show that the most general scalar potential can be derived from a superspace-Lagrangian given in terms of a (pseudo-) K\"ahler potential $\mathcal{K}$ together with the ordinary superpotential of the two-derivative theory and additional constraints. $\mathcal{K}$ depends not only on the chiral scalars, but also on the respective auxiliary fields. This quantity already appeared in \cite{Buchbinder:1994iw, Buchbinder:1995ideas} under the name of effective auxiliary field potential. In these theories the auxiliary fields can either remain algebraic or obtain kinetic terms. We discuss the equations of motion for the auxiliary fields in detail. On the one hand, we argue that even if they obtain kinetic terms, the auxiliary fields must still be treated as algebraic degrees of freedom, since they generically obtain masses at the cut-off scale $\Lambda$. On the other hand, we clarify the appearance of multiple solutions to the respective equations of motion. We extend the analysis of \cite{Ciupke:2015msa} and present a general argument that among the multiple on-shell theories there exists a unique physical theory compatible with the principles of effective field theory.

In special situations it does not suffice to discuss globally supersymmetric theories, but instead it is necessary to transcend to effective supergravities. In particular, these constitute the proper framework to discuss certain string compactifications such as Calabi-Yau orientifold compactifications of Type IIA/B. To attempt a full derivation of higher-derivative operators directly from string-theoretic corrections, as entertained for closed string modes in \cite{Ciupke:2015msa}, requires a prior systematic knowledge of higher-derivative operators in supergravity. In particular, this is due to the fact that the corrections to the 10D or 11D effective supergravities are not manifestly supersymmetric and so the supersymmetric 4D completion after compactification is not immediately clear. Thus, we are motivated to present a classification of higher-derivative operators in supergravity such that, in principle, a full matching becomes possible. As a byproduct such a classification also reveals the possible corrections to the scalar potential which can, in principle, emerge.

The general study of higher-derivative operators for chiral multiplets in supergravity is the topic of the second part of this work. Here we confine ourselves to the discussion in the context of old minimal supergravity following \cite{Wess:1992cp} and earlier work on the subject \cite{Baumann:2011nm}, see also \cite{Cecotti:1986jy, Cecotti:1986pk, Koehn:2012ar, Farakos:2012qu} for related studies. 
Our analysis is split in two parts: Firstly, we study certain conceptual and computational aspects of these higher-derivative operators. We collect these statements in an algorithm that allows to compute the linearized (in the coupling of the operator) on-shell action of a particular operator. 
In particular, we point out the importance of integrating out fields in the Einstein-frame and find that the linearized component action does not require solving the equations of motion of the auxiliary fields. 
In the second part of the analysis, we classify all leading and next-to-leading order higher-derivative operators and determine the component versions of those operators which induce four-derivative terms for the chiral scalars at the linearized level. This result is fully model-independent and is widely applicable to any analysis which seeks to determine the leading order corrections to the scalar potential or two- and four-derivative component terms for a specific (ungauged) matter-coupled supergravity.

In the last part of the paper we discuss two applications of the above results. Firstly, we investigate the structure of supersymmetric Minkowski and $AdS_4$-vacua. We pay special attention to the curvature constraints enforced by the necessary existence of Killing spinors and their compatibility with the general scalar potential of higher-derivative supergravity. Secondly, we discuss the form of the higher-derivative operators for the special case where the leading order (two-derivative) theory is given by a shift-symmetric no-scale model. These models appear in the context of the low-energy effective descriptions of certain string-compactifications. We find that the no-scale condition leads to the vanishing of many leading-order contributions and only those corrections remain, which are purely given by the chiral auxiliary. This leads to an a posteriori proof of respective claims made in \cite{Ciupke:2015msa}. 

This paper is organized as follows. In section~\ref{sec:flat_superspace_HD} we discuss higher-derivative operators for chiral multiplets in globally supersymmetric theories. After reviewing the basic formalism of flat $\mathcal{N}=1$ superspace in  sec.~\ref{sec:Preliminaries}, we present a superspace action for the general scalar potential in sec.~\ref{sec:HD_susy} and sec.~\ref{sec:multi_HD_susy}. A derivation of the action in sec.~\ref{sec:HD_susy} is presented in appendix~\ref{appendix:global}. 
Moreover, in sec.~\ref{on-shell-eft} and sec.~\ref{prop_aux} we discuss the equations of motion for the auxiliary fields. Afterwards, in section~\ref{Higher_deriv_sugra} we turn to the case of higher-derivative operators in $\mathcal{N}=1$ supergravity. In particular, in sec.~\ref{sec:curved_superspace} we review the formulation of old minimal supergravity in curved superspace and revisit the derivation of the two-derivative action for chiral multiplets coupled to supergravity in sec.~\ref{sugraaction}. Subsequently, in sec.~\ref{HD_Sugra_overview} we present an overview over higher-derivative operators in supergravity and, for the convenience of the reader, give a brief outline of the main results. In sec.~\ref{sec:higher_curvature_ops} we discuss the subclass of higher-derivative operators which induce also higher-curvature terms and, afterwards, in sec.~\ref{HCTIOF} we demonstrate the importance of integrating out fields in the Einstein frame. A supplementary demonstration in this context is given in appendix~\ref{More_on_prop_aux}. Sec.~\ref{sec:component_form} and appendix~\ref{component_identities} contain the computational tools which we then apply to determine in sec.~\ref{sec:two_der_operators} the leading order and in sec.~\ref{sec:four_derivative_operators} the next-to-leading order operators. The classification of the latter can be found in appendix~\ref{appendix:integration_by_parts}. In sec.~\ref{sec:vacua} and appendix~\ref{app:Killing_spinor} we investigate the vacuum structure of the general higher-derivative theory and in sec.~\ref{sec:no_scale} comment on implications for no-scale models. Finally, in sec.~\ref{sec:conclusions} we conclude this paper.
\section{Systematics of the Scalar Potential in Flat $\mathcal{N}=1$ Superspace}\label{sec:flat_superspace_HD}
We begin this section by reviewing the basic notions and formalism of flat superspace and globally supersymmetric theories for chiral multiplets. Readers familiar with these technical details can safely skip this first part and directly proceed to sec.~\ref{sec:HD_susy}. There we present the most general superspace action including all those higher-derivative operators, which manifestly contribute to the scalar potential for the chiral multiplets. Furthermore, we display the respective component version. Afterwards, we discuss two conceptual points. Firstly, in sec.~\ref{on-shell-eft} we tend to the equations of motion for the chiral auxiliary and discuss the appearance of multiple solutions thereof. Secondly, in sec.~\ref{prop_aux} we argue that if the auxiliary receives a kinetic term, this new degree of freedom is expected to be unphysical in a generic EFT description.
\subsection{Preliminaries}\label{sec:Preliminaries}
The effective action of any theory with global $\mathcal{N}=1$ supersymmetry in four spacetime dimensions can be expressed in flat $\mathcal{N}=1$ superspace. We begin by recapitulating the basics on flat superspace and theories with global supersymmetry, adopting the notations and conventions of \cite{Wess:1992cp}. Flat superspace is parametrized by the variables
\begin{equation}
 z^A = (x^a, \theta^\alpha, \bar{\theta}_{\dot{\alpha}}) \ , \qquad a=0,\dots,3 \ , \qquad \alpha , \dot{\alpha} = 0,1 \ .
\end{equation}
Henceforth, we use $a,b,c,\dots$ to label Minkowski space coordinates and $\alpha,\beta, \gamma, \dots$ (as well as $\dot{\alpha}, \dot{\beta}, \dot{\gamma}, \dots$) to label flat Grassmann space variables. Moreover, we abbreviate integration measures over flat superspace as 
\begin{equation}
 \d^8 z = \d^4 x \, \d^4 \theta  = \d^4 x \, \d^2 \theta \, \d^2 \bar{\theta} \ , \qquad \d^6 z =  \d^4 x  \, \d^2 \theta  \ , \qquad  \d^6 \bar{z} =  \d^4 x  \, \d^2 \bar{\theta} \ .
\end{equation}
Supersymmetry is realized via the superspace derivatives
\begin{equation}\label{flat_superspace_der_1}
 D_A = (\partial_a, D_{\alpha}, \bar{D}^{\dot{\alpha}}) \ ,
\end{equation}
where $\partial_a$ denote the conventional spacetime derivatives and the spinorial components are defined as 
\begin{equation}\label{flat_superspace_der_2}
 D_{\alpha} = \frac{\partial}{\partial \theta^{\alpha}} + i \sigma_{\alpha \dot{\alpha}}^a \bar{\theta}^{\dot{\alpha}} \frac{\partial}{\partial x^a} \qquad\text{and} \qquad\bar{D}_{\dot{\alpha}} = -\frac{\partial}{\partial \bar{\theta}^{\dot{\alpha}}} -i \theta^{\alpha}  \sigma_{\alpha \dot{\alpha}}^a  \frac{\partial}{\partial x^a}\ .
\end{equation}
Here $\sigma^a$ label the Pauli-matrices. Furthermore, the Pauli-matrices convert tensor superfields $V_{\alpha \dot{\alpha}}$ and vector superfields $V_a$ into each other
\begin{equation}
 V_a = - \tfrac{1}{2} \bar{\sigma}_a^{ \dot{\alpha}\alpha} V_{\alpha \dot{\alpha}} \ , \qquad V_{\alpha \dot{\alpha}} = \sigma^a_{\alpha \dot{\alpha}} V_a \ ,
\end{equation}
where $\bar{\sigma}_a^{ \dot{\alpha}\alpha } \equiv \epsilon^{\alpha \beta} \epsilon^{\dot{\alpha} \dot{\beta}} \sigma_a {}_{\beta \dot{\beta}}$. In the following we refer to $D_{\alpha}$ and $\bar{D}_{\dot{\alpha}}$ as (spinorial) superspace derivatives. The derivatives $D_A$ fulfill the following (anti-) commutation relations
\begin{equation}\label{anticom}
 \{D_{\alpha}, \bar{D}_{\dot{\beta}}\} = -2i \sigma_{\alpha \dot{\beta}}^a \partial_a \ , \qquad \{D_{\alpha}, D_{\beta} \} = \{ \bar{D}_{\dot{\alpha}} , \bar{D}_{\dot{\beta}} \} = [D_{\alpha},\partial_a] = [\bar{D}_{\dot{\alpha}} ,\partial_a] = 0 \ .
\end{equation}
In addition we use the convention
\begin{equation}
 D^2 = D^\alpha D_\alpha \ , \qquad \bar{D}^2 = \bar{D}_{\dot{\alpha}} \bar{D}^{\dot{\alpha}} \ .
\end{equation}
Moreover, from the commutation relations one deduces that the superspace derivatives obey the identities
\begin{align}\label{D30}
 &D_{\alpha} D_{\beta} D_{\gamma} = \bar{D}_{\dot{\alpha}} \bar{D}_{\dot{\beta}} \bar{D}_{\dot{\gamma}} =   0 \ , \\
 \label{DDisD2}
 &D_{\alpha} D_{\beta} = \tfrac{1}{2} \epsilon_{\alpha \beta} D^2 \ , \quad \bar{D}_{\dot{\alpha}} \bar{D}_{\dot{\beta}} = -\tfrac{1}{2} \epsilon_{\dot{\alpha}\dot{\beta}} \bar{D}^2  \ .
\end{align}
Let us now consider a theory with a chiral superfield $\Phi$ and the respective antichiral superfield $\Phi^\dagger$. The (anti-) chirality is defined by the following property
\begin{equation}\label{chiral}
 \bar{D}_{\dot{\alpha}} \Phi = D_{\alpha} \Phi^{\dagger} = 0 \ .
\end{equation}
This condition constrains the $\theta$-expansion for $\Phi$ to be  
\begin{equation}\label{chiral_superfield_exp}
 \Phi = A + \sqrt{2} \theta \chi + \theta^2 F + i \theta
 \sigma^a \bar{\theta} \partial_a A - \tfrac{i}{\sqrt{2}}
 \theta \theta \partial_a \chi \sigma^a \bar{\theta} + \frac{1}{4} \theta^2 \bar{\theta}^2 \Box A \ ,
\end{equation}
where $A$ is a complex scalar, $\chi$ a Weyl-fermion and $F$ an auxiliary complex scalar. We will abbreviate the component-expansion for chiral superfields as $\Phi = (A,\chi,F)$. 

In the following we will be interested in the following superfields
\begin{equation}
 \bar{\Psi} \equiv - \tfrac{1}{4} D^2 \Phi \ , \qquad \Psi  \equiv - \tfrac{1}{4} \bar{D}^2\bar{\Phi} \ .
\end{equation}
From eq.~\eqref{D30} we learn that $\Psi$ is a chiral and $\bar{\Psi}$ an anti-chiral superfield. More precisely, the superfield-expansion of $\bar{\Psi}$ reads
\begin{equation}\label{superfield_psi}
  \bar{\Psi} = F - \sqrt{2}i \partial_a \chi \sigma^a \bar{\theta} + \bar{\theta}^2 \Box A - i \theta \sigma^a \bar{\theta} \partial_a F + \tfrac{1}{\sqrt{2}} \bar{\theta}^2 \theta \Box \chi + \tfrac{1}{4}\theta^4 \Box F \ .
\end{equation}
Thus, the abbreviated forms of the component expansion are given by $\bar{\Psi}= (F,i\sigma^a \partial_a \chi,\Box A)$ and $\Psi = (\bar{F},-i\sigma^a \partial_a \bar{\chi},\Box \bar{A})$. Having introduced the relevant notation we can now proceed to discuss actions for the chiral multiplet.

The most general two-derivative Lagrangian for ungauged chiral multiplets reads
\begin{equation}\label{L_ord_susy}
 \mathcal{L}_0[\Phi, \bar{\Phi}] = \int \mathrm{d}^4 \theta K(\Phi,\bar{\Phi}) + \int \mathrm{d}^2 \theta \, W(\Phi) + h.c. \ ,
\end{equation}
where $K$ is a real function and can be understood as a K\"ahler potential of a respective K\"ahler manifold which the chiral scalars locally parametrize. The object $W$ is a chiral superfield and denoted as superpotential.
\subsection{Higher-Derivative Actions in Flat $\mathcal{N}=1, D=4$ Superspace}\label{sec:HD_susy}
So far we considered the general theory at the two-derivative level. In a generic EFT all operators consistent with the symmetries must be present and so generally an infinite tower of higher-derivative operators must be included. In the superspace formalism higher-derivatives are realized by acting with superspace-derivatives on superfields. In fact, it can be shown that $D_A$ are the only objects that anti-commute with the supersymmetry generators \cite{Buchbinder:1995ideas} and, therefore, the only required ingredient to study higher-derivatives in superspace. In conclusion, the general superspace effective action for a theory describing an ungauged chiral multiplet is of the form
\begin{equation}\begin{aligned}\label{Sgen}
 S_{\mathrm{gen}} &= \int \d^8 z \,\mathcal{K}(\Phi,\bar{\Phi}, D_A \Phi, D_B \bar{\Phi}, D_A D_B \Phi, \dots) \\
  & \quad + \int \d^6 z \mathcal{W}(\Phi, \partial_a \Phi, \bar{D}^2 \bar{\Phi}, \partial_a \partial_b \Phi, \dots) + h.c. \ ,
\end{aligned}\end{equation}
where the dots indicate a dependence on higher superspace-derivatives acting on $\Phi$ or $\bar{\Phi}$. The above action is supersymmetric, if $\mathcal{W}$ depends only on chiral superfields.\footnote{Indeed, $\partial_a \Phi$ and $\partial_a \partial_b \Phi$ are chiral superfields due to eq.~\eqref{anticom}.}

The general action in eq.~\eqref{Sgen} is rather involved and carries a dependence on infinitely many superfields. However, most higher-derivative operators in $S_{\mathrm{gen}}$ are purely kinetic at the component level. In this paper we focus on the subclass of higher-derivative operators which manifestly contribute to the scalar potential. More precisely, we are interested in the most general form of the off-shell superspace effective action under the condition that all terms in this action manifestly contribute to the scalar potential. We shall refer to this action as $S_{\text{eff}}$ from now on. $S_{\text{eff}}$ was already discussed and determined in \cite{Buchbinder:1994iw, Buchbinder:1995ideas} in the context of Wilsonian effective actions and was computed for a notable example, namely the one-loop Wess-Zumino model in \cite{Buchbinder:1994iw, Pickering:1996he, Kuzenko:2014ypa}. To determine $S_{\text{eff}}$ we evaluate $S_{\mathrm{gen}}$ at the supersymmetric condition
\begin{equation}\label{effectivepot}
 \partial_a \Phi = \partial_a \bar{\Phi} = 0 \ .
\end{equation}
Since $D_\alpha$ and $\bar{D}^{\dot{\alpha}}$ commute with $\partial_a$, all operators which we set to zero via eq.~\eqref{effectivepot} contribute only to the kinetic part of the Lagrangian and are, therefore, irrelevant in our analysis. When evaluating the general action in eq.~\eqref{Sgen} at the condition in eq.~\eqref{effectivepot} the resulting action $S_{\text{eff}}$ greatly simplifies \cite{Buchbinder:1994iw, Buchbinder:1995ideas}. Taking a bottom-up EFT perspective we give an alternative derivation of $S_{\text{eff}}$ in appendix~\ref{appendix:global}. There we also demand that $S_{\text{eff}}$ does not exhibit any redundancy in the operators that it includes. Briefly summarized appendix~\ref{appendix:global} consists of two parts. Firstly, we show that a dependence of the action on the superspace derivatives acting on $\Phi$ and $\bar{\Phi}$ is restricted by the (anti-)commutation relations in eq.~\eqref{anticom}. Secondly, using integration by parts identities we reduce the effective action even further, such that finally one is left with
\begin{equation}\begin{aligned}\label{Seff}
 S_{\mathrm{eff}} &= \int \d^8 z \, \mathcal{K}(\Phi, \bar{\Phi}, \Psi, \bar{\Psi}) + \int \d^6 z \, W(\Phi) + h.c. \ .
\end{aligned}\end{equation}
Up to some minor differences regarding possible redundancies of operators, this coincides with the action that was already obtained in \cite{Buchbinder:1994iw, Buchbinder:1995ideas}.

Let us pause a moment to clarify the physical meaning of the additional degree of freedom associated with $\Psi$. As displayed in eq.~\eqref{superfield_psi} the fermionic- and $\theta^2$-components of $\Psi$ are given by higher-derivatives of the chiral scalar and chiral fermion $\sigma^a \partial_a \bar{\chi}$ and $\Box \bar{A}$. In the sense of the Ostrogradski-procedure these higher-derivatives constitute additional degrees of freedom \cite{Ostro}. However, since we are discussing theories with off-shell supersymmetry the number of bosonic and fermionic degrees of freedom must match and, furthermore, allow a description in terms of an appropriate multiplet. This matching is achieved when taking into account the auxiliary field $\bar{F}$, such that the collection of component fields fits nicely into the chiral multiplet $\Psi$. 
\subsection{General Scalar Potential in Higher-Derivative Theory}\label{sec:multi_HD_susy}
It is now straightforward to generalize the above result in eq.~\eqref{Seff} to the case of $n_c$ chiral superfields $\Phi^i, i=1,\ldots,n_c$. We denote the respective components as $\Phi^i = (A^i,\chi^i,F^i)$ and 
\begin{equation}\label{psi_def}
 \Psi^i = - \tfrac{1}{4} \bar{D}^2 \bar{\Phi}^{i} = (\bar{F}^i,-i\sigma^a \partial_a \bar{\chi}^i,\Box \bar{A}^i) \ .
\end{equation}
The appropriate multi-field generalization of eq.~\eqref{Seff} reads 
\begin{equation}\label{Leffgeneral}
 \mathcal{L}_{\text{eff}} = \int \d^4 \theta \,\mathcal{K}(\Phi^i,\bar{\Phi}^{\bar{\jmath}},\Psi^k,\bar{\Psi}^{\bar{l}}) + \int \d^2 \theta\, W(\Phi^i) + h.c. \ ,
\end{equation}
which, strictly speaking, should be completed by adding a Lagrange multiplier whose equation of motion yields the constraints in eq.~\eqref{psi_def}.

Next we determine the component version of $\mathcal{L}_{\text{eff}}$. The Lagrangian in eq.~\eqref{Leffgeneral} can be understood as an ordinary theory with K\"ahler potential $\mathcal{K}$ and superpotential $W$, where the respective K\"ahler manifold is $2n_c$-dimensional, together with the constraints in eq.~\eqref{psi_def}. Therefore, the $\theta$-integration in eq.~\eqref{Leffgeneral} can be performed straightforwardly. We find
\begin{equation}\begin{aligned}\label{L_eff_rigid}
 \mathcal{L}_{\text{eff}} = &- \mathcal{K}_{,A^i \bar{A}^{\bar{\jmath}}}\, \partial_a A^i \partial^a \bar{A}^{\bar{\jmath}} - \mathcal{K}_{,A^i F^j}  \partial_a A^i \partial^a F^{j} - \mathcal{K}_{,\bar{A}^{\bar{\imath}} \bar{F}^{\bar{\jmath}}} \, \partial_a \bar{A}^{\bar{\imath}} \partial^a \bar{F}^{\bar{\jmath}} \\
 &-\mathcal{K}_{,F^i \bar{F}^{\bar{\jmath}}} \, \partial_a F^i \partial^a \bar{F}^{\bar{\jmath}} + \mathcal{K}_{,A^i F^j} F^i \Box A^j + \mathcal{K}_{,\bar{A}^{\bar{\imath}} \bar{F}^{\bar{\jmath}}} \, \bar{F}^{\bar{\imath}} \Box \bar{A}^{\bar{\jmath}}  \\
 & + \mathcal{K}_{,F^i \bar{F}^{\bar{\jmath}}} \, \Box A^i \Box \bar{A}^{\bar{\jmath}} + \mathcal{K}_{,A^i \bar{A}^{\bar{\jmath}}} \, F^i \bar{F}^{\bar{\jmath}} + F^i W_{,A^i} + \bar{F}^{\bar{\imath}} \bar{W}_{,\bar{A}^{\bar{\jmath}}}  \ ,
\end{aligned}\end{equation}
where the superfields $\mathcal{K}$ and derivatives thereof are understood as being evaluated at their respective scalar component. As before, we display only the bosonic terms here. 
Inspecting eq.~\eqref{L_eff_rigid} we observe that the auxiliary fields are now propagating degrees of freedom. The second derivatives of $\mathcal{K}$ control the kinetic terms for the scalars and, hence, determine whether the auxiliary fields are propagating and, in particular, ghostlike or not. We return to the discussion of the propagating auxiliary fields in sec.~\ref{prop_aux}. For now we do not make any further assumptions about the kinetic terms.

We are intrigued to find a geometric understanding of eq.~\eqref{L_eff_rigid} and eq.~\eqref{Leffgeneral}. A first guess would be that our theory is described by a $2n_c$-dimensional \hbox{(pseudo-)} K\"ahler geometry $\mathcal{M}_p$.\footnote{A pseudo-K\"ahler manifold obeys the same conditions as a K\"ahler manifold, but instead of being equipped with a positive definite metric it is endowed with an indefinite bilinear form. This situation occurs when the auxiliary fields are ghostlike or remain algebraic degrees of freedom.} However, the constraints in eq.~\eqref{psi_def} break the respective target-space invariance of the $2n_c$-dimensional geometry, and indeed it is easily seen that eq.~\eqref{L_eff_rigid} does not support a reparametrisation-invariance with respect to a $2n_c$-dimensional (pseudo-) K\"ahler geometry. Still, the ordinary reparametrisation invariance with respect to the chiral scalars $A^i, \bar{A}^{\bar{\jmath}}$ parametrising an $n_c$-dimensional complex manifold $\mathcal{M}_0$ must be maintained. Suppose that we are in a situation where we integrate out the auxiliary fields. We will justify this assumption later on in sec.~\ref{prop_aux}. Via the interactions in eq.~\eqref{L_eff_rigid} the solutions to the equations of motion for the auxiliaries formally read
\begin{equation}\label{F}
 F^i = F^i [A^i, \bar{A}^{\bar{\jmath}}, \partial_a A^i , \partial_a \bar{A}^{\bar{\jmath}}, \dots] \ ,
\end{equation}
where the dots indicate a dependence on higher spacetime-derivatives of the chiral scalars. Reparametrisation invariance with respect to $\mathcal{M}_0$ must, in particular, individually hold for the scalar potential in eq.~\eqref{L_eff_rigid}. 
To determine the scalar potential we have to truncate $F^i$ given in eq.~\eqref{F} so that only a dependence on $A^i, \bar{A}^{\bar{\jmath}}$ remains.\footnote{This truncation is achieved by evaluating eq.~\eqref{F} at the conditions $\partial_a A^i = \partial_a  \bar{A}^{\bar{\jmath}} = \dots = 0$ where the dots indicate all possible higher spacetime-derivatives of the chiral scalars.} Then target space-invariance necessarily requires that
\begin{equation}\label{F-term_one_form}
 \omega_F = F^i (A, \bar{A})\mathrm{d} A^i + \bar{F}^{\bar{\imath}}(A, \bar{A}) \mathrm{d} \bar{A}^{\bar{\imath}}
\end{equation}
constitutes a one-form on $\mathcal{M}_0$. For clarity, note that $F^i$ and $\bar{F}^{\bar{\imath}}$ in the above are the truncated versions of eq.~\eqref{F} for which we continue to use the same symbol. From eq.~\eqref{F-term_one_form} we learn that it is natural to discuss the scalar potential in eq.~\eqref{L_eff_rigid} in the context of the cotangent-bundle $T^* \mathcal{M}_0$. Target space invariance requires, furthermore, that $\mathcal{K}$ transforms as a scalar on the cotangent-bundle or in other words that it constitutes a zero-form on $T^* \mathcal{M}_0$. Note that this requirement is by no means automatically guaranteed but restricts the possible choices of $\mathcal{K}$. It is worth mentioning, that the aforementioned requirements are analogous to the respective requirements of the two-derivative theory. 

Checking the invariance of the kinetic terms in eq.~\eqref{L_eff_rigid} under reparametrisations of $A^i, \bar{A}^{\bar{\jmath}}$ is considerably more involved. In particular, the transformation behaviour of the auxiliaries must necessarily differ from eq.~\eqref{F-term_one_form} when including the dependence of $F^i$ on spacetime-derivatives of the chiral scalars which was indicated in eq.~\eqref{F}. The general discussion of these transformation properties is rather involved and we omit the details here.\footnote{For a generic $\mathcal{K}$ it is not guaranteed that one can always choose the transformation behaviour of the auxiliaries appropriately. There might exist situations where it is necessary to add further superspace higher-derivative operators which we excluded by means of eq.~\eqref{effectivepot}.}

One may wonder whether the manifold $\mathcal{M}_0$ is still endowed with a K\"ahler structure. In eq.~\eqref{L_eff_rigid} the kinetic terms for the chiral scalars are multiplied by a complicated metric which in general is not even hermitian. In the usual sense of a non-linear sigma model we, therefore, do not have a K\"ahler structure anymore. However, since we are interested in the most general scalar potential and not in the most general two- or higher-derivative component action, it is instructive to identify a geometric meaning of the scalar potential alone. Indeed, the scalar potential induced by $\mathcal{K}$ is given as the pseudo-norm of the one-form $\omega_F$ with respect to the bilinear form $\mathcal{K}_{,A^i \bar{A}^{\bar{\jmath}}}$. This object indeed defines a pseudo-K\"ahler structure on $\mathcal{M}_0$ and in the limit where all higher-derivative operators vanish this structure reproduces the K\"ahler structure of the ordinary two-derivative theory. 

Finally, let us make a comment regarding K\"ahler-invariance. Since eq.~\eqref{Leffgeneral} is a theory of chiral multiplets only, we observe that $\mathcal{L}_{\text{eff}}$ enjoys a respective extended K\"ahler invariance. The corresponding K\"ahler-transformations are of the form
  \begin{equation}\label{generlized_kahler_transf}
    \mathcal{K}(\Phi^i,\bar{\Phi}^{\bar{\jmath}},\Psi^k,\bar{\Psi}^{\bar{l}}) \longrightarrow \mathcal{K}(\Phi^i,\bar{\Phi}^{\bar{\jmath}},\Psi^k, \bar{\Psi}^{\bar{l}}) + G(\Phi^i,\Psi^j) + \bar{G}(\bar{\Phi}^{\bar{\jmath}},\bar{\Psi}^{\bar{l}}) \ ,
  \end{equation}
where $G$ is an arbitrary holomorphic function and $\bar{G}$ the respective anti-holomorphic function.
\subsubsection*{Alternative Higher-Derivative Lagrangian}
Let us mention a second higher-derivative Lagrangian, which is of interest. This Lagrangian also describes the most general effective scalar potential but it does not induce kinetic terms for the auxiliary fields. Moreover, the requirement of target space-invariance is, at least in parts, more easily realized here. We construct this Lagrangian from eq.~\eqref{Leffgeneral} via the following instruction. Within each operator in $\mathcal{K}$, which includes at least one factor of $\Psi^i \bar{\Psi}^{\bar{\jmath}}$, we replace this factor by a term $D^{\alpha} \Phi^i D_{\alpha} \Phi^{j} \bar{D}_{\dot{\alpha}} \bar{\Phi}^{\bar{k}} \bar{D}^{\dot{\alpha}}\bar{\Phi}^{\bar{l}}$ via the last identity in eq.~\eqref{integrationbyparts}. If this factor appears more than once we perform this procedure only for one of them. The resulting Lagrangian can then be cast into the form \footnote{The factor of $1/16$ is introduced purely for convenience here.}
\begin{equation}\begin{aligned}\label{Lmulti}
\mathcal{L}'_{\text{eff}} =& \int \d^4 \theta \,[K(\Phi,\bar{\Phi}) + \mathcal{F}(\Phi,\bar{\Phi},\Psi) + \bar{\mathcal{F}}(\Phi,\bar{\Phi},\bar{\Psi})]+ \int \d^2 \theta\, W(\Phi) + \int \d^2 \bar{\theta} \, \bar{W}(\bar{\Phi})  \\ 
 &+ \frac{1}{16}\int \d^4 \theta \, T_{ij\bar{k}\bar{l}}(\Phi,\bar{\Phi},\Psi,\bar{\Psi})\, D^{\alpha} \Phi^i D_{\alpha} \Phi^j \bar{D}_{\dot{\alpha}} \bar{\Phi}^{\bar{k}} \bar{D}^{\dot{\alpha}} \bar{\Phi}^{\bar{l}}  \ . 
\end{aligned}\end{equation}
The object $T_{ij\bar{k}\bar{l}}$ is a superfield and in order to support reparametrisation invariance it transforms as a tensor of $\mathcal{M}_0$ \cite{Koehn:2012ar}. From eq.~\eqref{Lmulti} we observe that 
\begin{equation}\label{T-tensor_properties}
 T_{ij\bar{k}\bar{l}} = T_{j i \bar{k}\bar{l}} = T_{j i \bar{l}\bar{k}} \ .
\end{equation}
Furthermore this object has to be a hermitian tensor to ensure reality of the Lagrangian. Let us emphasize that $\mathcal{L}'_{\text{eff}}$ and $\mathcal{L}_{\text{eff}}$ are in general distinct and only the respective scalar potentials coincide. More precisely, $\mathcal{L}'_{\text{eff}}$ and $\mathcal{L}_{\text{eff}}$ differ by purely kinetic superspace-operators.

For completeness we also determine the component form of $\mathcal{L}'_{\text{eff}}$. To this end it is necessary to compute the last integral in eq.~\eqref{Lmulti}. Using eq.~\eqref{flat_superspace_der_2} and the $\theta$-expansion for the (anti-) chiral superfield in eq.~\eqref{chiral_superfield_exp} one finds by direct computation
\begin{equation}\begin{aligned}
 & D^{\alpha} \Phi^i D_{\alpha} \Phi^{j} \bar{D}_{\dot{\alpha}} \bar{\Phi}^{\bar{k}} \bar{D}^{\dot{\alpha}}\bar{\Phi}^{\bar{l}}|_{\text{bos}} =\\
 & 16 \left[ (\partial_a A^i  \partial^a A^{j})( \partial_b \bar{A}^{\bar{k}} \partial^b \bar{A}^{\bar{l}})  - 2 F^i \bar{F}^{\bar{k}} (\partial_a A^{j} \partial^a \bar{A}^{\bar{l}})+ F^i F^j\bar{F}^{\bar{k}} \bar{F}^{\bar{l}} \right] \theta^2 \bar{\theta}^2\ ,
\end{aligned}\end{equation}
which, in turn, implies that the component version of $\mathcal{L}'_{\text{eff}}$ reads
\begin{equation}\begin{aligned}\label{Leffmulti}
 \mathcal{L}'_{\text{eff}} &= - (K_{,A^i \bar{A}^{\bar{\jmath}}} + \mathcal{F}_{,A^i \bar{A}^{\bar{\jmath}}}+\bar{\mathcal{F}}_{,A^i \bar{A}^{\bar{\jmath}}}) \partial_a A^i \partial^a \bar{A}^{\bar{\jmath}} - \bar{\mathcal{F}}_{,A^i F^j} \partial_a A^i \partial^a F^{j} \\ 
 & \quad - \mathcal{F}_{,\bar{A}^{\bar{\imath}} \bar{F}^{\bar{\jmath}} } \partial_a \bar{A}^{\bar{\imath}} \partial^a \bar{F}^{\bar{\jmath}} + \bar{\mathcal{F}}_{,A^i F^j} F^i \Box A^j + \mathcal{F}_{,\bar{A}^{\bar{\imath}} \bar{F}^{\bar{\jmath}}} \bar{F}^{\bar{\imath}} \Box \bar{A}^{\bar{\jmath}} \\
 &\quad + (K_{,A^i \bar{A}^{\bar{\jmath}}} + \mathcal{F}_{,A^i \bar{A}^{\bar{\jmath}}}+\bar{\mathcal{F}}_{,A^i \bar{A}^{\bar{\jmath}}}) F^i \bar{F}^{\bar{\jmath}} + F^i W_{,i}  + \bar{F}^{\bar{\imath}} \bar{W}_{,\bar{\imath}}\\
 & \quad  + T_{ij \bar{k} \bar{l}} \left[ (\partial_a A^i  \partial^a A^{j})( \partial_b \bar{A}^{\bar{k}}  \partial^b \bar{A}^{\bar{l}})  - 2 F^i \bar{F}^{\bar{k}} (\partial_a A^{j} \partial^a \bar{A}^{\bar{l}})+ F^i F^j\bar{F}^{\bar{k}} \bar{F}^{\bar{l}} \right]  \ .
\end{aligned}\end{equation}
Even though the above Lagrangian depends on $\partial_a F$, inside the equations of motion for the auxiliary fields this dependence cancels out and the resulting equations are purely algebraic. More precisely the equations of motion for the auxiliary fields read
\begin{equation}\begin{aligned}\label{Rigid_eom_F}
&  \mathcal{F}_{,\bar{A}^{\bar{\imath}} \bar{A}^{\bar{\jmath}} \bar{F}^{\bar{n}} } \left(\partial_a \bar{A}^{\bar{\imath}} \partial^a \bar{A}^{\bar{\jmath}} + \bar{F}^{\bar{\imath}} \Box \bar{A}^{\bar{\jmath}} \right) + (K_{,A^i \bar{A}^{\bar{n}}} + \mathcal{F}_{,A^i \bar{A}^{\bar{n}}}+\bar{\mathcal{F}}_{,A^i \bar{A}^{\bar{n}}}) F^i \\
& + 2 \mathcal{F}_{,\bar{A}^{\bar{\imath}} \bar{F}^{\bar{n}} } \Box \bar{A}^{\bar{\imath}}  + \bar{W}_{,\bar{n}} + \mathcal{F}_{,A^i \bar{A}^{\bar{\jmath}} \bar{F}^{\bar{n}}} F^i \bar{F}^{\bar{\jmath}} + 2 T_{ij\bar{k}\bar{l}} F^i \left(F^j\bar{F}^{\bar{l}}- \partial_a A^j \partial^a \bar{A}^{\bar{l}}\right) \\
& + T_{ij \bar{k} \bar{l},\bar{F}^{\bar{n}}} \left[ (\partial_a A^i  \partial^a A^{j})( \partial_b \bar{A}^{\bar{k}}  \partial^b \bar{A}^{\bar{l}})  - 2 F^i \bar{F}^{\bar{k}} (\partial_a A^{j} \partial^a \bar{A}^{\bar{l}})+ F^i F^j\bar{F}^{\bar{k}} \bar{F}^{\bar{l}} \right] = 0 \ .
\end{aligned}\end{equation}
In the next section we will analyse the equations of motion for the $F^i$ and the scalar potential more closely.
\subsection{On-shell Action and Effective Field Theory}\label{on-shell-eft}
So far we have discussed superspace actions for the general scalar potential which are off-shell by construction. Ultimately, one is interested in the on-shell action and, hence, in the solution to the equations of motion for the auxiliary fields. In this section we study the equations of motion for the auxiliaries in the context of effective field theory and, in particular, clarify the apparent presence of multiple on-shell theories. This study was already initiated in \cite{Ciupke:2015msa} for a special example and here we generalize the respective findings.

In the following, for the sake of simplicity and brevity we set $n_c = 1$ and perform the discussion using the Lagrangian in eq.~\eqref{L_eff_rigid}.The fact that this Lagrangian may include kinetic terms for $F$ is of no relevance here. We conduct a brief separate discussion of these kinetic terms in sec.~\ref{prop_aux}. To emphasize the difference between terms which are already present in the two-derivative theory and those are induced by the higher-derivatives we split $\mathcal{K}$ up as follows 
\begin{equation}\label{K=K+F}
 \mathcal{K}(\Phi,\bar{\Phi},\Psi,\bar{\Psi}) = K(\Phi,\bar{\Phi}) + \mathbb{F}(\Phi,\bar{\Phi},\Psi,\bar{\Psi}) \ ,
\end{equation}
such that $\mathbb{F}$ is at least linear in $\Psi$ and/or $\bar{\Psi}$.\footnote{The object $\mathbb{F}$ was already introduced in \cite{Buchbinder:1994iw, Buchbinder:1995ideas} and denoted as the effective auxiliary field potential (EAFP).} The respective off-shell scalar potential in eq.~\eqref{L_eff_rigid} reads
\begin{equation}\label{potgenHD}
 V = -K_{,A \bar{A}} \rvert F \rvert^2 - F W_{,A} - \bar{F} \bar{W}_{,\bar{A}} - \rvert F \rvert^2 \mathbb{F}_{,A\bar{A}}(A,\bar{A},\bar{F},F) \ .
\end{equation}
The on-shell scalar potential is obtained by determining the solution to the equation of motion for $F$, evaluating the solution at $\partial_a A = \partial_a \bar{A} = 0$ and finally inserting the result into the expression above. As we already mentioned we have to integrate out $F,\bar{F}$ even if a kinetic term for these fields is present. Thereby, terms proportional to $\partial_a F, \partial_a \bar{F}$ only induce additional kinetic operators for $A,\bar{A}$ after solving the equations of motion for $F,\bar{F}$ and, hence, do not have to be taken into account for the discussion of the scalar potential. To this end we can simply ignore the kinetic contributions inside the equations of motion for $F$ and $\bar{F}$. In particular, the equations of motion for $\bar{F}$ then read
\begin{equation}\label{eomaux}
 K_{,A \bar{A}} F + \bar{W}_{,\bar{A}} + F\mathbb{F}_{,A\bar{A}}(A,\bar{A},\bar{F},F) + \rvert F \rvert^2 \mathbb{F}_{,A\bar{A}\bar{F}}(A,\bar{A},\bar{F},F) = 0 \ .
\end{equation}
Contrary to the situation of the ordinary two-derivative theory where the equation of motion for $F$ is just a linear algebraic equation, in the general higher-derivative theory we have to deal with a general algebraic equation, which does not even have to be polynomial. This is due to the fact that in general $\mathbb{F}$ can be given by an infinite power series in $\Psi$ and $\bar{\Psi}$. However, in a local theory this sum has to be finite and therefore an upper bound on the number of derivatives must exist, let us denote it as $2N$, such that
\begin{equation}\label{Fbb}
 \mathbb{F} = \sum_{1 \leq n+m \leq N} \mathcal{T}_{nm}(\Phi, \bar{\Phi}) \Psi^n \bar{\Psi}^m \ ,
\end{equation}
where we assume that also the coefficient functions $\mathcal{T}_{ij}$ are truncated appropriately. When inserting eq.~\eqref{Fbb} into eq.~\eqref{eomaux}, we observe that the equations of motion for $F$ allow for up to $N+1$ solutions and, hence, we obtain up to $N+1$ distinct on-shell Lagrangians. In general these Lagrangians describe inequivalent dynamics for the scalar field (and chiral fermion) and, thus, we loose predictability of the classical dynamics. In \cite{Ciupke:2015msa} this issue was addressed for a special case and a unique physical on-shell theory was identified by demanding that the on-shell theory is analytic in the couplings of off-shell higher-derivative operators.\footnote{Additional insight into this problem can be gained in top-down situations \cite{Ciupke:2015msa}: In general one expects that the underlying UV theory for a given EFT is non-local and, thus, involves infinitely many higher-derivative operators.
This UV-theory should allow only for a single on-shell theory. In the local low-energy EFT we truncate this series and, thus, obtain multiple on-shell theories. The UV on-shell Lagrangian should consist of an infinite series of contributions with a certain correspondence to the infinite tower of higher-derivatives. This series can be truncated as well and should be reproduced at the appropriate order by one of the on-shell theories from the EFT-multiplet. In \cite{Ciupke:2015msa} this was shown to be the case for the example of the one-loop Wess-Zumino model \cite{Kuzenko:2014ypa}.} The same line of reasoning can directly be applied to the general case in eq.~\eqref{eomaux}.

In the following, for illustrative purposes we analyse the behaviour of the different on-shell Lagrangians taking a bottom-up perspective. Our findings support and emphasize the arguments put forward in \cite{Ciupke:2015msa}. Let us assume our theory is an effective field theory valid up to some cut-off scale $\Lambda$ and that all operators consistent with the symmetries are present. The relevant mass-dimensions of the fields and quantities in the Lagrangian are given by
\begin{equation}\begin{aligned}
 & [A] = [\bar{A}] = \Lambda \ , \qquad [F] = [\bar{F}] = \Lambda^2 \ , \\
 & [K] = [\mathbb{F}] = \Lambda^2 \ , \qquad [W] = [\bar{W}] = \Lambda^3 \ .
\end{aligned}\end{equation}
It is convenient to expand $K,\mathbb{F}$ and $W$ in inverse powers of this cut-off scale. The lowest order terms of this expansion read
\begin{equation}\begin{aligned}\label{exp_of_KWF}
 K(A,\bar{A}) &= \, \rvert A \rvert^2 + \mathcal{O}(\Lambda^{-1})  \\
 \mathbb{F}(A,\bar{A},\bar{F},F) &= \frac{\rvert A \rvert^2}{\Lambda^2}( T F +  \bar{T} \bar{F}) +  \dots + \mathcal{O}(\Lambda^{-3})   \\
 W(A) &= \Lambda^3 \left[ w_0 \frac{A}{\Lambda} + \frac{w_1}{2} \frac{A^2}{\Lambda^2} + \frac{w_2}{3} \frac{A^3}{\Lambda^3} + \mathcal{O}(\Lambda^{-4}) \right]\ ,
\end{aligned}\end{equation}
where for brevity we displayed only those leading order terms which contribute to the scalar potential. For instance, we did not explicitly display operators of order $\mathcal{O}(\Lambda^{-1})$ and $\mathcal{O}(\Lambda^{-2})$ in $\mathbb{F}$ which do not contribute in eq.~\eqref{eomaux} but modify the kinetic terms of the on-shell Lagrangian. We observe that when performing the limit $\Lambda \rightarrow \infty$ in the above off-shell Lagrangian, we recover the IR renormalisable two-derivative theory described by a canonical K\"ahler potential and a superpotential with at most cubic terms, just as demanded by the decoupling principle. Let us now analyse the behaviour of the respective on-shell theories.

After inserting eq.~\eqref{exp_of_KWF} into eq.~\eqref{eomaux} we expand the equation of motion in powers of $\Lambda$ and this expansion starts at order $\Lambda^2$. Therefore, the formal solution of eq.~\eqref{eomaux} can be written as 
\begin{equation}
 F = \Lambda^2 \sum_{n=0} \frac{F_{(n)}}{\Lambda^n} \ , 
\end{equation}
where $F_{(n)}$ has mass dimension $n$.  After inserting the above form of $F$ into eq.~\eqref{eomaux} we find that at leading order in $\Lambda$ we have to solve the following equation  
\begin{equation}\label{eomF0}
 F_{(0)} + w_0 + F_{(0)}\mathbb{F}_{,A\bar{A}}(0,0,\bar{F}_{(0)},F_{(0)}) + \rvert F_{(0)} \rvert^2 \, \mathbb{F}_{,A\bar{A}\bar{F}}(0,0,\bar{F}_{(0)},F_{(0)}) = 0 \ .
\end{equation}
This equation already suffices to conceptually understand the behaviour of the plethora of solutions to eq.~\eqref{eomaux}. For the remainder of this discussion we distinguish theories with and without tadpole-like terms in the superpotential:
\begin{itemize}
 \item Suppose that $w_0 \neq 0$, such that the ordinary two-derivative theory includes constant and tadpole-like terms in the on-shell Lagrangian. Then all $(N+1)$ solutions to eq.~\eqref{eomF0} are in general non-zero and the higher-derivatives induce $\mathcal{T}_{nm}(0,0)$-dependent corrections to the constant and tadpole term in the on-shell Lagrangian. In fact, every off-shell higher-derivative, regardless of the number of superspace-derivatives, contributes to these terms. Thus, we have a sensitivity to an infinite number of operators. 
 \item Let us now look at the theories with $w_0 = 0$. In this case eq.~\eqref{eomF0} admits one solution $F_{(0)} = 0$ as well as up to $N$ additional solutions with $F_{(0)} \neq 0$. For $F_{(0)} = 0$ we find a scalar potential which agrees with the scalar potential of the two-derivative theory up to operators of mass dimension three. More precisely, it reproduces the quadratic term of the two-derivative theory, but induces $\mathcal{T}_{nm}$-dependent corrections starting at cubic order in $A$ and $\bar{A}$ and, thus, can be regarded as contributing towards the on-shell Lagrangian at sub-leading order. The remaining $N$ solutions with $F_{(0)} \neq 0$, however, contribute constant and linear terms to the scalar potential, which in the infrared-regime are dominant over the quadratic terms of the two-derivative theory. In this sense they cannot be regarded as correcting the Lagrangian of the ordinary, two-derivative theory at sub-leading order and contradict the fact that the leading order IR-dynamics are captured by $K$ and $W$ and, therefore, violate the decoupling principle. To emphasize this further let us consider a superpotential $W(A) = A^n$ for some $n>3$. Then the scalar potential of the two-derivative theory consists purely of irrelevant mass dimension $(2n-2)$ operators, but the solutions with $F_{(0)} \neq 0$ still induce all possible relevant operators. In particular the IR-dynamics are then governed entirely by the higher-derivative contribution and not by the two-derivative part.
\end{itemize}
In summary, the above analysis shows that there always exists a unique theory given by $F_{(0)} = 0$ which is in agreement with the principles of EFT (unless the theory has spurious tadpole-like terms). The remaining theories are unphysical and should be regarded as artifacts of a truncation of an infinite sum of higher-derivatives. Hence, the physical on-shell theory can be determined by solving the equations of motion for $F$ order by order in the cut-off scale. 
\subsection{Propagating Auxiliary Fields}\label{prop_aux}
Finally, let us return to the possibility that the auxiliary fields obtain a kinetic term. More precisely, when inspecting the Lagrangians in eq.~\eqref{Leffgeneral} and eq.~\eqref{L_eff_rigid} we observe that depending on the signature of $\mathcal{K}_{,F^i \bar{F}^{\bar{\jmath}}}$ the auxiliary fields may constitute propagating degrees of freedom. 
The auxiliary fields are related via supersymmetry to higher-derivative fields given by $\sigma^a \partial_a \chi^i$ and $\Box A^i$. 
The latter are usually ghostlike degrees of freedom. However, in effective field theories they are spurious unphysical degrees of freedom and should be regarded as artifacts of a truncation of an infinite series of higher-derivative operators in the spirit of the previous section \cite{Simon:1990hd}. Thus, we are already inclined to interpret the propagating auxiliaries in the same way. 
Additionally, we now argue that independent of the signature of the kinetic terms the propagating auxiliary fields are generically unphysical in an EFT. This argument was already brought forward in \cite{Baumann:2011nm} for a particular operator of kinetic type, but applies equally in our discussion. For simplicity we consider theories with $n_c = 1$, a generalization to arbitrary $n_c$ follows immediately. Keeping track of the mass dimensions, a generic expansion of $\mathcal{K}$ reads
\begin{equation}\begin{aligned}
 \mathcal{K}(\Phi,\Phi^\dagger,\Psi,\Psi^\dagger) &= K(\Phi,\Phi^\dagger) +  \Bigl(\mathcal{T}_{10} (\Phi,\Phi^\dagger) \Psi  + \frac{1}{\Lambda^2} \, \mathcal{T}_{20}(\Phi,\Phi^\dagger) \Psi^2 + \text{h.c.} \Bigr) \\
 & \quad +  \frac{1}{\Lambda^2} \mathcal{T}_{11}(\Phi,\Phi^\dagger) \lvert \Psi \lvert^2 +  \dots \ ,
\end{aligned}\end{equation}
where the dots indicate higher-order terms in $\Psi, \Psi^\dagger$ and the superfields $\mathcal{T}_{10},\mathcal{T}_{11}$ and $\mathcal{T}_{20}$ are dimensionless. The last term on the r.h.s.~yields the first contribution to the kinetic term for $F$. In the component version given in eq.~\eqref{L_eff_rigid} it reads
\begin{equation}\label{eq:kin_F_terms}
 \mathcal{L}_{\text{eff}} \supset - \frac{1}{\Lambda^2} \mathcal{T}_{11}(\Phi,\Phi^\dagger) \partial_a F \partial^a \bar{F} + \dots \ .
\end{equation}
Independently of the details of $\mathcal{T}_{11}$ the canonically normalized scalar field, let us denote it by $\tilde{F}$, therefore picks up a factor of $1/\Lambda$. At leading order in the expansion of $\mathcal{K}$ in $\Psi$ and $\Psi^\dagger$ this reads $\Lambda \tilde{F} \sim F$. Thus, when recasting the scalar potential as given in eq.~\eqref{potgenHD} in terms of $\tilde{F}$ we pick up an additional factor of $\Lambda^2$ in the terms that are at least quadratic in $F$ and $\bar{F}$ respectively. Generically, we therefore expect $\tilde{F}$ to have a mass of order $\Lambda$. This is in contradiction with the assumption that we are dealing with a low-energy effective field theory describing physics below the cut-off $\Lambda$ and indicates that $\tilde{F}$ is not a physical degree of freedom and should be integrated out. Hence, we must continue to treat the auxiliary fields as algebraic degrees of freedom. In this case the kinetic terms in eq.~\eqref{eq:kin_F_terms} yield kinetic and higher-derivative terms for $A,\bar{A}$ after eliminating $F,\bar{F}$ via their respective equations of motion. 

We now end the discussion of the effective scalar potential in global supersymmetry and turn to the case of local supersymmetry.
\section{Higher-Derivatives for Chiral Matter in $\mathcal{N}=1$ Curved Superspace}\label{Higher_deriv_sugra}
\subsection{Superfields in Curved Superspace}\label{sec:curved_superspace}
So far we have considered theories with global $\mathcal{N}=1$ supersymmetry which are formulated in flat superspace. We now turn to $\mathcal{N}=1$ supergravity which is the correct framework to study effective actions obtained from string compactifications. Before we can discuss higher-derivative theories of supergravity we first have to establish the basics conventions and notation, which will be the subject of this section. In the next section we then review the construction of the two-derivative action for supergravity coupled to chiral multiplets. Afterwards, we turn to the investigation of higher-derivative operators.

Over the years several formalisms to construct supergravity actions have been engineered including superspace-techniques as well as superconformal methods. Even though more general and elegant versions of curved superspace, such as $U(1)$-superspace \cite{Binetruy:2000zx} or conformal superspace \cite{Butter:2009cp}, exist, we continue to use the ordinary Wess and Bagger superspace-formalism and, hence, adopt the conventions and notations of \cite{Wess:1992cp}. This has the advantage that we can directly compare our results to the existing literature on higher-derivative supergravity, in particular to \cite{Baumann:2011nm}, but also to the results of the rigid theory in the preceding sections. The superspace-formulation of supergravity is highly reminiscent of the construction of ordinary gravity and involves studying the differential geometry of curved superspace. In this section we begin by reviewing the formalism and basic notions of the differential geometry of curved superspace along the lines of \cite{Wess:1992cp}.

We choose curved superspace to be locally parametrized by the variables
\begin{equation}
 z^M = (x^m, \theta^\mu, \bar{\theta}_{\dot{\mu}})  \ , \qquad m=0,\dots,3 \ , \qquad \mu , \dot{\mu} = 1,2 \ .
\end{equation}
In the following $m,n,\dots$ denote curved spacetime indices and $\mu,\nu, \dots$ ($\dot{\mu},\dot{\nu}, \dots$) curved Grassmannian indices. The convention of summing over superspace indices reads
\begin{equation}
 \mathrm{d} z^M \omega_M = \mathrm{d} x^m \omega_m + \mathrm{d} \theta^\mu \omega_\mu + \mathrm{d} \bar{\theta}_{\dot{\mu}} \omega^{\dot{\mu}} \ .
\end{equation}
The geometry of curved superspace is described by a super-vielbein $E_M^A$ together with a connection $\Omega_{MA} {}^B$. The vielbein converts curved superspace indices to local flat superspace indices. Moreover, the connection allows us to introduce a super-covariant derivative, which for instance acts on a vector field $V^A$ as
\begin{equation}
 \mathcal{D}_M V^A = \partial_M V^A + (-1)^{mb}V^B \Omega_{MB} {}^A \ ,
\end{equation}
where $m,b$ take values $0 (1)$ if $M,B$ are vector (spinor) indices. This covariant derivative is the curved superspace analogue of the flat superspace derivative in eqs.~\eqref{flat_superspace_der_1},~\eqref{flat_superspace_der_2}. Naturally, we can define a torsion
\begin{equation}\label{torsion}
 T_{MN} {}^A = \mathcal{D}_N E_M^A - (-1)^{nm} \mathcal{D}_M E_N^A \ ,
\end{equation}
and, similarly, the super-curvature tensor
\begin{equation}\begin{aligned}\label{super_Riemann}
 R_{NMA} {}^B &= \partial_N \Omega_{MA} {}^B - (-1)^{nm} \partial_M \Omega_{NA} {}^B + (-1)^{n(m+a+c)}\Omega_{MA} {}^C \Omega_{NC} {}^B \\
 & \quad- (-1)^{m(a+c)}\Omega_{NA} {}^C \Omega_{MC} {}^B \ .
\end{aligned}\end{equation}
The torsion and curvature are the relevant objects required to express any geometric quantity of curved superspace. In particular they determine the curved superspace analogue of the (anti-) commutation relations in eq.~\eqref{anticom}. More precisely, we have
\begin{equation}\label{curved_anticom}
 (\mathcal{D}_C \mathcal{D}_B - (-1)^{bc} \mathcal{D}_B \mathcal{D}_C ) V^A = (-1)^{d(c+b)} V^D R_{CBD} {}^A - T_{CB} {}^D \mathcal{D}_D V^A \ .
\end{equation}
The algebra of super-covariant derivatives plays an important role later in the discussion of higher-derivatives and we will make extensive use of it. 

In analogy to ordinary gravity, the vielbein and the connection and, hence, the torsion and curvature describe the gravitational degrees of freedom. These objects contain a large number of component superfields, which can be reduced to a minimal set of superfields by imposing constraints on the torsion. These constraints have to be chosen, such that they reproduce the flat SUSY algebra in the rigid limit and such that they allow to consistently define covariantly chiral superfields. Here we follow the conventions of old minimal supergravity, see \cite{Wess:1992cp} for the constraints on the torsion. The next step involves solving the respective Bianchi identities for the curvature and torsion. Certain components of the torsion and curvature remain non-vanishing, some of which are displayed in appendix~\ref{component_identities}, and they can be expressed entirely in terms of the vielbein as well as the following superfields
\begin{equation}\label{superfields_supergravity}
 R \ , \quad \bar{R} \ , \quad G_a \ , \quad W_{\alpha \beta \gamma} \ , \quad \bar{W}_{\dot{\alpha} \dot{\beta} \dot{\gamma}} \ .
\end{equation}
The superfields $R$ and $W_{\alpha \beta \gamma}$ are covariantly chiral, that is they satisfy
\begin{equation}
 \bar{\mathcal{D}}_{\dot{\alpha}} R =  \bar{\mathcal{D}}_{\dot{\alpha}}W_{\alpha \beta \gamma} = 0 \ ,
\end{equation}
while $\bar{R}$, $\bar{W}_{\dot{\alpha} \dot{\beta} \dot{\gamma}}$ denote their conjugate superfields, which are covariantly anti-chiral. Moreover, $G_a$ is a real superfield $\bar{G}_a = G_a$. Note also that $W_{\alpha \beta \gamma}$ and $\bar{W}_{\dot{\alpha} \dot{\beta} \dot{\gamma}}$ are completely symmetric tensors.

As explained in \cite{Wess:1992cp} one may utilize the gauge symmetry to partially gauge-fix the superfields in eq.~\eqref{superfields_supergravity} as well as the super-vielbein. For instance, we may fix the higher components of $E_M^A$, but leave the $\theta = \bar{\theta}= 0$ component unfixed. This procedure yields
\begin{equation}
 E_M^A \rvert_{\theta = \bar{\theta}=0} = \begin{pmatrix}
e_m^a & 
\tfrac{1}{2} \psi_m^\alpha& \tfrac{1}{2} \bar{\psi}_{m\dot{\alpha}}  \\
0 & \delta_\mu^\alpha & 0 \\
0 & 0 & \delta^{\dot{\mu}}_{\dot{\alpha}}  \\
\end{pmatrix} \ ,
\end{equation}
where $e_m^a$ denotes the graviton and $\psi_m^\alpha$ the gravitino. Similarly, gauge-fixing the superfields $R$ and $G_a$ leaves only their $\theta = \bar{\theta}= 0$ components as degrees of freedom, which are denoted as
\begin{equation}\label{R_G_a_0}
 R\rvert \,= - \tfrac{1}{6} M \ , \qquad G_a \rvert \,= - \tfrac{1}{3} b_a \ ,
\end{equation}
where from now on we use the convention $R\rvert \, \equiv R\rvert_{\theta = \bar{\theta}=0}$ for any superfield.\footnote{Correspondingly, the lowest component of $W_{\alpha \beta \gamma}$ is given as a symmetrised combination of the gravitino \cite{Wess:1992cp}.} In eq.~\eqref{R_G_a_0} we find a complex auxiliary scalar $M$ as well as a real auxiliary vector $b_a$. These auxiliary fields are necessary in order to match the off-shell counting of bosonic and fermionic degrees of freedom. Altogether the gravitational multiplet encompasses the component fields $(e_m^a, \psi_m^\alpha,M,b_a)$.

In the following we are interested in the coupling of covariantly chiral multiplets to supergravity. In analogy to flat superspace, covariantly chiral multiplets are defined by the condition $\bar{\mathcal{D}}_{\dot{\alpha}} \Phi^i = 0$ and have the following components
\begin{equation}\label{ComponentsofPhi}
 A^i = \Phi^i \rvert \ , \qquad \chi_{\alpha}^i = \tfrac{1}{\sqrt{2}} \mathcal{D}_{\alpha} \Phi^i \rvert \ , \qquad F^i = - \tfrac{1}{4} \mathcal{D}^2 \Phi^i \rvert \ ,
\end{equation}
where $\mathcal{D}^2 = \mathcal{D}^\alpha \mathcal{D}_\alpha$. Similarly, we use the notation $\bar{\mathcal{D}}^2 = \bar{\mathcal{D}}_{\dot{\alpha}} \bar{\mathcal{D}}^{\dot{\alpha}}$. It is convenient to introduce a new Grassmann variable $\Theta_\alpha$, such that
\begin{equation}
 \Phi^i = A^i + \sqrt{2} \Theta^\alpha \chi_{\alpha}^i + \Theta^\alpha \Theta_\alpha F^i \ .
\end{equation}
Correspondingly we can introduce a differentiation and integration with respect to $\Theta_\alpha$. In particular we have that
\begin{equation}\label{def_Theta_int}
 \int \mathrm{d}^2 \Theta \, f \equiv  - \tfrac{1}{4} \mathcal{D}^2 f \rvert
\end{equation}
for any superfield $f$. Via the Bianchi identities one can determine the $\Theta$-expansion of the superfield $R$. Displaying only the bosonic terms it reads
 \begin{equation}\label{R}
 R = - \tfrac{1}{6}\left[M + \Theta^2\left(-\tfrac{1}{2}\mathcal{R} + \tfrac{2}{3}\rvert M \rvert^2 + \tfrac{1}{3} b_a b^a - i e_a^m \mathcal{D}_m b^a \right) \right] \ ,
\end{equation}
where $\mathcal{R}$ denotes the scalar (spacetime) curvature. The formalism of this section provides the tools to compute the component form for actions of matter-coupled supergravity. In particular it will be necessary to determine components of superfields with several covariant derivatives. The general rule here is to iteratively apply the (anti-) commutation relations in eq.~\eqref{curved_anticom} until the number of covariant derivatives has reduced enough such that the respective component can be related to already known objects such as the components in eqs.~\eqref{ComponentsofPhi},~\eqref{R}. In appendix~\ref{component_identities} we apply this algorithm to several superfields, whose components are important in the following sections. 
\subsection{Ordinary Matter-coupled Supergravity}\label{sugraaction}
An action for supergravity coupled to chiral matter can be constructed following the methods of flat superspace. We begin by reviewing the two-derivative Lagrangian in order to familiarize ourselves with the derivation of the respective component action. Furthermore, we revert back to some key formulas later on during the discussion of higher-derivative operators. 

Given some scalar superfield $U$, we can construct a super-diffeomorphism invariant action via the integral
\begin{equation}\label{S=EU}
 S_U =  \frac{1}{\kappa^2} \int \mathrm{d}^8 z E U \ ,
\end{equation}
where $E$ denotes the super-determinant of the super-vielbein, $\mathrm{d}^8 z$ the measure on curved superspace and $\kappa = M_{p}^{-1}$ the inverse Planck mass, which in the following we set to one.\footnote{If one is interested in the precise mass-scales appearing, it is straightforward to reintroduce $\kappa$ at any point.} To ensure that $S_{U}$ is real, we have to consider a real scalar superfield $U$. Matter-coupled supergravity is constructed from the following superfield 
\begin{equation}\label{U}
 U_{(0)} = -3 \mathrm{e}^{-K(\Phi,\bar{\Phi})/3} + \frac{W(\Phi)}{2R} +  \frac{\bar{W}(\bar{\Phi})}{2\bar{R}} \ ,
\end{equation}
where $K$ is the K\"ahler potential and $W$ the superpotential. For the purpose of computing the component form of the action it is convenient to rewrite the action by expressing it via a chiral integral. In flat superspace we can replace the $\mathrm{d}^2 \bar{\theta}$ measure by the object $\bar{D}^2$, which projects an arbitrary superfield onto a chiral superfield. Due to the more complicated algebra of derivatives in eq.~\eqref{curved_anticom} the object $\bar{\mathcal{D}}^2$ does not project onto chiral superfields. Instead the proper curved space generalization is given by $(\bar{\mathcal{D}}^2 - 8R)$, which indeed has the property
\begin{equation}\label{chiral_projector}
 \bar{\mathcal{D}}_{\dot{\alpha}} (\bar{\mathcal{D}}^2 - 8R) \mathcal{S} = 0
\end{equation}
for any scalar superfield $S$.\footnote{Eq.~\eqref{chiral_projector} still holds for tensor superfields of the type $\mathcal{S}_{\alpha_1 \dots \alpha_n}$.} Owing to this property the object $(\bar{\mathcal{D}}^2 - 8R)$ is denoted as chiral projector. Hence, to construct an action for supergravity we have to replace the $\mathrm{d}^2 \bar{\theta}$ integration by the chiral projector. Altogether, the action can be rewritten as
\begin{equation}\label{SugraSeff}
 S_{(0)} =  \int \d^4 x \int \d^2 \Theta \, \mathcal{E} \left[ \tfrac{3}{8}(\bar{\mathcal{D}}^2-8R)\mathrm{e}^{-\frac{1}{3} K(\Phi,\bar{\Phi})} +  W(\Phi) \right] + h.c. \ ,
\end{equation}
where the object $\mathcal{E}$, denoted as chiral density, is a chiral superfield and enjoys the expansion
\begin{equation}\label{E}
 \mathcal{E} = e(1-\Theta^2 \bar{M}) \ ,
\end{equation}
where $e$ denotes the determinant of the vielbein $e_m^a$ and we displayed only the bosonic components. 

We now review the derivation of the component version of eq.~\eqref{SugraSeff}. In the following, we ignore the fermionic terms to shorten notation. Using eq.~\eqref{def_Theta_int}, eq.~\eqref{E} as well as the components in eq.~\eqref{ComponentsofPhi} and eq.~\eqref{R} one finds
\begin{equation}\begin{aligned}
  \mathcal{L}_{(0)} / e = - &\tfrac{3}{32} \mathcal{D}^2 \bar{\mathcal{D}}^2 \mathrm{e}^{- K/3} \rvert -\tfrac{3}{32}  \bar{\mathcal{D}}^2 \mathcal{D}^2 \mathrm{e}^{- K/3} \rvert- \tfrac{1}{2} \bar{M}\bar{\mathcal{D}}^2 \mathrm{e}^{- K/3} \rvert - \tfrac{1}{2} M \mathcal{D}^2 \mathrm{e}^{- K/3} \rvert \\ 
 & + W_{,i} F^i + \bar{W}_{,\bar{\jmath}} \bar{F}^{\bar{\jmath}} - W \bar{M} - \bar{W} M + \mathrm{e}^{- K/3}\rvert \left(-\tfrac{1}{2}\mathcal{R}-\tfrac{1}{3}\rvert M \rvert^2 + \tfrac{1}{3} b_a b^a \right) \ .
\end{aligned}\end{equation}To proceed, we have to compute the respective $\theta= \bar{\theta}=0$-components of the superfields which appear in the above Lagrangian.
Furthermore, we observe that $\mathcal{L}_{(0)}$ is not expressed in the Einstein frame. To obtain an Einstein frame action we perform a Weyl-rescaling of the metric as follows
\begin{equation}\label{Weyl_rescaling}
  g_{mn} \longrightarrow \tilde{g}_{mn} =  g_{mn}  \,\mathrm{e}^{- K/3}\rvert \ .
\end{equation}
After replacing the superfields with their component versions the resulting Weyl-transformed Lagrangian reads
\begin{equation}\begin{aligned}\label{L2p}
  \mathcal{L}_{(0)} / e =& - \tfrac{1}{2}\mathcal{R}  -\tfrac{3}{4}e^{2 K/3} \partial_m (e^{- K/3}) \partial^m (e^{- K/3}) +  \tfrac{1}{3} b_a b^a + \text{total derivative} \\ 
 & +  e^{2 K/3}(-\bar{M}W-M\bar{W}+W_i F^i +\bar{W}_{\bar{\jmath}} \bar{F}^{\bar{\jmath}})  +  K_{i\bar{\jmath}} \, e^{ K/3} F^i \bar{F}^{\bar{\jmath}} \\
 & - \tfrac{1}{3} e^{ K/3}(M +  K_{\bar{\jmath}}\bar{F}^{\bar{\jmath}})(\bar{M} +  K_i F^i) -  (K_{i\bar{\jmath}} - \tfrac{1}{3} K_i K_{\bar{\jmath}}) \partial_m A^i \partial^m \bar{A}^{\bar{\jmath}} \\
 & - \tfrac{i}{3} b^a e^m_a (K_i \partial_m A^i -  K_{\bar{\jmath}} \partial_m \bar{A}^{\bar{\jmath}})\ .
\end{aligned}\end{equation}
Next we successively integrate out the auxiliary fields. The equations of motion for the auxiliary vector yield
\begin{equation}\label{b_0}
 b_{(0)}^a = \tfrac{i }{2} \eta^{ab} e^m_b (K_i \partial_m A^i - K_{\bar{\jmath}} \partial_m \bar{A}^{\bar{\jmath}}) \ .
\end{equation}
Inserting this into the component Lagrangian we find
\begin{equation}\begin{aligned}
 \mathcal{L}_{(0)} / e = &- \tfrac{1}{2 } \mathcal{R} - K_{i\bar{\jmath}} \, \partial_m A^i \partial^m \bar{A}^{\bar{\jmath}}  - \tfrac{1}{3} e^{ K/3}(M +  K_{\bar{\jmath}}\bar{F}^{\bar{\jmath}})(\bar{M} +  K_i F^i) \\
 &\quad +  K_{i\bar{\jmath}} \, e^{ K/3} F^i \bar{F}^{\bar{\jmath}} + e^{2 K/3}(-\bar{M}W-M\bar{W}+W_i F^i +\bar{W}_{\bar{\jmath}} \bar{F}^{\bar{\jmath}}) \ .
\end{aligned}\end{equation}
The equations of motion for $M$ are solved by
\begin{equation}\label{M_0}
 \bar{M}_{(0)} = -  ( K_i F^i + 3 \bar{W} e^{ K/3}) \ ,
\end{equation}
which, when inserted back into $\mathcal{L}_{(0)}$, yields
\begin{equation}\begin{aligned}\label{L0}
 \mathcal{L}_{(0)} / e = &- \tfrac{1}{2} \mathcal{R} - K_{i\bar{\jmath}} \, \partial_m A^i \partial^m \bar{A}^{\bar{\jmath}} +  K_{i\bar{\jmath}} \,e^{K/3} F^i \bar{F}^{\bar{\jmath}} +3\lvert W \lvert^2 e^K  \\
 &\quad + e^{2K/3}(D_i W F^i +D_{\bar{\jmath}} \bar{W} \bar{F}^{\bar{\jmath}}) \ .
\end{aligned}\end{equation}
Lastly the equations of motion for the chiral auxiliaries read
\begin{equation}\label{F_0}
 \bar{F}_{(0)}^{\bar{\jmath}} = - e^{K/3} K^{i \bar{\jmath}} D_i W \ , 
\end{equation}
which results in the familiar scalar potential
\begin{equation}\label{V0}
 V_{(0)} = e^K (K^{i \bar{\jmath}} D_i W D_{\bar{\jmath}} \bar{W} - 3 \lvert W \rvert^2) \ ,
\end{equation}
where $D_i W = W_i +  K_i W$ denotes the K\"ahler-covariant derivative.
\subsection{Higher-Derivative Supergravity: Preliminaries}\label{HD_Sugra_overview}
We now proceed to discuss higher-derivative operators in superspace. To shorten notation we consider only a single chiral field $\Phi$ from now. The multi-field generalization is straightforward and can always be performed in the final component Lagrangian. A generic supergravity Lagrangian including higher-derivative operators can be constructed from a superspace-integral of the form in eq.~\eqref{S=EU}. More precisely, we have
\begin{equation}\begin{aligned}\label{LHD}
 \mathcal{L}_{hd} = \int \mathrm{d}^4 \theta E &\Bigl[-3U(\Phi, \bar{\Phi},\mathcal{D}_A \Phi, \mathcal{D}_A \bar{\Phi},\mathcal{D}_A \mathcal{D}_B\Phi,\dots , R , \bar{R}, G_a ,W_{\alpha \beta \gamma} , \bar{W}_{\dot{\alpha} \dot{\beta} \dot{\gamma}} , \mathcal{D}_A R , \dots) \\
  & + \frac{1}{2R} W(\Phi, R, \dots) + \frac{1}{2R^\dagger} \bar{W}(\bar{\Phi},\bar{R},\dots) \Bigr] \ ,
\end{aligned}\end{equation}
where the superpotential $W$ is allowed to depend on those higher-derivative superfields which are chiral and the dots indicate further super-covariant derivatives acting on the relevant superfields. Firstly, let us emphasize the importance of the dependence of $U$ and $W$ on the gravitational superfields $R, \bar{R}, G_a, W_{\alpha \beta \gamma},\bar{W}_{\dot{\alpha} \dot{\beta} \dot{\gamma}}$ and derivatives thereof. Even if we only care about higher-derivatives for the chiral multiplets, these superfields must be included in the Lagrangian as they are related to higher-derivative operators for the multiplets $\Phi, \bar{\Phi}$ via the algebra of super-covariant derivatives. A simple illustrative example is given by the operator $\mathcal{D}^4 \Phi$, which by means of eq.~\eqref{chiral_projector} and, hence, implicitly eq.~\eqref{anticom}, we can rewrite as $ \mathcal{D}^4 \Phi = 8 \bar{R} \mathcal{D}^2 \Phi$.

As before in the global case, we are interested in higher-derivative operators that contribute to the effective scalar potential for chiral multiplets. The result of flat superspace, that the general scalar potential can be derived from the superspace-Lagrangian in eq.~\eqref{Leffgeneral}, does not hold in curved superspace. The reason is the existence of the additional complex auxiliary field $M$, which can give rise to new corrections to the scalar potential.\footnote{Due to the index structure of $b_a$, the auxiliary vector purely couples to kinetic terms and, hence, does not affect the scalar potential.} The complex auxiliary field $M$ is also (and together with the Weyl-rescaling) responsible for the difference between $V_{(0)}$ in eq.~\eqref{V0} and its rigid limit. An attempt to classify these new corrections to the scalar potential would eventually have to overcome the complications that are induced by the algebra of super-covariant derivatives. More precisely, 
for all operators which we discuss in the next sections, the algebra of super-covariant derivatives induces relations between these higher-derivative operators and corrections to the scalar potential.\footnote{There exist operators which do not contribute to $V$ and are obtained by taking linear combinations of the operators which we discuss later on.} As a particular example, consider the (flat superspace) operator $D_a \Phi D^a \bar{\Phi}$. From eq.~\eqref{effectivepot} we know that this operator does not correct the scalar potential. However, the appropriate curved superspace generalization $\mathcal{D}_a \Phi \mathcal{D}^a \bar{\Phi}$ does contribute to the scalar potential as shown in \cite{Baumann:2011nm}.

In conclusion, we will not attempt to derive the most general form of the scalar potential in curved superspace. Nevertheless, let us conjecture a possible Lagrangian that should yield the most general scalar potential. Naively, all we have to do is to allow for a generic dependence on $F$, $\bar{F}$ but also on $M$ and $\bar{M}$. Therefore, we consider the Lagrangian in eq.~\eqref{LHD} with
\begin{equation}\label{conjectured_L}
 U = \mathrm{e}^{-\tfrac{1}{3} K(\Phi, \bar{\Phi}, \Psi, \bar{\Psi}, R , \bar{R})} \ , \qquad \Psi = (\bar{\mathcal{D}}^2 - 8R) \bar{\Phi} \ , \qquad \bar{\Psi} = (\mathcal{D}^2 - 8\bar{R}) \Phi \ .
\end{equation}
Note that we could also choose $K$ to be a function of $\mathcal{D}^2 \Phi, \bar{\mathcal{D}}^2 \bar{\Phi}$ instead of $\Psi, \bar{\Psi}$. This way the dependence of $F$ and $\bar{F}$ is more clean, but the price to pay is that $\mathcal{D}^2 \Phi, \bar{\mathcal{D}}^2 \bar{\Phi}$ are not covariantly (anti-) chiral which in turn complicates the computation of the component action. 
We leave the verification/falsification of the above Lagrangian to future research. Instead we now turn our attention to particular higher-derivative operators. More precisely, the goal of the following sections is to classify the leading order and next-to-leading order higher-derivative operators and determine their component forms. We are also motivated to identify operators which contribute four-derivative terms for the chiral scalar off-shell. These are of particular interest for the analysis in \cite{Ciupke:2015msa} for which we need to know the most general combination of higher-derivate operators of this type.

Let us first demonstrate that we do not have to include higher-derivatives in the superpotential. To see this we rewrite eq.~\eqref{LHD} as a chiral integral in the spirit of eq.~\eqref{SugraSeff}, which yields
\begin{equation}\label{comp_form_easier}
 \mathcal{L}_{hd} = \int \d^2 \Theta \, \mathcal{E} \left[ \tfrac{3}{8}(\bar{\mathcal{D}}^2-8R)U + W \right] + h.c. \ .
\end{equation}
Higher-derivative superfields appearing in the superpotential must necessarily be covariantly chiral. However, an arbitrary chiral superfield $\mathcal{C}$ can always be written as $\mathcal{C} = (\bar{\mathcal{D}}^2 - 8R) \mathcal{S}$ for an appropriate scalar superfield $\mathcal{S}$ and, hence, be absorbed into a term inside $U$. Thus, unless explicitly stated otherwise we discuss higher-derivative operators as contributing to $U$.\footnote{However, it is sometimes convenient to analyse certain operators via corrections to $W$, we will later on turn to explicit examples of this.}

From now on we consider a generic effective supergravity with a cut-off scale $\Lambda \leq M_p$. There are two different situations of interest
\begin{equation}
 (1): \ \Lambda = M_p \ , \qquad (2): \ \Lambda < M_p \ .
\end{equation}
In the first case the effective operators are generated by Planckian physics. For instance they may be induced directly from string theory. The second case could, for example, correspond to an effective supergravity where heavy fields are integrated out whose mass is not much smaller than the Planck mass. In this situation the terms in the higher-derivative Lagrangian are suppressed by $\Lambda$ and/or $M_p$. In the following we are more interested in scenario $(1)$ and, hence, we do not distinguish between $\Lambda$ and $M_p$ any further, but just collectively assume that operators are $\Lambda$-suppressed. Should one be interested in case $(2)$, then the proper mass scales can always be reintroduced at a later stage. In particular, operators involving the gravitational superfields should always be $M_p$-suppressed. 
In conclusion, we expand the superfield $U$ in eq.~\eqref{LHD} in inverse powers of $\Lambda$ and truncate this series at an appropriate order. In effective supergravities descending from string compactifications the K\"ahler potential is typically given by a non-local function. This, in turn, suggests that the couplings of the higher-derivative operators may also be given by non-local functions. Therefore, in the following we distinguish higher-derivative operators only by their dependence on 
\begin{equation}\label{eq:DRGW}
 (\mathcal{D}_A, \quad R, \quad \bar{R}, \quad G_a, \quad W_{\alpha \beta \gamma}, \quad\bar{W}_{\dot{\alpha} \dot{\beta} \dot{\gamma}})  \ ,
\end{equation}
but we explicitly do not distinguish operators that differ by a dependence on $\Phi$ and $\bar{\Phi}$ alone. 
To make this statement more clear, let us define a higher-derivative operator of order $\Delta$ as an operator where the collective mass-dimension of the objects in eq.~\eqref{eq:DRGW} appearing in this operators is given by $\Lambda^{\Delta/2}$. The mass-dimensions of the individual objects read
\begin{equation}\begin{aligned}
 & [\mathcal{D}_\alpha] =  [\bar{\mathcal{D}}_{\dot{\alpha}}] = \Lambda^{1/2} \ , \qquad [\mathcal{D}_a] = [R] = [\bar{R}] = [G_a] = \Lambda \ , \\
 & \qquad  \qquad  \qquad [W_{\alpha \beta \gamma}]=[\bar{W}_{\dot{\alpha} \dot{\beta} \dot{\gamma}}] = \Lambda^{3/2} \ .
\end{aligned}\end{equation}
For the convenience of the reader we now give an outline of the results of the remaining sections. In the following we simplify our discussion by explicitly distinguishing between those operators which induce higher-curvature terms and, hence, higher-derivatives for the gravitational sector and those that do not. Let us from now on refer to the former class of operators as higher-curvature operators. The higher-curvature operators will be the subject of the next section. Afterwards, we turn to the analysis of higher-derivative operators which are not higher-curvature operators. We classify the $\Delta=2$ and $\Delta=4$ operators of this type and determine their component actions in sec.~\ref{sec:two_der_operators} and sec.~\ref{sec:four_derivative_operators}. For the sake of clarity we provide a brief outline in tab.~\ref{tab:survey} which includes references to the respective sections, the number of independent operators which result from our analysis as well as key formulas. 
\begin{table}[!htb]
\onehalfspacing
 \centering
 \begin{tabular}{| l | c | l | l |}\hline
 Class & $\#$ Operators & Section & Form \\ \hline 
 $\Delta=2$ & $2$ & sec.~\ref{sec:two_der_operators} & eq.~\eqref{two_derivative_operators} \\ 
 $\Delta=4$, HC & $3^*$ & sec.~\ref{sec:higher_curvature_ops} & eq.~\eqref{Sprime1} \\
 $\Delta=4$, EH & $9$ & sec.~\ref{sec:four_derivative_operators} & tab.~\ref{list_of_four_superspace_ops} \\ \hline
 \end{tabular}
  \caption{Overview of different classes of operators. Here HC stands for higher curvature operators, EH for operators which induce purely Einstein-Hilbert ordinary gravity. Visible are also the sections in which these classes are discussed and the equations respectively the table in which the form of these operators is displayed. The number of independent higher curvature operators depends on whether matter-coupling is taken into account or not. If not then the super Gauss-Bonnet theorem shown in eq.~\eqref{super_gauss_bonnet} reduces the number of operators further.}
  \label{tab:survey}
\end{table}
\subsection{Higher-Curvature Superspace Operators}\label{sec:higher_curvature_ops}
We begin our analysis by discussing first those higher-derivative operators that include higher-curvature terms in their component forms. Note that these have been studied in the past \cite{PhysRevD.17.3179, Howe1978138, Gates:1983nr, Theisen1986687, PhysRevD.33.2504} and we briefly summarize the essential information on them here. In the later sections, the only exception being appendix~\ref{app:Killing_spinor}, we do not include them in the analysis anymore.

In ordinary gravity the leading order four-derivative corrections to the Einstein-Hilbert term are of the form \cite{Stelle:1976gc, Stelle:1977ry}
\begin{equation}\label{S1}
 S_{\mathcal{R}^2} = \int \mathrm{d}^4 x \, e  \left(\lambda_1 \mathcal{R}^2 + \lambda_2 \mathcal{R}_{mn} \mathcal{R}^{mn} + \lambda_3 \mathcal{R}_{mnpq} \mathcal{R}^{mnpq} + \lambda_4 \Box \mathcal{R} \right) \ ,
\end{equation}
where $\mathcal{R}_{mn}$ and $\mathcal{R}_{mnpq}$ denote the (spacetime-) Ricci and Riemann tensors. The last term is a total derivative and, hence, can be ignored here. Using the Gauss-Bonnet theorem, which relates a particular linear combination of the above operators to a topological invariant, one can in fact simplify the four-derivative action to include merely the $\mathcal{R}^2$ and $\mathcal{R}_{mn} \mathcal{R}^{mn}$ operators. A generalization of $S_{\mathcal{R}^2}$ to supergravity is given by the action \cite{Theisen1986687}
\begin{equation}\label{Sprime1}
 S'_{\mathcal{R}^2} = \int \mathrm{d}^8 z E \left(c_1 R \bar{R} + c_2 G^a G_a +  \frac{c_3}{R} W_{\alpha \beta \gamma} W^{\alpha \beta \gamma} + h.c. \right) \ .
\end{equation}
The component version of the first operator includes $\mathcal{R}^2$-terms, while that of the second includes $\mathcal{R}^2$- and $\mathcal{R}_{mn} \mathcal{R}^{mn}$-terms and that of the third is given by the square of the Weyl-tensor.\footnote{A matter-coupled version of eq.~\eqref{Sprime1} was investigated in \cite{PhysRevD.33.2504}.} In particular, one can demonstrate the superspace-version of the Gauss-Bonnet theorem \cite{Howe1978138, Gates:1983nr, Theisen1986687}
\begin{equation}\label{super_gauss_bonnet}
 \int \mathrm{d}^8 z E \left(16 R \bar{R} + 8 G^a G_a +  \frac{2}{R} W_{\alpha \beta \gamma} W^{\alpha \beta \gamma} + h.c. \right) = 32 \pi^2 \chi \ ,
\end{equation}
where $\chi$ denotes the Euler number.

Furthermore, one may also consider operators with derivatives of the gravitational superfields, such as $\mathcal{D}^2 R$ and $\mathcal{D}_a G^a$.\footnote{In fact these two operators are related via the identity $\mathcal{D}^2 R - \bar{\mathcal{D}}^2 \bar{R} = 4 i \mathcal{D}_a G^a $.} However, these operators do not include higher-curvature terms in their respective component expressions. We return to the discussion of this type of operators in the context of higher-derivative operators for the covariantly chiral matter, in particular in appendix~\ref{appendix:integration_by_parts}.
\subsection{Non-Minimal Coupling and Integrating Out Fields}\label{HCTIOF}
Next we investigate an important conceptual question which naturally arises in the context of effective field theories with non-minimal couplings to gravity and higher-curvature terms. More precisely, we discuss the difference between those higher-curvature terms which arise off-shell and those that arise by integrating out heavy propagating fields or auxiliary fields. So far the higher-curvature superspace operators, in particular those in eq.~\eqref{Sprime1}, induce higher-curvature terms purely off-shell. These have to be contrasted to those operators which induce higher-curvature terms after integrating out auxiliary fields. To illustrate what this means, consider as a first example a (not necessarily supersymmetric) theory of a collection of scalar fields $\phi_1, \dots, \phi_n$ coupled to gravity subject to the Lagrangian
\begin{equation}\label{ToyL}
 \mathcal{L}/e = - \tfrac{1}{2} \mathcal{R} + \tilde{\mathcal{L}}(\phi_1, \dots, \phi_n) \ .
\end{equation}
Now, suppose one of the scalars, say $\phi_1$, has a mass $M$ much larger than the masses of the remaining scalars, that is $M \gg m_2 , \dots, m_n$, and we want to integrate it out to obtain an effective theory for physics at scales much smaller than $M$. We can integrate $\phi_1$ out in any frame of our choice, such as the Einstein frame or any other frame, in which $\phi_1$ couples non-minimally to $\mathcal{R}$. However, while integrating out in the Einstein-frame results in a Lagrangian of the form
\begin{equation}
 \mathcal{L}_{EFT}/e = - \tfrac{1}{2} \mathcal{R} +  \mathcal{L}_{\text{effective}} (\phi_2, \dots, \phi_n) \ ,
\end{equation}
in another frame with non-minimal coupling between $\phi_1$ and $\mathcal{R}$ we find
\begin{equation}
 \mathcal{L}_{EFT}'/e = f(\mathcal{R}) + \mathcal{L}_{\text{effective}}' (\mathcal{R}, \phi_2, \dots, \phi_n) \ ,
\end{equation}
for some particular function $f(\mathcal{R})$. It is well-known that such a theory of gravity can be recast into the form of an Einstein-Hilbert term minimally coupled to a real scalar by performing a Weyl-transformation. One might wonder, how this additional degree of freedom emerged, since all we did was to choose a different frame prior to integrating out $\phi_1$. In fact, this additional degree of freedom is nothing but $\phi_1$, which is reintroduced into the spectrum and by performing the Weyl-transformation we retain the original theory in eq.~\eqref{ToyL}. This example shows that degrees of freedom should always be integrated out in the Einstein-frame, as otherwise one might obtain $f(\mathcal{R})$ theories, which are merely dual descriptions of the to-be-integrated-out degrees of freedom.

Let us now analyse a second example, where the previous issue arises in the context of integrating out auxiliary fields, and which is relevant in the discussion of higher-derivative supergravity. Consider the following action
\begin{equation}\label{S_F}
 S_g = \int \mathrm{d}^8 z E (g(R) + \bar{g}(\bar{R})) \ .
\end{equation}
At the superspace level one can demonstrate that this action is equivalent to an ordinary Einstein-Hilbert superspace action coupled to a covariantly chiral superfield $\Sigma$ \cite{CECOTTI198786, Hindawi:1995qa}, thus, displaying a superspace generalization of the duality between $f(\mathcal{R})$ theories of gravity and Einstein-Hilbert gravity coupled to a real scalar field. This equivalence can also be understood at the component level. The off-shell component form of $S_g$ does not contain any higher-curvature terms, but includes a coupling of the auxiliary field $M$ to the scalar curvature. 

On the one hand, we may rewrite the action in the Einstein-frame by performing a Weyl-transformation. Thereby the auxiliary field $M$, since it enters in the Weyl-factor, picks up a kinetic term and, hence, constitutes the complex scalar of the new chiral superfield $\Sigma$. On the other hand, we could choose to integrate out $M$ before performing a Weyl-transformation. This way we obtain a particular $f(\mathcal{R})$-theory. Both procedures agree with each other after using the duality between $f(\mathcal{R})$ theories and ordinary general relativity coupled to a real scalar. In this sense we have a situation similar to the previous example in eq.~\eqref{ToyL}, i.e.~higher-curvature terms emerge when auxiliary fields are integrated out in a frame in which they are non-minimally coupled to $\mathcal{R}$. Again we interpret these $f(\mathcal{R})$-theories as dual descriptions, which encode the dynamics of the auxiliary fields. At this point we could stop our analysis if these propagating auxiliary fields would be part of the physical spectrum. However, quite similar to the discussion in sec.~\ref{prop_aux}, these propagating auxiliaries are, despite having the correct sign of the kinetic terms, unphysical in the context of effective field theory as they generically have a mass at the cutoff-scale of the EFT.\footnote{Note also that the fermionic component of $\Sigma$ is given by a higher-derivative of the gravitino \cite{Hindawi:1995qa}, which constitutes an additional degrees of freedom via the standard Ostrogradski procedure. This should alert us, that care has to be taken with the proper interpretation of our theory. As we have mentioned, in the EFT-context degrees of freedom that are associated with (usually ghost-like) higher-derivatives are unphysical since they emerge from truncating a more fundamental non-local theory that contains an infinite series of higher-derivatives.} We demonstrate this explicitly in appendix~\ref{More_on_prop_aux}.

Altogether, we conclude that we should avoid the description of (unphysical) propagating auxiliary degrees of freedom via on-shell higher-curvature terms. This can most conveniently be done by choosing to perform a Weyl-transformation to the Einstein frame before any auxiliary field is integrated out. 
\subsection{Component Form of Operators and On-Shell Results}\label{sec:component_form}
After discussion of the former conceptual points we now turn to the analysis of those higher-derivative operators which do not induce higher-curvature terms. To begin with it is necessary to introduce some computational tools and formulae, which comprise the topic of this section. In particular, we present an algorithm that allows to determine the final on-shell component version of a given operator. In the upcoming sections we then apply these formulae to determine the component forms of the $\Delta=2$ and $\Delta=4$ higher-derivative operators. 

In the following we consider a particular higher-derivative operator $\mathcal{O}$ coupled to the general two-derivative theory given in eq.~\eqref{SugraSeff}. To this end we regard the following Lagrangian
\begin{equation}\begin{aligned}\label{L_total}
 \mathcal{L}_{\mathcal{O}} = \mathcal{L}_{(0)} + \hat{\mathcal{L}}_{\mathcal{O}} \ , \qquad \text{where} \qquad \hat{\mathcal{L}}_{\mathcal{O}}  &= \frac{3}{4} \int \d^2 \Theta \,   \mathcal{E} (\bar{\mathcal{D}}^2-8R) \, \mathcal{O} + h.c. \ .
\end{aligned}\end{equation}
To reduce the computational effort for determining the component Lagrangian of a specific operator, it is convenient to note that \cite{CREMMER1979105}
\begin{equation}\label{O=Odagger}
  \int \d^2 \Theta \, \mathcal{E} (\bar{\mathcal{D}}^2-8R) \mathcal{O} + h.c. = \int \d^2 \Theta \, \mathcal{E} (\bar{\mathcal{D}}^2-8R) \mathcal{O}^\dagger + h.c. + \text{total derivative} \ .
\end{equation}
In turn, this implies that it is not necessary to consider $\mathcal{O}$ as a real operator in eq.~\eqref{L_total}, but that instead it is sufficient to consider a complex operator $\mathcal{O}$. Therefore, in the following we always simply use complex operators $\mathcal{O}$ without adding $\mathcal{O}^\dagger$. With this in mind, we rewrite the higher-derivative contribution in eq.~\eqref{L_total} as
\begin{equation}\begin{aligned}\label{Su}
  \hat{\mathcal{L}}_{\mathcal{O}} / e 
 = - &\tfrac{3}{16} \mathcal{D}^2 \bar{\mathcal{D}}^2 \mathcal{O} \rvert - \tfrac{3}{4} \bar{M} \bar{\mathcal{D}}^2 \mathcal{O}\rvert - \tfrac{1}{4} M \mathcal{D}^2 \mathcal{O}\rvert \\
 &  + \mathcal{O}\rvert \left(-\tfrac{1}{2}\mathcal{R} - \tfrac{1}{3} \rvert M \rvert^2 + \tfrac{1}{3} b_a b^a - i \mathcal{D}_m b^m \right) + h.c. \ .
\end{aligned}\end{equation}
Next let us provide an algorithm to determine the respective component Lagrangian, which respects the principles of EFT.
\begin{enumerate}
 \item We begin by computing the respective components version of the objects appearing within eq.~\eqref{Su}.
 \item Perform the Weyl-rescaling of the spacetime-metric. Depending on the operator $\mathcal{O}$, the Weyl-rescaling can be affected by the terms in eq.~\eqref{Su} and, thus, differ from the two-derivative version in eq.~\eqref{Weyl_rescaling}.
 \item Then we integrate out the auxiliary fields following the methods explained in sec.~\ref{on-shell-eft},~\ref{prop_aux}. More precisely one has to expand the solution in inverse powers of $\Lambda$ and solve the equations of motion order by order in $\Lambda^{-1}$. If some auxiliary field receives a kinetic term, we continue to treat it as an algebraic degree of freedom, inspired by the results in sec.~\ref{prop_aux} and sec.~\ref{HCTIOF}, and again determine the solution to the equations of motion by applying perturbation theory in $\Lambda^{-1}$. 
 \item It is convenient to truncate the solutions for the auxiliary fields at the highest order in $\Lambda^{-1}$, which appears within eq.~\eqref{Su}. After insertion of the truncated solution back into the Lagrangian, terms which exceeds the maximal mass dimension should be neglected.
\end{enumerate}
Let us make some remarks regarding this algorithm. Firstly, we should emphasize once more the importance of performing the Weyl-rescaling before integrating out the auxiliary fields. This point was already stressed and exemplified in the previous section. Furthermore, in principle we would have to apply this algorithm also to the two-derivative part of the Lagrangian. In particular, this would include an appropriate truncation of the expressions for $K$ and $W$. However, in the context of effective supergravities describing the low-energy 4D dynamics of string compactifications, it is useful to keep $K$ and $W$ arbitrary and, furthermore, to allow higher-derivative operators to be multiplied by arbitrary functions of the chiral fields. 

In the next sections we discuss the $\Delta=2$ and $\Delta=4$ operators. The operators we consider are multiplied by an arbitrary coupling function $T$ which carries a dependence on $\Phi, \bar{\Phi}$ and on the cut-off scale $\Lambda$. More precisely, the coupling carries an overall mass dimension $T \sim \Lambda^{-\Delta/2-\mathcal{I}}$ with $\Delta=2,4$ and $\mathcal{I}$ denotes the mass dimension of the collection of $\Phi, \bar{\Phi}$ which supercovariant derivatives act on. According to the above algorithm it suffices to consider the component Lagrangian for a particular operator only up to order $T$, since terms of order $\mathcal{O}(T^2)$ would receive corrections from operators of higher order. Therefore, in the following we restrict ourselves to determine the on-shell form of these operators only at linear order in $T$. In this case the computation of the on-shell component Lagrangian simplifies considerably, the reason being a special property of the theory in eq.~\eqref{L_total} which we explain now. Generically the presence of $\hat{\mathcal{L}}_{\mathcal{O}}$ in eq.~\eqref{L_total} affects the solutions of the equations of motion of the auxiliary fields. We expand the auxiliary fields in powers of $T$ such that
\begin{equation}\begin{aligned}\label{exp_of_aux}
 b^a &= b_{(0)}^a + b_{(1)}^a + \mathcal{O}(\Lambda^{-\Delta})  \\
  M &= M_{(0)} + M_{(1)} + \mathcal{O}(\Lambda^{-\Delta})  \\
  F &= F_{(0)} + F_{(1)} + \mathcal{O}(\Lambda^{-\Delta})  \ ,
\end{aligned}\end{equation}
where $b_{(0)}^a, M_{(0)}$ and $F_{(0)}$ are given in eqs.~\eqref{b_0},~\eqref{M_0},~\eqref{F_0} and $b_{(1)}^a, M_{(1)}, F_{(1)}$ are linear in $T$ and depend on the details of $\hat{\mathcal{L}}_{\mathcal{O}}$. The aforementioned special property of the theory can now by stated as follows: 
\begin{framed}
 The corrections to the auxiliary field induced by $\mathcal{O}$ cancel out at order $\mathcal{O}(T)$ when inserted in the full Lagrangian $\mathcal{L}_{\mathcal{O}}$. Thus, for the computation of the linearized on-shell action it suffices to insert $b_{(0)}^a, M_{(0)}$ and $F_{(0)}$ in eq.~\eqref{L_total}. 
\end{framed}
Expressed differently, at the level of the linearized on-shell we do not obtain corrections involving $b_{(1)}^a, M_{(1)}$ or $F_{(1)}$. Therefore, the terms of order $\mathcal{O}(T)$ in the on-shell Lagrangian arise only from $\hat{\mathcal{L}}_{\mathcal{O}}$ or via the Weyl-rescaling in $\mathcal{L}_{(0)}$. This statement can be understood from the structure of the ordinary two-derivative Lagrangian. Let us now explicitly demonstrate this for the terms involving the auxiliary vector, the argument follows immediately also for the remaining auxiliary fields. Generically the presence of $\hat{\mathcal{L}}_{\mathcal{O}}$ corrects the Weyl-factor $\Omega$ as follows
\begin{equation}\label{Omega+deltaOmega}
 \Omega = \mathrm{e}^{-K/3} + \Lambda^{-\Delta} \, \delta \Omega(M,\bar{M},b_a,F,\bar{F}, \Phi, \bar{\Phi}) \ ,
\end{equation}
where $\delta \Omega$ depends on the details of the higher-derivative operator $\mathcal{O}$. Now the terms in $\mathcal{L}_{(0)}$ displayed in eq.~\eqref{L2p} that involve the auxiliary vector can, after performing the Weyl-transformation with respect to $\Omega$ given in eq.~\eqref{Omega+deltaOmega}, be rewritten as
\begin{equation}\label{b0b1=0}
 \mathcal{L}_{(0)} \supset \frac{1}{3 \Omega} \mathrm{e}^{-K/3} (b_a b^a - 2b_a b_{(0)}^a) + \mathcal{O}(\delta \Omega)\ ,
\end{equation}
where the $\mathcal{O}(\delta \Omega)$ terms indicate the contributions to the Lagrangian that carry a dependence on the Weyl-factor. These terms appear only when $\delta \Omega$ explicitly depends on $b_a$. Inserting the expansion in eq.~\eqref{exp_of_aux} into eq.~\eqref{b0b1=0} we find that the terms involving $b_{(1)}^a$ precisely cancel and, hence, the only corrections linear in $T$ are obtained via $\Omega$. 

Again let us emphasize that this argument can be made for $M$ and $F$ in precisely the same way. Ultimately, the reason for this cancellation is the quadratic form of $\mathcal{L}_{(0)}$. The above observation greatly simplifies the computation of the on-shell action, since we do not have to determine the solution to the equations of motion for the auxiliary fields, but merely insert $b_{(0)}^a, M_{(0)}$ and $F_{(0)}$ in $\hat{\mathcal{L}}_{\mathcal{O}}$ and in the corrections induced by $\delta\Omega$ in $\mathcal{L}_{(0)}$. In fact, since it is straightforward to obtain the on-shell theories, in the following we display most of the component forms off-shell.
\subsection{Operators of Two Super-Covariant-Derivative Order}\label{sec:two_der_operators}
Following the general discussion of the previous section, we now turn to a systematic study of higher-derivative operators. In the previous sections we introduced the concept of a higher-derivative operator of order $\Delta$. The lowest possible order is $\Delta=2$. There exist three distinct operators we can construct at this order and they read
\begin{equation}\begin{aligned}\label{two_derivative_operators}
 \mathcal{O}_{(1)} &= T \mathcal{D}_\alpha \Phi \mathcal{D}^\alpha \Phi  \ , \qquad \mathcal{O}_{(2)} = T \mathcal{D}^2 \Phi  \ , \qquad \mathcal{O}_{(3)} = T R \ ,
\end{aligned}\end{equation}
where the coupling function $T = T(\Phi/\Lambda, \bar{\Phi}/\Lambda)$ is an arbitrary function of the chiral and anti-chiral superfields and has mass-dimension $T \sim \Lambda^{-1-\mathcal{I}}$ with $\mathcal{I}=2,1,0$ for $\mathcal{O}_{(1)}, \mathcal{O}_{(2)}, \mathcal{O}_{(3)}$. It turns out that, similar to their rigid counterparts, the operators $\mathcal{O}_{(1)}$ and $\mathcal{O}_{(2)}$ are equivalent to each other. This equivalence can be demonstrated via integration by parts identities which we display in appendix~\ref{appendix:integration_by_parts}. We now illustrate the general algorithm of the previous section at the example of $\mathcal{O}_{(1)}$. Later on we also turn to the third operator $\mathcal{O}_{(3)}$.

Let us now follow the general prescription of the previous section to compute the component form of $\mathcal{O}_{(1)}$ together with the ordinary two-derivative Lagrangian $\mathcal{L}_{(0)}$ given in eq.~\eqref{SugraSeff}. Here we explicitly include the kinetic terms into the analysis. Following the algorithm presented in the previous section, as a first step we need to compute the relevant quantities within eq.~\eqref{Su}. The $\theta=\bar{\theta}=0$-component of $\mathcal{O}_{(1)}$ is purely fermionic. Furthermore, we need the following quantities
\begin{equation}\begin{aligned}
 \mathcal{D}^2 \mathcal{O}_{(1)} \rvert  = 16T F^2 \ , \qquad \bar{\mathcal{D}}^2 \mathcal{O}_{(1)}\rvert =  - 16 T  (\partial A )^2 \ ,
\end{aligned}\end{equation}
where we made use of eq.~\eqref{DbDPhi}. Furthermore, it is necessary to determine the $\mathcal{D}^2 \bar{\mathcal{D}}^2$-component of $\mathcal{O}_{(1)}$. Direct computation ignoring fermionic terms yields
\begin{equation}\begin{aligned}
  \mathcal{D}^2 \bar{\mathcal{D}}^2 \mathcal{O}_{(1)} \rvert &= \Bigl( T_{\bar{\Phi}} \bar{\mathcal{D}}^2 \bar{\Phi} (\mathcal{D}^2 \Phi)^2 + 4 T_{\bar{\Phi}} \mathcal{D}^2 \Phi \mathcal{D}_\alpha \bar{\mathcal{D}}^{\dot{\alpha}} \bar{\Phi} \bar{\mathcal{D}}_{\dot{\alpha}} \mathcal{D}^\alpha \Phi - 2 T \mathcal{D}^2 \Phi \mathcal{D}_\alpha \bar{\mathcal{D}}^2 \mathcal{D}^\alpha \Phi \\
 & \quad + 2 T_{\Phi} \mathcal{D}^2 \Phi \bar{\mathcal{D}}^{\dot{\alpha}} \mathcal{D}_\alpha \Phi \bar{\mathcal{D}}_{\dot{\alpha}} \mathcal{D}^\alpha \Phi + 4 T \mathcal{D}^2 \bar{\mathcal{D}}^{\dot{\alpha}} \mathcal{D}_\alpha \Phi \bar{\mathcal{D}}_{\dot{\alpha}} \mathcal{D}^\alpha \Phi  \Bigr) \Bigl\rvert \ .
\end{aligned}\end{equation}
Inserting the component expressions in eqs.~\eqref{DbDPhi},~\eqref{D4_comp} and \eqref{Da_comp} in the above formula we find
\begin{equation}\begin{aligned}\label{DDDbDbO1}
 \tfrac{1}{16} \mathcal{D}^2 \bar{\mathcal{D}}^2 \mathcal{O}_{(1)} \rvert &= - 4 T_{\bar{A}} \lvert F \rvert^2 F   - \tfrac{8}{3} TMF^2 + 8 T \partial_m F \partial^m A + \tfrac{4}{3} T\bar{M}\partial_m A \partial^m A \\
 & \quad   + \tfrac{16}{3}i TF b_a e_a^m \partial_m A  + 8 T_{\bar{A}} F \partial_m A \partial^m \bar{A} + 4 T_A F \partial_m A \partial^m A   \ .
\end{aligned}\end{equation}
We are now equipped with the necessary quantities and can proceed to determine the overall component form of $\mathcal{O}_{(1)}$. It is convenient to decompose the result as follows
\begin{equation}\label{component_example_1}
 \hat{\mathcal{L}}_{\mathcal{O}_{(1)}} / e = - V_{\mathcal{O}_{(1)}} + \mathcal{L}_{\mathcal{O}_{(1)}}^{\text{(2-der)}} \ ,
\end{equation}
where the individual parts of the Lagrangian are given by
\begin{equation}\begin{aligned}
V_{\mathcal{O}_{(1)}} &= - 4 T M F^2  - 12 \lvert F \lvert^2 T_{\bar{A}} F + \text{h.c.} \ , \\ 
 \mathcal{L}_{\mathcal{O}_{(1)}}^{\text{(2-der)}} &=  8 T [\bar{M} (\partial A)^2 - 3 \partial F \partial A    - 2 i b_m F  \partial^m A ] - 24 T_{\bar{A}} F \lvert \partial A  \lvert^2  - 12 T_A F (\partial A)^2  + \text{h.c.} \ .
\end{aligned}\end{equation}
Following the general prescription of sec.~\ref{sugraaction} let us now determine the linearized on-shell Lagrangian. To begin with we rewrite the Lagrangian in the Einstein frame. Since $\hat{\mathcal{L}}_{\mathcal{O}}$ does not contain couplings to the scalar curvature, the Weyl transformation continues to be given by eq.~\eqref{Weyl_rescaling}. 
The next step consists of integrating out the auxiliary fields. As demonstrated in the previous section, we obtain the linearized on-shell action by simply replacing the auxiliaries within $\hat{\mathcal{L}}_{\mathcal{O}_{(1)}}$ by the solutions of the ordinary two-derivative theory given in eqs.~\eqref{b_0},~\eqref{M_0} and \eqref{F_0}. 
Altogether, the final linearized on-shell Lagrangian reads
\begin{equation}\begin{aligned}\label{O1_component}
 \mathcal{L}_{\mathcal{O}_{(1)}}  &= e (\mathcal{L}_{(0)} - V_{(1)} - \mathcal{L}_{\text{der}} )\ , \\
 V_{(1)} &= 12 \mathrm{e}^{5K/3} (K^{A\bar{A}})^2 ( D W )^2 \Bigl[ K^{A\bar{A}} \bar{D} \bar{W} ( \bar{T}_A - \tfrac{1}{3}\bar{T} K_A) +  \bar{T} \bar{W} \Bigr] + h.c.\\
   \mathcal{L}_{\text{der}} &= 24 \mathrm{e}^{2K/3} K^{A \bar{A}}   \Bigl[ (\partial A)^2 \left( \bar{D} \bar{W} (\tfrac{1}{3}T K_A + \tfrac{1}{2} T_A) + K_{A \bar{A}} \, T \bar{W} \right) + T \bar{D} \bar{W}  \, \Box A\Bigr] + \text{h.c.} \ . 
\end{aligned}\end{equation}
Here we introduced the abbreviation $\Box = \mathcal{D}^m \mathcal{D}_m$ and $DW = W_A + K_A W$ and $\bar{D}\bar{W}$ denote the K\"ahler-covariant derivatives. 
In the above $\mathcal{L}_{(0)}$ is given by the kinetic terms inside eq.~\eqref{L0} together with the scalar potential in eq.~\eqref{V0}. Let us make a few remarks regarding the above result. The linearized Lagrangian contains only two-derivative terms. However, when using the full solution to the equations of motion for $F$ one finds a non-local theory, which in particular includes an infinite sum of higher-derivatives. Due to the mixing between $M$ and $F$ in eq.~\eqref{component_example_1} the linear correction to the scalar potential looks rather involved. Besides the corrections of the type $\lvert F_{(0)}\rvert^2 F_{(0)}$, which survive the rigid limit and can also be inferred from the lowest order contribution to eq.~\eqref{L_eff_rigid}, we, furthermore, find terms of the type $F_{(0)}^2 \bar{W}$. Roughly speaking, these can be read as describing a mixing between the gravitational piece and the F-term piece of the ordinary (two-derivative) scalar potential.

It remains to discuss the operator $\mathcal{O}_{(3)}$ in eq.~\eqref{two_derivative_operators}. 
It is conceptually straightforward to perform the computation and, therefore, we display only the final result here. Altogether, we find the following off-shell component expression
\begin{equation}\begin{aligned}
 \hat{\mathcal{L}}_{\mathcal{O}_{(3)}} / e = - \tfrac{1}{2} \mathcal{R} \Omega_{\mathcal{O}_{(3)}} -  V_{\mathcal{O}_{(3)}} + \mathcal{L}_{\mathcal{O}_{(3)}}^{\text{(2-der)}} \ ,
\end{aligned}\end{equation}
where we introduced the abbreviations
\begin{equation}\begin{aligned}
 \Omega_{\mathcal{O}_{(3)}} &= -\tfrac{1}{3}TM + \tfrac{1}{2} T_{\bar{A}} \bar{F} + h.c. \ , \\
 V_{\mathcal{O}_{(3)}} &= -\tfrac{1}{6} \lvert M \rvert^2( T_{\bar{A}} \bar{F}- \tfrac{1}{3} TM ) - \tfrac{1}{2} T_{A\bar{A}} M \lvert F \rvert^2 + \tfrac{1}{6} T_A M^2 F  + h.c. \ , \\
 \mathcal{L}_{\mathcal{O}_{(3)}}^{\text{(2-der)}} &= (\tfrac{1}{3} b_a b^a - i \mathcal{D}_m b^m)(\tfrac{1}{2} T_{\bar{A}} \bar{F}-\tfrac{1}{3}TM) + \tfrac{1}{2}M T_{\bar{A}\bar{A}} (\partial \bar{A})^2 \ , \\
  &\quad + \tfrac{1}{2} M T_{\bar{A}}(\Box \bar{A} + \tfrac{2}{3} i b^m \partial_m \bar{A})) + h.c. \ .
\end{aligned}\end{equation}
Note that setting $T$ constant reproduces the result obtained in \cite{Hindawi:1995qa}. In particular, we observe that the Einstein-Hilbert term is modified by the presence of $\mathcal{O}_{(3)}$. This, in turn, implies that the Weyl-rescaling is affected. The on-shell Lagrangian can now be obtained readily, but note that now one has to take into account the modified Weyl-factor. The resulting on-shell form is rather lengthy and, hence, we do not display it here.
\subsection{Operators of Four Super-Covariant-Derivative Order}\label{sec:four_derivative_operators}
\begin{table}[htb!]
\onehalfspacing
 \centering
 \begin{tabular}{| l | l | l | l |}
  \hline
  Label & Operator & Real & $\partial^4$-Terms \\\hline
  $\mathcal{O}_{(4 \rvert 2)}$ & $\mathcal{D}_\alpha \Phi \mathcal{D}^\alpha \Phi \bar{\mathcal{D}}_{\dot{\alpha}} \bar{\Phi}  \bar{\mathcal{D}}^{\dot{\alpha}} \bar{\Phi} $ & $\checkmark$ & $\checkmark$ \\ 
  $\mathcal{O}_{(3 \rvert 1)}$ & $\mathcal{D}^2 \Phi \mathcal{D}_\alpha \Phi \mathcal{D}^\alpha \Phi $ &  &  \\
  $\mathcal{O}_{(3 \rvert 3)}$ & $\bar{\mathcal{D}}_{\dot{\alpha}} \mathcal{D}_\alpha \Phi  \mathcal{D}^\alpha \Phi \bar{\mathcal{D}}^{\dot{\alpha}} \bar{\Phi}  $ &  & $\checkmark$ \\
  $\mathcal{O}_{(2 \rvert 1)}$ & $ \mathcal{D}^2 \Phi \bar{\mathcal{D}}^2 \bar{\Phi} $ & $\checkmark$ & $\checkmark$\\
  $\mathcal{O}_{(2 \rvert 2)}$ & $(\mathcal{D}^2 \Phi )^2 $ &  & \\
  $\mathcal{O}_{(2 \rvert 3)}$ & $\mathcal{D}_a \Phi \mathcal{D}^a \bar{\Phi} $& $\checkmark$ & $\checkmark$ \\
  $\mathcal{O}_{(R \rvert 1)}$ &  $R \mathcal{D}_\alpha \Phi \mathcal{D}^\alpha \Phi $ &  & $\checkmark$\\
  $\mathcal{O}_{(R \rvert 2)}$ &  $R \mathcal{D}^2 \Phi$ &  & $\checkmark$\\
  $\mathcal{O}_{(R \rvert 3)}$ &  $R^2  $  &   &\\\hline
 \end{tabular}
  \caption{Particular minimal choice of four-superspace-derivative operators which are mutually distinct and cannot be related to each other. The individual operators are understood as being multiplied by a superfield $T(\Phi,\bar{\Phi})$ and $\bar{T}(\Phi,\bar{\Phi})$ for their conjugate parts. In the last two rows we indicated whether the operator is real- or complex-valued and whether it contributes four-derivatives terms for the chiral scalar in the linearized on-shell Lagrangian.}
  \label{list_of_four_superspace_ops}
\end{table}
For $\Delta=4$, that is at the four superspace-derivative level, the amount of possible operators increases significantly. We perform the explicit classification of these operators in appendix~\ref{appendix:integration_by_parts}. Let us briefly summarize the content of the latter appendix now. We begin by writing down the list of all allowed operators in tab.~\ref{four_der_operators}. However, many operators in this list are redundant and can be recast into combinations of other operators. The main tools to identify equivalences between operators are, on the one hand, the algebra of covariant derivatives in eq.~\eqref{anticom} and, on the other hand, integration by parts identities in curved superspace. These are introduced and explained in detail in appendix~\ref{appendix:integration_by_parts}. We then use these identities to determine a minimal set of mutually inequivalent operators. More precisely, this set contains 9 operators. This minimal set of operators can be chosen in several different but equivalent ways, only the total number of operators is fixed. Here we make a particular choice of these operators which is displayed in tab.~\ref{list_of_four_superspace_ops}. In this list we also indicate whether the operators are real-valued, that is $\mathcal{O} = \bar{\mathcal{O}}$, and whether they induce four-derivative terms for the chiral scalar in the linearized on-shell Lagrangian. 

Operators which induce $\partial^4$-terms off-shell and/or at the linearized on-shell level are of particular interest to us and, therefore, from now on we constrain our discussion to this subclass only. In tab.~\ref{list_of_four_superspace_ops} we find six operators of this type. We determine the component form of all these operators and display the results below. Note that for brevity we omit the details of intermediate results, such as for instance the $\mathcal{D}^2 \bar{\mathcal{D}}^2$-components of the various operators. However, in appendix~\ref{component_identities} we collect all component identities which are required to compute these $\mathcal{D}^2 \bar{\mathcal{D}}^2$-components. In our survey below we indicate which component-identities are necessary for determining $\mathcal{D}^2 \bar{\mathcal{D}}^2 \mathcal{O} \rvert$. For those operators which are real we take the coupling function $T$ to be real-valued as well. For operators which are not real we assume that $T$ is complex-valued.
\subsubsection*{Operator $\mathcal{O}_{(4 \rvert 2)}$:}
This operator was already studied in several papers \cite{Cecotti:1986jy, Koehn:2012ar, Farakos:2012qu}. The respective component version is particularly simple and reads
\begin{equation}
 \tfrac{1}{48} \hat{\mathcal{L}}_{\mathcal{O}_{(4 \rvert 2)}} / e = T \lvert F \rvert^4 + T (\partial A)^2 (\partial \bar{A})^2 - 2T \lvert F \rvert^2  \lvert \partial A \rvert^2 \ .
\end{equation}
\subsubsection*{Operator $\mathcal{O}_{(3 \rvert 3)}$:}
Computation of this operator requires the component identities in eq.~\eqref{D4_comp} and eq.~\eqref{Da_comp}. Altogether, we find
\begin{equation}\begin{aligned}
   \hat{\mathcal{L}}_{\mathcal{O}_{(3 \rvert 3)}} / e &=  - V_{\mathcal{O}_{(3 \rvert 3)}} + \mathcal{L}_{\mathcal{O}_{(3 \rvert 3)}}^{(\text{4-der})}  +   \mathcal{L}_{\mathcal{O}_{(3 \rvert 3)}}^{(\text{2-der})}   \ , \\
   V_{\mathcal{O}_{(3 \rvert 3)}} &= - 16  \lvert F \lvert^2 (T M F + \bar{T} \bar{M} \bar{F})  \ ,\\
  \mathcal{L}_{\mathcal{O}_{(3 \rvert 3)}}^{(\text{4-der})} &= - 24T_{\bar{A}} (\partial A)^2 (\partial \bar{A})^2  - 24 T(\partial A)^2(\Box \bar{A} + \tfrac{2}{3} i b^m \partial_m \bar{A}) + \text{h.c.} \\
   \mathcal{L}_{\mathcal{O}_{(3 \rvert 3)}}^{(\text{2-der})}  &= - 16 TMF  \lvert  \partial A \lvert^2 - 48 T \bar{F} (\partial_m A \partial^m F + \tfrac{2i}{3} F b^m \partial_m A ) \\
   & \qquad - 24 T_A \lvert F \lvert^2 (\partial A)^2 + \text{h.c.}  \ .
\end{aligned}\end{equation}
\subsubsection*{Operator $\mathcal{O}_{(2 \rvert 1)}$:}
This operator additionally requires new components displayed in eq.~\eqref{D6_comp}. The final result reads 
\begin{equation}\begin{aligned}\label{eq:O21}
  \hat{\mathcal{L}}_{\mathcal{O}_{(2 \rvert 1)}} / e &=  - \tfrac{1}{2} \mathcal{R} \Omega_{\mathcal{O}_{(2 \rvert 1)}} - V_{\mathcal{O}_{(2 \rvert 1)}} + \mathcal{L}_{\mathcal{O}_{(2 \rvert 1)}}^{(\text{4-der})} + \mathcal{L}_{\mathcal{O}_{(2 \rvert 1)}}^{(\text{2-der})}    \ , \\
  \tfrac{1}{16} \Omega_{\mathcal{O}_{(2 \rvert 1)}} &= - 2 T \lvert F \lvert^2  \ , \\
  \tfrac{1}{16} V_{\mathcal{O}_{(2 \rvert 1)}} &= 6T_{A\bar{A}} \lvert F \lvert^4 +  4\lvert F \lvert^2 ( T_A M F + T_{\bar{A}} \bar{M} \bar{F}) + 8 T \lvert M \lvert^2 \lvert F \lvert^2 \ , \\
  \tfrac{1}{16} \mathcal{L}_{\mathcal{O}_{(2 \rvert 1)}}^{(\text{4-der})} &= - 6 T(\Box A - \tfrac{2}{3} i b^m \partial_m A)(\Box \bar{A} +\tfrac{2}{3} i b^m \partial_m \bar{A}) \ , \\
  \tfrac{1}{16} \mathcal{L}_{\mathcal{O}_{(2 \rvert 1)}}^{(\text{2-der})} &=   - 3 T_{AA} \lvert F \lvert^2 (\partial A)^2 - 9 T_{\bar{A}} \lvert F \lvert^2 (\Box {\bar{A}} + \tfrac{2i}{3} b^m \partial_m \bar{A})   \\
   &  \quad  - 4 T_{\bar{A}} ( MF (\partial \bar{A})^2 + \tfrac{3}{2} \bar{F} \partial_m F \partial^m \bar{A}) - \tfrac{1}{3} T \lvert F \rvert^2 b_a b^a  \\
   &  \quad + \tfrac{2}{3} T \lvert M\rvert^2 \lvert \partial A \rvert^2 - 4TMF(\Box \bar{A} + \tfrac{2i}{3} b^m \partial_m \bar{A})  \\
   &  \quad -4 TM \partial_m F \partial^m \bar{A} -2i T \bar{F} b^m \partial_m F - 3 T\bar{F} \Box F + h.c. \ .
\end{aligned}\end{equation}
Note that here we applied partial integration to terms of the form $\partial_m \bar{M} \partial^m A$. 
\subsubsection*{Operator $\mathcal{O}_{(2 \rvert 3)}$:}
For this operator we find the following component form
\begin{equation}\begin{aligned}
   \hat{\mathcal{L}}_{\mathcal{O}_{(2 \rvert 3)}} / e &= - \tfrac{1}{2} \mathcal{R} \Omega_{\mathcal{O}_{(2 \rvert 3)}} - V_{\mathcal{O}_{(2 \rvert 3)}} + \mathcal{L}_{\mathcal{O}_{(2 \rvert 3)}}^{(\text{4-der})} + \mathcal{L}_{\mathcal{O}_{(2 \rvert 3)}}^{(\text{2-der})}    \ , \\
   \Omega_{\mathcal{O}_{(2 \rvert 3)}} &=  4 T \lvert \partial A \rvert^2 \\
   V_{\mathcal{O}_{(2 \rvert 3)}} &= -\tfrac{4}{3} T \lvert F \rvert^2 \lvert M \rvert^2 \\
   \mathcal{L}_{\mathcal{O}_{(2 \rvert 3)}}^{(\text{4-der})} &= - 3 T_{\bar{A}} \left[ \lvert  \partial A \rvert^2 (\Box \bar{A} + \tfrac{2}{3} i b^m \partial_m \bar{A}) + 2 \partial^m \bar{A} \, \partial^n A \, \mathcal{D}_m \mathcal{D}_n \bar{A} \right]  \\
   & \quad -3 T_{\bar{A} \bar{A}} \lvert  \partial A \rvert^2 (\partial \bar{A})^2 + 3  T   \partial^m A \partial^n \bar{A} \left[ \mathcal{R}_{mn} + \tfrac{2}{9} b_m b_n + \tfrac{2}{3}i \mathcal{D}_n b_m \right] \\
   & \quad   - 3  T \partial_m A \, \mathcal{D}^m (\Box \bar{A} + \tfrac{2}{3}i b^n \partial_n \bar{A})  + \text{h.c.} \\
   \mathcal{L}_{\mathcal{O}_{(2 \rvert 3)}}^{(\text{2-der})} &=  T_{\bar{A}} \left[ MF (\partial \bar{A})^2 + \bar{M} \bar{F} \lvert  \partial A \rvert^2 - 6 \bar{F} \partial_m F \partial^m \bar{A} - 4i \lvert F \rvert^2 b^m \partial_m \bar{A} \right] \\
   & \quad - 3 T_{A\bar{A}} \lvert F \lvert^2 \lvert  \partial A \rvert^2  - T \left( \tfrac{1}{3}\lvert  \partial A \rvert^2 \lvert M \rvert^2 + \tfrac{4}{3} \lvert F \rvert^2 b_a b^a + 3 \lvert \partial F \rvert^2 \right) \\
   & \quad  + T(FM \Box \bar{A} + M \partial_m F \partial^m \bar{A} - F \partial_m M \partial^m \bar{A}) \\
   & \quad + \tfrac{4}{3} T i b^m(FM \partial_m \bar{A} + 3\bar{F} \partial_m F) + \text{h.c.} \ .
\end{aligned}\end{equation}
Note that $\mathcal{O}_{(2 \rvert 3)}$ was already studied in \cite{Baumann:2011nm} for the special case with $T$ being a constant. Here we displayed the component form for the generalized operator where $T$ is an arbitrary function of $\Phi$ and $\bar{\Phi}$. The computation of the above result requires knowledge of several additional superfield component identities. These are displayed in eq.~\eqref{Da_comp} and eq.~\eqref{D4_Da_comp}. While these identities were already computed in \cite{Baumann:2011nm} we recalculated them as a cross-check.\footnote{In particular, we found a minus-sign difference and a typo in that reference.} 

Let us compare the above result with \cite{Baumann:2011nm}. Overall we find a remarkable agreement, the only difference with the latter reference being a minus sign in the scalar potential that can be traced back to a minus-sign difference in eq.~\eqref{Da_comp}. 
\subsubsection*{Operator $\mathcal{O}_{(R \rvert 1)}$:}
\begin{equation}\begin{aligned}
   \hat{\mathcal{L}}_{\mathcal{O}_{(R \rvert 1)}} / e &= - \tfrac{1}{2} \mathcal{R} \Omega_{\mathcal{O}_{(R \rvert 1)}} - V_{\mathcal{O}_{(R \rvert 1)}} + \mathcal{L}_{\mathcal{O}_{(R \rvert 1)}}^{(\text{4-der})} + \mathcal{L}_{\mathcal{O}_{(R \rvert 1)}}^{(\text{2-der})}    \ , \\
   \Omega_{\mathcal{O}_{(R \rvert 1)}} &= 2T (\partial A)^2 + 2 \bar{T} (\partial \bar{A})^2  \\
   V_{\mathcal{O}_{(R \rvert 1)}} &= \tfrac{2}{3} T M^2 F^2 + 2T_{\bar{A}} M \lvert F \lvert^2 F +  \text{h.c.}\\
   \mathcal{L}_{\mathcal{O}_{(R \rvert 1)}}^{(\text{4-der})} &= 2T (\partial A)^2 ( \tfrac{1}{3} b_a b^a - i \mathcal{D}_m b^m) +  \text{h.c.} \\
   \mathcal{L}_{\mathcal{O}_{(R \rvert 1)}}^{(\text{2-der})} &=  4 T_{\bar{A}} MF \lvert \partial A \lvert^2 + 2 T_A MF (\partial A)^2 + 4 TM  \partial_m F \partial^m A  \\
   & \quad +   \tfrac{8i}{3} T M F b^m \partial_m A  + \text{h.c.}
\end{aligned}\end{equation}
\subsubsection*{Operator $\mathcal{O}_{(R \rvert 2)}$:}
%
%
%
\begin{equation}\begin{aligned}
   \hat{\mathcal{L}}_{\mathcal{O}_{(R \rvert 2)}} / e &= - \tfrac{1}{2} \mathcal{R} \Omega_{\mathcal{O}_{(R \rvert 2)}} - V_{\mathcal{O}_{(R \rvert 2)}} + \mathcal{L}_{\mathcal{O}_{(R \rvert 2)}}^{(\text{4-der})} + \mathcal{L}_{\mathcal{O}_{(R \rvert 2)}}^{(\text{2-der})}    \ , \\
    \Omega_{\mathcal{O}_{(R \rvert 2)}} &= - 2 T_{\bar{A}} \lvert F \rvert^2 - 2T (  \Box A  - \tfrac{2i}{3} b^m \partial_m A + \tfrac{1}{3}  M F) + \text{h.c.} \\
    V_{\mathcal{O}_{(R \rvert 2)}} &= 2 M F \left( T_{A \bar{A}} \lvert F \rvert^2  + \tfrac{1}{3} T_A M F + \tfrac{2}{3} T_{\bar{A}} \bar{M} \bar{F} + \tfrac{2}{9} T \lvert M \rvert^2 \right) + \text{h.c.} \\
    \mathcal{L}_{\mathcal{O}_{(R \rvert 2)}}^{(\text{4-der})} &= -2 T(\Box A - \tfrac{2i}{3} b^m \partial_m A)(\tfrac{1}{3} b_a b^a - i\mathcal{D}_m b^m) + \text{h.c.}  \\
    \mathcal{L}_{\mathcal{O}_{(R \rvert 2)}}^{(\text{2-der})} &=  - 2  MF \left[ T_A (\Box A - \tfrac{2i}{3} b^m \partial_m A) +T_{\bar{A}}(\bar{A} +\tfrac{2i}{3} b^m \partial_m \bar{A}) \right] \\
    &\quad - 2 T_{\bar{A}} \left[ \lvert F \rvert^2 (\tfrac{1}{3} b_a b^a - i\mathcal{D}_m b^m) - \tfrac{2}{3} \lvert M \rvert^2 \lvert \partial A \rvert^2  + 2 M \partial_m F \partial^m \bar{A} \right] \\
    &\quad -2 MF \left[ T_{\bar{A} \bar{A}}  (\partial \bar{A})^2 + \tfrac{1}{3} T (\tfrac{1}{3} b_a b^a - i\mathcal{D}_m b^m ) \right] - 2T M \Box F \\
    &\quad +\tfrac{4}{3} T \left[ M \partial_m \bar{M} \partial^m A - i M b^m \partial_m F + \tfrac{2}{3}i \lvert M \rvert^2 b^m \partial_m A  \right]  + \text{h.c.}
\end{aligned}\end{equation}
\subsubsection*{Comments and Remarks}
Let us make a few remarks regarding the above list of component forms for the operators as well as the result of the classification in tab.~\ref{list_of_four_superspace_ops}. Firstly, out of the six operators there are two which do not have a rigid counterpart. More precisely, these are given by $\mathcal{O}_{(R \rvert 1)}$ and $\mathcal{O}_{(R \rvert 2)}$ which indeed identically vanish in the rigid limit $M_p \rightarrow 0$. 
The rigid limit of the remaining four operators is given by 
\begin{equation}\begin{aligned}
 &\mathcal{O}_{(4 \rvert 2)} \longrightarrow D_\alpha \Phi D^\alpha \Phi \bar{D}_{\dot{\alpha}} \bar{\Phi}  \bar{D}^{\dot{\alpha}} \bar{\Phi} \\
 &\mathcal{O}_{(3 \rvert 3)} \longrightarrow \bar{D}_{\dot{\alpha}} D_\alpha \Phi  D^\alpha \Phi \bar{D}^{\dot{\alpha}} \bar{\Phi}  \\
 &\mathcal{O}_{(2 \rvert 1)} \longrightarrow D^2 \Phi \bar{D}^2 \bar{\Phi} \\
 &\mathcal{O}_{(2 \rvert 3)} \longrightarrow D_a \Phi D^a \bar{\Phi} \ .
\end{aligned}\end{equation}
The component version of each of these operators contains four-derivative terms, for $\mathcal{O}_{(4 \rvert 2)}$ and $\mathcal{O}_{(2 \rvert 1)}$ this is also clear from eq.~\eqref{Leffmulti} and eq.~\eqref{L_eff_rigid}. At the level of global supersymmetry these operators are no longer independent. Using eqs.~\eqref{list_redudancies_algebra} together with tab.~\ref{integration_by_parts} one finds that in the global limit one operator is redundant leaving three independent operators. This is in perfect agreement with the result of \cite{Khoury:2010gb}.

Secondly, one may perform a consistency check of the component operators with the expectations from linearized supergravity. In that case the Lagrangian is described via a coupling to a Ferrara-Zumino multiplet, see e.g.~\cite{Ferrara:1974pz, Komargodski:2010rb, Dumitrescu:2011iu, Festuccia:2011ws}. The authors of \cite{Baumann:2011nm} already showed that the linearized version (in an expansion in $M$ and $b_a$) of $\mathcal{O}_{(2 \rvert 3)}$ (for $T$ constant) agrees with the expectations from linearized gravity. From the results in \cite{Baumann:2011nm} we can also perform this check for the linearized version of $\mathcal{O}_{(2 \rvert 1)}$ after setting $T$ to be constant. Indeed we find that eq.~\eqref{eq:O21} matches with a certain Ferrara-Zumino multiplet. Note that for the remaining operators we cannot apply the formulae in \cite{Baumann:2011nm} and, therefore, it would be necessary to recalculate the general form of the Ferrara-Zumino multiplet.

Furthermore, let us make some comments regarding target-space and K\"ahler-invariance of the operators we discussed so far. Firstly, analogous to the discussion in sec.~\ref{sec:multi_HD_susy} ensuring target-space invariance for the operators in eq.~\eqref{two_derivative_operators} and tab.~\ref{list_of_four_superspace_ops} is non-trivial. While target-space invariance for the contributions to the scalar potential can be made manifest more easily, the kinetic terms are harder to understand. We omit the details of this discussion here for the following reason. Ultimately, we are interested in situations where we compute higher-derivative operators directly from UV-physics and, hence, the operators must be target-space invariant by construction. For instance, we may integrate out heavy fields or compute quantum corrections. More specifically, we will be interested in KK-compactifications of ten-dimensional IIB supergravity where 10D higher-derivative corrections source 4D higher-derivative operators. In that case, the 10D action is not manifestly supersymmetric and the 4D supersymmetric completion is far from obvious. Here our results are of particular importance, since now a full matching to parts of the component Lagrangian can be performed and thereby, at least in principle, a full set of manifestly supersymmetric operators inferred. Naturally, target-space invariance is always guaranteed in that case. However, the discussion of target-space invariance is necessary in situations where we take a purely bottom-up EFT approach and attempt to write down all possible operators consistent with the symmetries. 

Secondly, we now briefly discuss K\"ahler-invariance for the higher-derivative operators. Recall that an important feature of the two-derivative theory given in eq.~\eqref{S=EU} and eq.~\eqref{U} is an invariance under a combination of a super-Weyl and K\"ahler transformation \cite{Howe1978138, Binetruy:2000zx}. We observe that for instance the component form in eq.~\eqref{O1_component} does not exhibit this ordinary K\"ahler-invariance. Generically theories of supergravity with higher-derivative operators allow for a larger set of transformations which leave the action invariant, one might refer to these as generalized K\"ahler transformations. In some form this was already visible in the context of flat superspace in eq.~\eqref{generlized_kahler_transf}. A concrete example in supergravity was discussed in \cite{Baumann:2011nm}, where an explicit improvement term given by $\mathcal{O}_{(G\rvert 2)}$ in tab.~\ref{four_der_operators} was added to the particular higher-derivative Lagrangian given by $\mathcal{O}_{(2\rvert 3)}$. In general we do not expect that a particular higher-derivative operator happens to be invariant under the restricted, two-derivative K\"ahler-transformations by itself. In fact this applies only to those operators, which are either super-Weyl invariant or have an explicit dependence on $K$ and $W$ in the coupling function $T$, that allows for a cancellation against their respective super-Weyl weight. For instance, the operator $\mathcal{O}_{(4 \rvert 2)}$ is super-Weyl-invariant \cite{Farakos:2012qu} and, hence, supports the ordinary K\"ahler-invariance. However, note that since the two-derivative K\"ahler invariance allows to cast the component Lagrangian in a rather simple form, it is interesting to find higher-derivative operators with this property. One may for instance consider combinations of the operators we discussed here. We leave this to future research.

\section{Supersymmetric Vacua in Higher-Derivative Theories}\label{sec:vacua}
The form of the scalar potential determines the structure of the vacua of the theory. Equipped with $S_{\text{eff}}$ in eq.~\eqref{Leffgeneral} for the case of rigid $\mathcal{N}=1$ and the results of sec.~\ref{Higher_deriv_sugra} for the case of supergravity, it is interesting to study the effect that the higher-derivatives have on the vacua of the theory. We initiate this discussion by turning first to supersymmetric vacua. Supersymmetry preservation requires additional conditions on the structure of the backgrounds that can be used to derive properties thereof.
\subsection{Supersymmetric Vacua in Rigid $\mathcal{N}=1$}
Let us start by discussing supersymmetric vacua of rigid $\mathcal{N}=1$. The off-shell supersymmetry transformation of the chiral fermions read \cite{Wess:1992cp}
\begin{equation}
 \delta_{\zeta} \chi^i = i\sqrt{2} \sigma^a \bar{\zeta} \partial_a A^i + \sqrt{2} \zeta F^i \ ,
\end{equation}
where $\zeta$ is the parameter of the supersymmetry transformation.\footnote{Naturally, the off-shell supersymmetry transformations of the chiral multiplet are unaffected by the presence of higher-derivatives. These transformations are defined entirely by the supersymmetry transformation of a general superfield together with the chirality constraint in eq.~\eqref{chiral}.} Supersymmetric vacua, therefore, are defined by
\begin{equation}
 \langle \partial_a A^i \rangle = 0 \ , \qquad \langle F^i \rangle = 0 \ .
\end{equation}
The equations of motion for the chiral auxiliaries derived from eq.~\eqref{L_eff_rigid} after evaluating at $ \langle \partial_a A^i \rangle =  \langle \partial_a F^i \rangle = 0$ read
\begin{equation}
 - \mathcal{K}_{A^i \bar{A}^{\bar{\jmath}}} \bar{F}^{\bar{\jmath}} - \mathcal{K}_{A^k \bar{A}^{\bar{\jmath}} F^i} F^k \bar{F}^{\bar{\jmath}}  - W_{A^i} = 0 \ .
\end{equation}
Evaluating the above equation at the supersymmetric condition $\langle F^i \rangle = 0$ and assuming that the derivatives of $\mathcal{K}$ in the above equation are regular at the supersymmetric point, we find that
\begin{equation}
  \langle W_{A^i} \rangle= 0 \ , \qquad \langle V \rangle = 0 \ .
\end{equation}
In other words the supersymmetric vacua of the general higher-derivative theory are identical to the vacua of the two-derivative theory. Note that this was also shown based on a conjectured form of the higher-derivative scalar potential in \cite{Cecotti:1986jy}. Note furthermore, that the associated moduli spaces of the supersymmetric backgrounds for the general higher-derivative and for the ordinary two-derivative theories precisely coincide.
\subsection{Supersymmetric Vacua in $\mathcal{N}=1$ Supergravity}\label{sec:sugra_vacua}
We now turn to supersymmetric vacua of $\mathcal{N}=1$ supergravity. The supersymmetry variation of the chiral fermion now reads \cite{Wess:1992cp}
\begin{equation}
 \delta_{\zeta} \chi^i = i\sigma^m \bar{\zeta} (\sqrt{2} \partial_m A^i - \psi_m \chi) + \sqrt{2} \zeta F^i \ .
\end{equation}
Therefore, again we find the condition that $\langle F^i \rangle = 0$. In addition, the supersymmetry variation of the gravitino \cite{Wess:1992cp}
\begin{equation}\label{gravitino_var}
 \delta_{\zeta} \psi_m^\alpha = -2 \mathcal{D}_m \zeta^\alpha + ie_m^c \left[\frac{1}{3} M(\epsilon \sigma_c \bar{\zeta})^\alpha+b_c \zeta^\alpha+\frac{1}{3}b^d(\zeta \sigma_d \bar{\sigma}_c)^\alpha \right] 
\end{equation}
has to vanish in a supersymmetric background. This requires that the spacetime background admits four independent Killing spinors. In particular, maximally symmetric spacetimes allow for the existence of four Killing spinors \cite{Freedman}. In this case, the supersymmetric backgrounds are either $M_4$, in which case $M=0$, or $AdS_4$, in which case $M \neq 0$ \cite{Freedman}. $dS_4$ on the other hand does not allow for Killing spinors and, hence, supersymmetry is always broken. Since we are discussing off-shell theories, this result holds regardless of the structure of the Lagrangian and, hence, applies both to the ordinary two-derivative as well as to the generic higher-derivative theory. Furthermore, note that there exist three further possibilities for maximally supersymmetric backgrounds given by $\mathbb{R} \times S^3$, $AdS_3 \times \mathbb{R}$ and pp-waves \cite{Festuccia:2011ws, Kuzenko:2012vd}. These backgrounds require $\langle b_a \rangle \neq 0$ and have recently been constructed in the context of pure higher-derivative supergravity \cite{Kuzenko:2016nbu}. Here we do not further consider these possibilities as they do not arise as solutions from perturbed Einstein-Hilbert gravity and do not fall into the class of effective field theories we have been discussing so far. 

The aforementioned Killing spinors also determine the curvature of the background. This can be seen easily by analysing the supersymmetry variations of the superfield $R$. In particular, since $R$ is covariantly chiral, preservation of supersymmetry requires that the component $R\rvert_{\Theta^2}$ in eq.~\eqref{R} vanishes. This is precisely the case if 
\begin{equation}\label{Susy_Lambda}
 \langle \mathcal{R} \rangle = \tfrac{4}{3} \langle \lvert M \rvert^2 \rangle \ .
\end{equation}
The above equation can also be derived from the integrability condition for the Killing spinors.\footnote{Note that the existence of a single Killing spinor already demands that the background is an Einstein-manifold.} Moreover, eq.~\eqref{Susy_Lambda} establishes a relation between the cosmological constant and the auxiliary scalar $M$ in the vacuum. For the ordinary two-derivative theory, the on-shell scalar potential, which in turn sets the value of the cosmological constant, precisely agrees with eq.~\eqref{Susy_Lambda} at the level of the Einstein equations.
\subsubsection*{Minkowski Vacua}
The $M_4$ vacua are characterised by the conditions
\begin{equation}\label{sugra_M4_min}
 \langle F^i \rangle = 0 \ , \qquad \langle M \rangle = 0 \ .
\end{equation}
For a generic higher-derivative theory the scalar potential is of the form
\begin{equation}\begin{aligned}
 V &= V_{\text{off}}^{(0)} + V_{hd}( M,\bar{M}, F^i, \bar{F}^{\bar{\jmath}},A^i,\bar{A}^{\bar{\jmath}})  \ ,  \\
 V_{\text{off}}^{(0)} &=  e^{2 K/3}(\bar{M}W + M\bar{W} - W_i F^i - \bar{W}_{\bar{\jmath}} \bar{F}^{\bar{\jmath}})  -  K_{i\bar{\jmath}} \, e^{ K/3} F^i \bar{F}^{\bar{\jmath}} \\
 & + \tfrac{1}{3} e^{ K/3}(M +  K_{\bar{\jmath}}\bar{F}^{\bar{\jmath}})(\bar{M} +  K_i F^i) \ ,
\end{aligned}\end{equation}
which, in turn, yields the following equations of motion for $\bar{F}^{\bar{\jmath}}$ 
\begin{equation}\begin{aligned}\label{eom_aux_mink}
 K_{i\bar{\jmath}} \, e^{ K/3} F^i + e^{2 K/3}\bar{W}_{\bar{\jmath}} - \frac{1}{3} e^{ K/3} K_{\bar{\jmath}}(\bar{M} +  K_i F^i) + \frac{\partial}{\partial \bar{F}^{\bar{\jmath}}}V_{hd} = 0 \ .
\end{aligned}\end{equation}
Now, from our results in sec.~\ref{Higher_deriv_sugra} we know that $V_{hd}$ is at least cubic in a combined expansion in powers of $M,\bar{M}, F^i$ and $\bar{F}^{\bar{\jmath}}$. Therefore, eq.~\eqref{sugra_M4_min} implies that $\langle \frac{\partial}{\partial \bar{F}^{\bar{\jmath}}}V_{hd} \rangle = 0$ and, hence, eq.~\eqref{eom_aux_mink} at the supersymmetric vacuum reads
\begin{equation}
 \langle \bar{W}_{\bar{\jmath}} \rangle = 0 \ .
\end{equation}
Similarly, the equations of motion for $M$ lead us to the condition that \hbox{$\langle W \rangle = 0$}. Note that, automatically we also find that $\langle \frac{\partial}{\partial A^i} V \rangle = \langle \frac{\partial}{\partial \bar{A}^{\bar{\jmath}}} V \rangle = 0$. In total, the supersymmetric Minkowski vacua in eq.~\eqref{sugra_M4_min} are equivalently defined by the conditions 
\begin{equation}\label{eq:susy_mink_vacua}
 \langle \bar{W}_{\bar{\jmath}} \rangle = \langle W \rangle = 0 \ .
\end{equation}
Therefore, the supersymmetric $M_4$ vacua as well as their corresponding moduli spaces of the general higher-derivative and of the ordinary two-derivative theories are identical. Again these results are in agreement with \cite{Cecotti:1986jy}. 
Note that from eq.~\eqref{eq:susy_mink_vacua} it follows that the moduli space is defined by a set of holomorphic equations. Hence, the moduli space is given by a complex sub-manifold of the K\"ahler manifold and, in turn, K\"ahler itself.  
\subsubsection*{Anti-de Sitter Vacua}
Let us now turn to the $AdS_4$ vacua. Compared to the Minkowski case the $AdS_4$ vacua require more effort to understand. We characterise these vacua by the conditions
\begin{equation}\label{Ads4}
 \langle F^i \rangle = 0 \ , \qquad \langle M \rangle \neq 0 \ , \qquad \langle \mathcal{R} \rangle = \tfrac{4}{3} \langle \lvert M \rvert^2 \rangle \ , \qquad \langle V_i \rangle = \langle V_{\bar{\jmath}} \rangle = 0 \ .
\end{equation}
In the ordinary two-derivative theory these properties are equivalent to the conditions $\langle D_i W \rangle = 0$ and $\langle W \rangle \neq 0$. 

In the higher-derivative theory it is not a priori clear whether the conditions in eq.~\eqref{Ads4} can still simultaneously be satisfied or whether some of the conditions are redundant meaning they are implied by some of the others expressions (such as in the two-derivative theory). Let us begin by investigating the curvature constraint in eq.~\eqref{Susy_Lambda}. In appendix~\ref{app:Killing_spinor} we demonstrate that eq.~\eqref{Susy_Lambda} can still be automatically satisfied for a general higher-derivative supergravity after integrating out $M$ and after solving the higher-curvature Einstein equations. The explicit analysis is performed for a Lagrangian supporting the most general scalar potential and an $\mathcal{R} + \mathcal{R}^2$ gravity.\footnote{In the same appendix we argue that further higher-curvature terms will not affect the conclusions.} This serves as evidence pointing towards the statement that eq.~\eqref{Susy_Lambda} is redundant and does not have to be included in the list of conditions in eq.~\eqref{Ads4}.

Let us now investigate the remaining conditions in eq.~\eqref{Ads4} after integrating out the auxiliary fields. Firstly, the equation of motion for $\bar{M}$ after evaluating at eq.~\eqref{Ads4} reads
\begin{equation}\label{eom_mbar}
 \left\langle e^{2 K/3} W + \frac{1}{3}e^{K/3} M + \frac{\partial}{\partial \bar{M}}V_{hd} \right\rangle = 0 \ .
\end{equation}
Furthermore, the equations of motion for the chiral auxiliaries as given in eq.~\eqref{eom_aux_mink} can be simplified by using eq.~\eqref{eom_mbar} and are of the form
\begin{equation}\begin{aligned}\label{eom_fbar}
 \left\langle  D_{\bar{\jmath}} \bar{W}  - \mathrm{e}^{-2K/3} \left(K_{\bar{\jmath}} \frac{\partial}{\partial M}- \frac{\partial}{\partial \bar{F}^{\bar{\jmath}}}\right) V_{hd} \right\rangle = 0 \ .
\end{aligned}\end{equation}
The equations of motion in eq.~\eqref{eom_mbar} and in eq.~\eqref{eom_fbar} have several consequences. Firstly, the value of the cosmological constant is determined by $\langle M \rangle$ via eq.~\eqref{Susy_Lambda} and, hence, differs from the two-derivative result. 
Secondly, eq.~\eqref{eom_fbar} is in general satisfied for $\langle D_i W \rangle \neq 0$. Therefore, the position of the supersymmetric vacuum is shifted. More importantly, in general we expect $\langle V_i \rangle = \langle V_{\bar{\jmath}} \rangle = 0$ to be independent conditions and no longer satisfied just by means of eqs.~\eqref{eom_mbar},\eqref{eom_fbar}. In turn, generically the vacua should not admit any flat directions. In particular, in situations where $V_{hd}$ constitutes a small correction to the ordinary scalar potential and, hence, the supersymmetric vacuum is shifted to a nearby position, we expect that the supplementary conditions $\langle V_i \rangle = \langle V_{\bar{\jmath}} \rangle = 0$ lift any flat direction which may have existed in the two-derivative theory. This observation is also in agreement with the existing literature on $(\mathcal{N}=1, D=3)$ superconformal field theories (SCFT) which are expected to be dual to $AdS_4$-supergravities via the $AdS$/CFT correspondence \cite{Maldacena:1997re}. The moduli space of the $AdS_4$ vacua corresponds to the space of deformations of exactly marginal operators in the respective SCFT. In particular, as stated in \cite{Cordova} generically one expects that there are no such deformations in the $(\mathcal{N}=1, D=3)$ SCFTs and, therefore, no moduli spaces in the dual $AdS_4$ vacua.

Of course the above arguments merely describe the general expectation and do not represent strict bounds on the moduli space. It would be interesting to check whether rigorous statements about the moduli spaces can be made. In particular, one may try to derive a bound on the dimension, similar to the discussion in the ordinary two-derivative theory where the moduli space has dimension $\leq n_c$ \cite{deAlwis:2013jaa}. To perform such a discussion it would be necessary to make an explicit analysis for the most general higher-derivative theory (including also the dependence on the chiral auxiliary fields) which is outside the scope of this paper. We leave these issues to future research. 

In summary, supersymmetric $AdS_4$ vacua are defined by eq.~\eqref{eom_fbar} and eq.~\eqref{eom_mbar} together with the conditions $\langle V_i \rangle = \langle V_{\bar{\jmath}} \rangle = 0$ and $\langle M \rangle \neq 0$. As in the ordinary theory the curvature constraint in eq.~\eqref{Susy_Lambda} is automatically satisfied on-shell. 
\section{Application to Stringy No-Scale Models}\label{sec:no_scale}
The main result of sec.~\eqref{Higher_deriv_sugra} is a catalog of higher-derivative operators and respective component identities. We may now proceed and study their relevance for the particular examples studied in \cite{Ciupke:2015msa}. More precisely, we want to study the higher-derivative operators for the situation where $\mathcal{L}_{(0)}$ is a no-scale supergravity. No-scale models are special examples, where extra corrections to the scalar potential are generally of interest, since they are defined by the property that $V_{(0)} = 0$ \cite{Cremmer:1983bf}. In string compactifications one often deals with no-scale supergravities which additionally possess a Peccei-Quinn shift-symmetry. These no-scale supergravities have recently been classified in \cite{Ciupke:2015ora}. In particular, we are motivated by the low-energy effective $\mathcal{N}=1$ supergravity obtained from orientifold-compactifications of type IIB string theory with background fluxes. Among the chiral multiplets of those theories are the K\"ahler moduli, which are at leading order described by such a no-scale model.

The shift-symmetry implies that up to K\"ahler transformations we can choose the superpotential to be given by a constant. The defining condition for a no-scale model then reads
\begin{equation}
  K^{i \bar{\jmath}} K_i K_{\bar{\jmath}} = 3 \ .
\end{equation}
Using eq.~\eqref{M_0} the above equation implies that
\begin{equation}\label{M=0}
 M_{(0)} = 0 \ .
\end{equation}
Let us now turn our attention to the minimal list of higher-derivative operators in tab.~\ref{list_of_four_superspace_ops}. Eq.~\eqref{M=0} implies that the component forms simplify considerably and, in particular, that at leading order many operators do not contribute to the scalar potential. More specifically, among the operators which contribute four-derivative terms at the linearized level only $\mathcal{O}_{(4 \rvert 2)}$ and $\mathcal{O}_{(2 \rvert 1)}$ induce new terms in the scalar potential. In fact, both operators lead to the same correction, which is of the form 
\begin{equation}
 \lvert F \lvert^4 \ .
\end{equation}
This type of correction was already studied in detail in \cite{Ciupke:2015msa} where it was claimed that $\lvert F \lvert^4$ is the only correction to the scalar potential that can be linked via supersymmetry to four-supercovariant derivative operators which also induce four-derivative terms (at the linearized level). We have now explicitly demonstrated this statement.\footnote{Let us emphasize that this observation goes beyond the particular analysis in \cite{Ciupke:2015msa}, where higher-derivatives induced by 10D $(\alpha')^3$-corrections were investigated, but would equally hold if other leading order corrections to the scalar potential coming from higher-derivative corrections of different origin, such as $g_s$-corrections, were determined.}
\section{Conclusions}\label{sec:conclusions}
Higher-derivative operators are universally present in effective field theories. In supersymmetric theories these operators require corrections to the scalar potential which, in turn, modify the vacuum structure of the theory but are also relevant for instance in the context of inflation. In addition, these corrections are already present in globally supersymmetric theories and, thus, dominate over the Planck-suppressed operators appearing in the scalar potential of supergravity. Here we initiated a systematic study of higher-derivative operators for theories in $\mathcal{N}=1,D=4$ flat and curved superspace with particular emphasis on the scalar potential.

In flat superspace we displayed a superspace action for the most general scalar potential in eq.~\eqref{Leffgeneral}. This action is determined by a $2n_c$-dimensional (pseudo-) K\"ahler potential and superpotential together with the supplementary constraints in eq.~\eqref{psi_def}. The additional degrees of freedom are chiral (higher-derivative) multiplets and their scalar components are given by the chiral auxiliary fields which are now generically propagating. However, note that care must be taken with the proper geometric understanding of this theory, since the constraints in eq.~\eqref{psi_def} are incompatible with a $2n_c$-dimensional target-space invariance. If we constrain ourselves to the discussion of the scalar potential alone, then the action is geometrically understood in the context of the cotangent-bundle over the manifold of the chiral scalars. In eq.~\eqref{Lmulti} and eq.~\eqref{Leffmulti} we also displayed an alternative higher-derivative Lagrangian which, contrary to the aforementioned theory, does not induce kinetic terms for the chiral auxiliary fields.

The theories in eq.~\eqref{Leffgeneral} and eq.~\eqref{Leffmulti} are off-shell by construction. Obtaining the on-shell theory is a daunting task, since the equations of motion for the auxiliary fields are now arbitrary algebraic equations. The algebraic nature of the auxiliary fields remains true even if they obtain kinetic terms, since as we discussed in sec.~\ref{prop_aux} their masses sit at the cut-off scale of the EFT. For local theories the equations of motion for $F^i$ are polynomial equations and, hence, induce a multiplet of on-shell theories. We extended the analysis of the equations of motion in the context of effective field theories initiated in \cite{Ciupke:2015msa} and illustrated that there exists only a single solution that produces a physically viable Lagrangian. We interpret the remaining solutions as artifacts of a truncation of an infinite sum of higher-derivative operators. 

For $\mathcal{N}=1$ old-minimal supergravity we conjectured a possible extension of eq.~\eqref{Leffgeneral} given in eq.~\eqref{conjectured_L}. However, due to the complicated form of the algebra of super-covariant derivatives we did not attempt to prove this conjecture. Instead, we classified the leading order (denoted as $\Delta=2$) and next-to-leading order ($\Delta=4$) higher-derivative operators for the chiral multiplets including also a brief review of higher-curvature operators. The classification of the $\Delta=4$ operators performed in appendix~\ref{appendix:integration_by_parts} was substantially more involved, the results being displayed in tab.~\ref{list_of_four_superspace_ops}. To compute the component actions of higher-derivative operators we developed several tools. Firstly, we provided a catalogue of component identities for higher-derivative superfield in appendix~\ref{component_identities} extending the results of \cite{Baumann:2011nm}. Secondly, we developed an algorithm to compute the (on-shell) component action. In particular, we emphasized the importance of performing the Weyl-transformation to the Einstein-frame before integrating out the auxiliary fields. Moreover, we demonstrated that for the computation of the linearized on-shell action it suffices to insert the leading order solutions for the auxiliary fields in the Lagrangian. Thus, in this context it is not necessary to solve the equations of motion for the auxiliary fields. Henceforth, we constrained ourselves to give only the off-shell component results, as the respective linearized on-shell theories are obtained readily. The component forms of the $\Delta=2$ operators and a subclass (more precisely all operators which induce four-derivative terms for the component fields) of $\Delta=4$ are displayed in sec.~\ref{sec:two_der_operators} and sec.~\ref{sec:four_derivative_operators}. These results are universally applicable for computing leading-order higher-derivative corrections to a generic supergravity with chiral multiplets. 

In the last sections we discussed two particular applications of the aforementioned results. Firstly, we investigated the vacuum structure of the (general) higher-derivative theories. On the one hand, the supersymmetric Minkowski vacua of the higher-derivative theory coincide with the two-derivative vacua, in agreement with the earlier investigation \cite{Cecotti:1986jy}. On the other hand, the higher-derivative operators modify the structure of the supersymmetric $AdS_4$-vacua. While we found that the Killing spinor equation is still automatically satisfied, we expect that the supersymmetric points are not necessarily extremal anymore. Therefore, generically we anticipate that there is no moduli space. This expectation also agrees with corresponding observations for the dual three-dimensional SCFTs \cite{Cordova}. Note also, that as already discussed in \cite{Hindawi:1995qa}, supersymmetry-breaking no longer requires the non-vanishing of the $F$-terms, but can be induced by higher-curvature operators alone. Finally, we investigated the form of the higher-derivative corrections for shift-symmetric no-scale models. 

Let us outline possible future directions and open questions that remain. Firstly, the discussion of higher-derivative operators may be generalized to gauged theories with vector multiplets. Furthermore, it would be interesting to understand how the target space invariance, in particular for the theory in eq.~\eqref{Leffgeneral}, is restored at the level of the full Lagrangian. Moreover, being equipped with tab.~\ref{list_of_four_superspace_ops} and the component forms of the operators displayed in sec.~\ref{sec:four_derivative_operators} one may now perform a full matching of UV-derived four-derivative terms with the operators in tab.~\ref{list_of_four_superspace_ops}. For the particular case of \cite{Ciupke:2015msa} four-derivative terms for the chiral scalars were inferred by performing a Kaluza-Klein reduction of ten-dimensional higher-derivative operators and matched to $\mathcal{O}_{(4\rvert 2)}$ alone. In turn, the operator was only computed up to a constant. However, via a complete matching to the operators in tab.~\ref{list_of_four_superspace_ops} one would be able to fully determine the supersymmetric completion of these four-derivative component terms. This will be performed in a separate future paper.\footnote{There has also been recent work on determining supersymmetric embeddings of the Dirac-Born-Infeld action which describes the effective action of D-branes \cite{Khoury:2010gb, Koehn:2012ar, Bielleman:2016grv, Aoki:2016cnw}. The operators displayed in sec.~\ref{sec:four_derivative_operators} might also be relevant for a matching in this context.}
Finally, one might also study the relevance of the higher-derivative operators in inflationary dynamics. However, such an investigation would have to determine the target-space invariant versions of the higher-derivative operators presented here.
\section*{Acknowledgements}
I would like to thank Daniel Baumann, Benedict Broy, Daniel Green, Jan Louis, Severin L\"{u}st, Constantin Muranaka, Alexander Westphal, Clemens Wieck and Lucila Zarate for comments and discussions. This work has been supported by the ERC Consolidator Grant STRINGFLATION under the HORIZON 2020 contract no. 647995 and by the German Science Foundation (DFG) within the Collaborative Research Center 676 ”Particles, Strings and the Early Universe”.
\appendix
\section{Derivation of Superspace Action for General Scalar Potential}\label{appendix:global}
In this appendix we provide the derivation of the superspace effective scalar potential which we presented in sec.~\ref{sec:HD_susy}. More precisely our goal is to simplify eq.~\eqref{Sgen} assuming that we only consider operators in the action that manifestly contribute to the scalar potential. To this end we evaluate the general action at the supersymmetric condition in eq.~\eqref{effectivepot}. Therefore, we can ignore all operators in $S_{\text{gen}}$ which involve the spacetime-derivatives $\partial_a$. Moreover, the condition in eq.~\eqref{effectivepot} also restricts the dependence on the spinorial component of the superspace-derivatives. To begin with, all mixed-type combinations of spinorial superspace-derivatives acting on $\Phi$ or $\bar{\Phi}$ vanish when evaluated at eq.~\eqref{effectivepot}. With mixed-type we mean that the combination involves at least one power of $D_{\alpha}$ as well as $\bar{D}_{\dot{\beta}}$. This can be seen iteratively. For two superspace-derivatives the possible combinations are given by $D_{\alpha} \bar{D}_{\dot{\beta}} \bar{\Phi}$ and $ \bar{D}_{\dot{\beta}} D_{\alpha} \Phi$. Making use of eq.~\eqref{anticom} one indeed finds that these terms vanish at the condition in eq.~\eqref{effectivepot}. The possible terms with three superspace-derivatives are given by
\begin{equation}\begin{aligned}\label{list}
 D_{\alpha} \bar{D}_{\dot{\beta}} \bar{D}_{\dot{\gamma}} \bar{\Phi} \ , \quad & D_{\alpha} D_{\beta} \bar{D}_{\dot{\gamma}} \bar{\Phi}  \ , \quad  \bar{D}_{\dot{\alpha}} D_{\beta} \bar{D}_{\dot{\gamma}} \bar{\Phi} \ , \quad \bar{D}_{\dot{\alpha}} \bar{D}_{\dot{\beta}} D_{\gamma} \Phi \ , \\
    &D_{\alpha} \bar{D}_{\dot{\beta}} D_{\gamma} \Phi  \ , \quad  \bar{D}_{\dot{\alpha}} D_{\beta} D_{\gamma} \Phi \ .
\end{aligned}\end{equation}
Again via eq.~\eqref{anticom} one finds that these terms vanish after inserting eq.~\eqref{effectivepot}. From eq.~\eqref{D30} we learn that terms with more than three superspace-derivatives are simply further superspace derivatives acting on the terms in eq.~\eqref{list} and, thus, the claim holds in general. 
Furthermore, eq.~\eqref{D30} implies that we have to consider only terms with at most two superspace-derivatives acting on $\Phi$ or $\bar{\Phi}$. Finally, eq.~\eqref{DDisD2} shows that it suffices to consider a dependence of $S_{\mathrm{eff}}$ on $D_{\alpha} \Phi$, $\bar{D}_{\dot{\alpha}} \bar{\Phi}$ as well as $D^2 \Phi$ and $\bar{D}^2 \bar{\Phi}$.

After the above considerations we can express the effective superspace action evaluated at eq.~\eqref{effectivepot} as follows
\begin{equation}\begin{aligned}\label{Seff1}
 S_{\mathrm{eff}} &= \int \d^8 z \, \mathcal{K}(\Phi,\bar{\Phi}, D^2 \Phi, D_{\alpha} \Phi D^\alpha \Phi, \bar{D}^2 \bar{\Phi},\bar{D}_{\dot{\alpha}} \bar{\Phi} \bar{D}^{\dot{\alpha}} \bar{\Phi})  \\
  &\quad + \int \d^6 z \, W(\Phi,\bar{D}^2 \bar{\Phi}) + \int \d^6 \bar{z} \,\bar{W}(\bar{\Phi}, D^2 \Phi) \ .
\end{aligned}\end{equation}
Let us turn more closely to $W$. We consider the superpotential to be a power series in $\Phi$ and $\bar{D}^2 \bar{\Phi}$. An arbitrary term in this series is of the form
\begin{equation}
 \Phi^k (\bar{D}^2 \bar{\Phi})^l = \bar{D}^2 (\Phi^k (\bar{D}^2 \bar{\Phi})^{l-1} \bar{\Phi}) \ , \qquad k,l \in \mathbb{N}_0 \ ,
\end{equation}
where equality holds due to eq.~\eqref{chiral} and eq.~\eqref{D30}. Using the following identity 
\begin{equation}
 \int \d^6 z \bar{D}^2 f(x,\theta,\bar{\theta}) = -4 \int \d^8 z f(x,\theta,\bar{\theta}) \ ,
\end{equation}
where $f$ is an arbitrary superfield, we observe that the dependence of $W$ on $\bar{D}^2 \bar{\Phi}$ can be entirely absorbed into $\mathcal{K}$. Therefore, we confine the discussion of $S_{\mathrm{eff}}$ to $\mathcal{K}$. Moreover, we focus on the bosonic part of $S_{\mathrm{eff}}$ from now on. 
In this case we can simplify $\mathcal{K}$ even further. To see this, as a first step we compute 
\begin{equation}\label{DPDPDPDP}
 (D_{\alpha} \Phi D^{\alpha} \Phi)^2\Bigl\rvert_{\text{bos}} = 0 \ .
\end{equation}
The above identity can either be computed directly or alternatively be derived from integration by parts identities in superspace such as eq.~\eqref{integrationbyparts}. Consequently within $\mathcal{K}$ we only have to consider terms with up to a single factor of $D_{\alpha} \Phi D^{\alpha} \Phi$ and $\bar{D}_{\dot{\alpha}} \bar{\Phi} \bar{D}^{\dot{\alpha}} \bar{\Phi}$. Up to boundary terms as well as mixed type superspace-derivative terms, which yield purely kinetic contributions, the following integration by parts identities hold for arbitrary functions $T$ \footnote{These integration by parts identities can also be understood as arising in the rigid limit of the curved superspace identities in appendix~\ref{appendix:integration_by_parts}.}
\begin{equation}\begin{aligned}\label{integrationbyparts}
  &\int \d^8 z  D^2 \Phi \, T(\Phi,\bar{\Phi},D^2 \Phi, \bar{D}^2 \bar{\Phi}) \\
  & =  \int \d^8 z \, ( D_{\alpha} \Phi  D^{\alpha} \Phi ) \, \partial_{\Phi} T(\Phi,\bar{\Phi},D^2 \Phi, \bar{D}^2 \bar{\Phi})  \ , \\
  &\int \d^8 z \, \bar{D}^2 \bar{\Phi} \, T(\Phi,\bar{\Phi},D^2 \Phi, \bar{D}^2 \bar{\Phi}) \\
  & =  - \int \d^8 z \, ( \bar{D}_{\dot{\alpha}} \bar{\Phi} \bar{D}^{\dot{\alpha}} \bar{\Phi} ) \, \partial_{\bar{\Phi}} T(\Phi,\bar{\Phi},D^2 \Phi, \bar{D}^2 \bar{\Phi})  \ , \\
  &\int \d^8 z \, D^2 \Phi \bar{D}^2 \bar{\Phi} \, T(\Phi,\bar{\Phi},D^2 \Phi, \bar{D}^2 \bar{\Phi})   \\
  &=- \int \d^8 z \, ( D_{\alpha} \Phi  D^{\alpha} \Phi )( \bar{D}_{\dot{\alpha}} \bar{\Phi} \bar{D}^{\dot{\alpha}} \bar{\Phi} ) \, \partial_{\Phi} \partial_{\bar{\Phi}} T(\Phi,\bar{\Phi},D^2 \Phi, \bar{D}^2 \bar{\Phi})  \ .
\end{aligned}\end{equation}
Thus, we infer that the factors $D_{\alpha} \Phi D^{\alpha} \Phi$ and $\bar{D}_{\dot{\alpha}} \bar{\Phi} \bar{D}^{\dot{\alpha}} \bar{\Phi}$ in $\mathcal{K}$ can always be recast into additional factors of $D^2 \Phi$ and $\bar{D}^2 \bar{\Phi}$ respectively. The equivalence between the superfields $D_{\alpha} \Phi D^{\alpha} \Phi$ and $\bar{D}_{\dot{\alpha}} \bar{\Phi} \bar{D}^{\dot{\alpha}} \bar{\Phi}$ as well as $D^2 \Phi$ and $\bar{D}^2 \bar{\Phi}$ can also be understood from the fact that while $D^2 \Phi$ ($\bar{D}^2 \bar{\Phi}$) are anti-chiral (chiral) superfields, $D_{\alpha} \Phi D^{\alpha} \Phi$ and $\bar{D}_{\dot{\alpha}} \bar{\Phi} \bar{D}^{\dot{\alpha}} \bar{\Phi}$ are complex linear superfields, that is they satisfy
\begin{equation}
 D^2(D_{\alpha} \Phi D^{\alpha} \Phi) = \bar{D}^2 (\bar{D}_{\dot{\alpha}} \bar{\Phi} \bar{D}^{\dot{\alpha}} \bar{\Phi}) = 0 \ .
\end{equation}
In total we find that, without loss of generality, the general superspace action for the effective scalar potential is of the form
\begin{equation}\begin{aligned}\label{Seff_final}
 S_{\mathrm{eff}} = \int \d^8 z \, \mathcal{K}(\Phi,\bar{\Phi}, D^2 \Phi, \bar{D}^2 \bar{\Phi})  + \int \d^6 z \, W(\Phi) + \int \d^6 \bar{z} \,\bar{W}(\bar{\Phi}) \ .
\end{aligned}\end{equation}
In sec.~\ref{sec:multi_HD_susy} we proceed to generalize this action to the multi-field case and to discuss its respective component version.
\section{More on Propagating Auxiliary Fields}\label{More_on_prop_aux}
In this appendix we consider the action
\begin{equation}\label{S_g_app}
 S_g = \int \mathrm{d}^8 z E \,g(R) + \text{h.c.} \ ,
\end{equation}
and study the dynamics of the auxiliary $M$ described by $S_g$. Quite similar to the discussion in sec.~\ref{prop_aux} $M$ is a propagating degree of freedom. Here we want to demonstrate that $M$, despite having the correct sign of the kinetic terms, is unphysical in the context of effective field theory since it generically has a mass at the cutoff-scale of the EFT. Firstly, in a generic EFT we expect $g(R)$ to be an expansion
\begin{equation}\label{fR}
 g(R) = -3 (1+a R + \dots) \ .
\end{equation}
The superfield $R$ has mass dimension one and so, if our EFT is subject to the cut-off $\Lambda$, we expect that $a \sim \Lambda^{-1}$. The component version of eq.~\eqref{S_g_app} with eq.~\eqref{fR} reads \cite{Hindawi:1995qa}
\begin{equation}\begin{aligned}\label{LFR}
  \mathcal{L}_g / e &= -\tfrac{1}{2}\mathcal{R} (1 - \tfrac{a}{3}(M+\bar{M}))  + \tfrac{1}{3} b_a b^a ( 1- \tfrac{a}{3}(M+\bar{M})) - \tfrac{i}{3} a  b^a \partial_a (M-\bar{M}) \\
  & \quad +\tfrac{1}{3} \lvert M \lvert^2 (1 - \tfrac{a}{6}(M+\bar{M})) + \dots\ .
\end{aligned}\end{equation}
It is convenient to split the complex auxiliary $M$ into real and imaginary parts as follows
\begin{equation}
 \mu \equiv \frac{M+\bar{M}}{2} \ , \qquad \nu \equiv \frac{M - \bar{M}}{2i} \ .
\end{equation}
Let us take the description of the new degree of freedom by performing first a Weyl transformation in eq.~\eqref{LFR}
\begin{equation}
 g_{mn} \rightarrow \Omega g_{mn} \ , \qquad \Omega = 1 - \tfrac{2a}{3} \mu \ .
\end{equation}
The resulting Einstein-frame Lagrangian is of the form
\begin{equation}\begin{aligned}\label{LFR2}
  \mathcal{L}_g / e = - \frac{1}{2}\mathcal{R} - \frac{a^2}{3 \Omega^2} (\partial \mu)^2 + \frac{1}{3} b_a b^a  + \frac{2}{3 \Omega} a  b^a \partial_a \nu + \frac{\mu^2 + \nu^2}{3\Omega^2}(1-\tfrac{a}{3}\mu) \ ,
\end{aligned}\end{equation}
where we introduced the abbreviation $(\partial \mu)^2 \equiv \partial_m \mu \partial^m \mu$. Integrating out the auxiliary $b_a$ we find
\begin{equation}
 b_a = -\frac{a}{\Omega} \partial_a \nu \ , 
\end{equation}
such that the on-shell Lagrangian reads
\begin{equation}\begin{aligned}\label{LFR3}
  \mathcal{L}_g / e = - \frac{1}{2}\mathcal{R} - \frac{a^2}{3 \Omega^2} \left[ (\partial \mu)^2  + (\partial \nu)^2 \right] + \frac{\mu^2 + \nu^2}{3\Omega^2}(1-\tfrac{a}{3}\mu) \ .
\end{aligned}\end{equation}
We proceed to canonically normalize the scalars $\mu$ and $\nu$. The relation between $M$ and the canonical field $\phi$ is given by
\begin{equation}\label{can_norm}
 M = \frac{1}{a} \tilde{M}(\phi) \ ,
\end{equation}
where $\tilde{M}(\phi)$ is a dimensionless function. The scalar potential for $\phi$ in eq.~\eqref{LFR2} has a global factor of $a^2 \sim \Lambda^2$ after replacing $M$ via eq.~\eqref{can_norm}, which demonstrates the mass of the new degree of freedom is generically at the cut-off scale $\Lambda$ and, hence, this new degree of freedom is unphysical and should be integrated out.
\section{Component Identities for Superfields in Curved Superspace}\label{component_identities}
In this appendix we provide the necessary formulae to compute the component actions for matter coupled supergravity in sec.~\ref{sugraaction}, in particular for the four covariant derivative operators in sec.~\ref{sec:four_derivative_operators}. We start by giving a catalog of component identities for higher super-covariant derivatives acting on the superfields $(\Phi ,\bar{\Phi}, R, \bar{R}, G_{\alpha \dot{\alpha}})$. A useful list of component identities that goes beyond the formulas presented in \cite{Wess:1992cp} was already given in \cite{Baumann:2011nm}. Our results below partially overlap with that reference, but we also compute new identities which are required for the discussion in sec.~\eqref{sec:four_derivative_operators}. Here we derive all components starting from the solution to the Bianchi identities and the algebra of super-covariant derivatives. As a cross-check we also re-derive component identities which appeared in \cite{Baumann:2011nm}. We find some minor disagreements with the results in that reference on some component identities which we indicate explicitly later on.

The tool for the computation of component identities are the (anti-) commutation relations in eq.~\eqref{curved_anticom}, which relate the covariant derivatives to the torsion defined in eq.~\eqref{torsion} and the super-curvature tensor in eq.~\eqref{super_Riemann}. After imposing the constraints of old minimal supergravity on the torsion and solving the Bianchi identities, the only non-zero components of the torsion are given by \cite{Wess:1992cp}
\begin{equation}\begin{aligned}\label{torsion_comp}
 T_{\alpha \dot{\alpha}} {}^a & = T_{\dot{\alpha} \alpha} {}^a = 2i \sigma_{\alpha \dot{\alpha}}^a \\
 T_{\dot{\alpha} a} {}^\alpha &= - T_{ a \dot{\alpha}} {}^\alpha =  - i R \epsilon_{\dot{\alpha} \dot{\beta}} \bar{\sigma}_a^{\dot{\beta} \alpha} \\
 T_{\alpha a} {}^{\dot{\alpha}} &= - T_{ a \alpha} {}^{\dot{\alpha}} =  - i \bar{R} \epsilon_{\alpha \beta} \bar{\sigma}_a^{\dot{\alpha} \beta} \\
 T_{\beta a} {}^\alpha &= - T_{ a \beta} {}^\alpha =  \tfrac{i}{8} \bar{\sigma}_a^{\dot{\gamma} \gamma} ( \delta_\gamma^\alpha G_{\beta \dot{\gamma}} - 3 \delta_\beta^\alpha G_{\gamma \dot{\gamma}} +3 \epsilon_{\beta \gamma} G^\alpha_{\dot{\gamma}} ) \\ 
 T_{\dot{\beta} a} {}^{\dot{\alpha}} &= - T_{ a \dot{\beta}} {}^{\dot{\alpha}} =  \tfrac{i}{8} \bar{\sigma}_a^{\dot{\gamma} \gamma} ( \delta_{\dot{\gamma}}^{\dot{\alpha}} G_{\gamma \dot{\beta}} - 3 \delta_{\dot{\beta}}^{\dot{\alpha}} G_{\gamma \dot{\gamma}} +3 \epsilon_{\dot{\beta} \dot{\gamma}} G^{\dot{\alpha}}_{\gamma} ) \ ,
 \end{aligned}\end{equation}
as well as the components $T_{a b}^\alpha$ and $T_{a b}^{\dot{\alpha}}$ which we do not display here as they are not required. Additionally, we need certain components of the curvature tensor.
The following list contains some of the frequently used components \footnote{The last component below is not displayed explicitly in \cite{Wess:1992cp}. However, it can be computed directly from the Bianchi identities given in that reference.}
\begin{equation}\begin{aligned}\label{curvature_comp}
 R_{\delta \gamma \beta \alpha} &= 4 (\epsilon_{\delta \beta} \epsilon_{\gamma \alpha} + \epsilon_{\gamma \beta} \epsilon_{\delta \alpha}) \bar{R} \\ 
 R_{\dot{\delta} \dot{\gamma} \dot{\beta} \dot{\alpha}} &= 4 (\epsilon_{\dot{\delta} \dot{\beta}} \epsilon_{\dot{\gamma} \dot{\alpha}} + \epsilon_{\dot{\gamma} \dot{\beta}} \epsilon_{\dot{\delta} \dot{\alpha}}) R \\ 
 R_{\delta \dot{\gamma}  \beta \alpha} &=  R_{\dot{\gamma} \delta  \beta \alpha} = - \epsilon_{\delta \beta} G_{\alpha \dot{\gamma}} - \epsilon_{\delta \alpha} G_{\beta \dot{\gamma}} \\
 R_{\delta \dot{\gamma}  \dot{\beta} \dot{\alpha}} &=  R_{\dot{\gamma} \delta \dot{\beta} \dot{\alpha}} = - \epsilon_{\dot{\gamma} \dot{\beta}} G_{\delta  \dot{\alpha}} - \epsilon_{\dot{\gamma}  \dot{\alpha}} G_{\delta \dot{\beta}} \\
 R_{\dot{\delta}c \dot{\beta} \dot{\alpha}} &= - R_{c \dot{\delta} \dot{\beta} \dot{\alpha}} =- \tfrac{1}{2} \bar{\sigma}_c^{\dot{\gamma} \gamma} \Bigl[ i(\epsilon_{\dot{\delta} \dot{\beta}} \epsilon_{\dot{\gamma} \dot{\alpha}} + \epsilon_{\dot{\delta} \dot{\alpha}} \epsilon_{\dot{\gamma} \dot{\beta}}) \bar{\mathcal{D}}_{\dot{\epsilon}} G_{\gamma}^{\dot{\epsilon}} + \tfrac{i}{2}(\epsilon_{\dot{\delta} \dot{\gamma}} \bar{\mathcal{D}}_{\dot{\beta}} + \epsilon_{\dot{\delta} \dot{\beta}} \bar{\mathcal{D}}_{\dot{\gamma}})G_{\gamma \dot{\alpha}} \\ 
 & \qquad \qquad \qquad \qquad \quad + \tfrac{i}{2} (\epsilon_{\dot{\delta} \dot{\gamma}} \bar{\mathcal{D}}_{\dot{\alpha}} + \epsilon_{\dot{\delta} \dot{\alpha}} \bar{\mathcal{D}}_{\dot{\gamma}})G_{\gamma \dot{\beta}} \Bigr] \\
 R_{\alpha \dot{\alpha} a b} &=   2i \epsilon_{abcd} \, \sigma_{\alpha \dot{\alpha}}^d G^c 
\end{aligned}\end{equation}
It is also useful to note the following relations
\begin{equation}\label{DG}
 \mathcal{D}^\alpha G_{\alpha \dot{\alpha}} = \bar{\mathcal{D}}_{\dot{\alpha}} \bar{R} \ , \qquad \bar{\mathcal{D}}^{\dot{\alpha}} G_{\alpha \dot{\alpha}} = \mathcal{D}_\alpha R \ . 
\end{equation}
From the components of the torsion and curvature we also derive the useful identities
\begin{equation}
 T_{\dot{\alpha} a} {}^{\dot{\alpha}} = T_{\alpha a} {}^\alpha = 2i G_a \ , \qquad R_{\dot{\alpha} \beta \alpha} {}^{\beta} = 3 G_{\alpha \dot{\alpha}} \ .
\end{equation}
Furthermore, by using the Bianchi identities in \cite{Wess:1992cp} we deduce the following equations
\begin{align} \label{DG_antisymmetric}
 \mathcal{D}_\alpha G_{\beta \dot{\beta}} - \mathcal{D}_\beta G_{\alpha \dot{\beta}} &=  \epsilon_{\alpha \beta} \bar{\mathcal{D}}_{\dot{\beta}} \bar{R} \ , \\
 \label{DbG_antisymmetric} \bar{\mathcal{D}}_{\dot{\alpha}} G_{\beta \dot{\beta}} - \bar{\mathcal{D}}_{\dot{\beta}} G_{\beta \dot{\alpha}} &= \epsilon_{\dot{\alpha} \dot{\beta}} \mathcal{D}_\beta R \ .
\end{align}
We now proceed to present a catalog of those component identities for the chiral superfields which are required for the computation of the component forms of the higher-derivative operators in eq.~\eqref{two_derivative_operators} and tab.~\ref{list_of_four_superspace_ops}. The lowest order components for $\Phi$ and $\bar{\Phi}$ can be found from their definition in eq.~\eqref{ComponentsofPhi}. Since we are interested only in those bosonic components, we start by identifying purely fermionic terms. In general, these are components with an odd number of spinorial covariant derivatives acting on $\Phi$ or $\bar{\Phi}$. A simple example is given by
\begin{equation}\label{DDDferm}
 \tfrac{1}{\sqrt{2}} \mathcal{D}_{\alpha} \mathcal{D}^2 \Phi \rvert = - \tfrac{4}{3} \chi_{\alpha} \bar{M} \ ,
\end{equation}
which can be checked by means of eq.~\eqref{chiral_projector}.

The bosonic terms are those with an even number of spinorial covariant derivatives. In the remainder of this appendix we display only the bosonic terms of the components. In particular, at the level of two-superspace derivatives we find
\begin{equation}\label{DbDPhi}
 \bar{\mathcal{D}}_{\dot{\alpha}} \mathcal{D}_{\alpha} \Phi \rvert = \{\bar{\mathcal{D}}_{\dot{\alpha}}, \mathcal{D}_{\alpha} \}\Phi \rvert = - T_{\alpha \dot{\alpha}} {}^A \mathcal{D}_A \Phi \rvert = -2i \sigma_{\alpha \dot{\alpha}}^a \mathcal{D}_a \Phi \rvert = -2i \sigma_{\alpha \dot{\alpha}}^a e_a^m \partial_m A\ ,
\end{equation}
where we used eq.~\eqref{torsion_comp} and we displayed only the bosonic terms. It is also useful to note the following identities
\begin{equation}\label{DD=D^2}
 \mathcal{D}_\alpha \mathcal{D}_\beta \Phi = \tfrac{1}{2} \epsilon_{\alpha \beta} \mathcal{D}^2 \Phi \ , \qquad  \bar{\mathcal{D}}_{\dot{\alpha}} \bar{\mathcal{D}}_{\dot{\beta}} \bar{\Phi} = - \tfrac{1}{2} \epsilon_{\dot{\alpha} \dot{\beta}} \bar{\mathcal{D}}^2 \bar{\Phi} \ .
\end{equation}
\subsubsection*{Components of $\mathcal{O}(\mathcal{D}^4)$ acting on $\Phi,\bar{\Phi}$}
Furthermore, there are several components involving four spinorial super-covariant derivatives which are of relevance for the computation of the component action of the operators in sec.~\ref{sec:four_derivative_operators}. Using eq.~\eqref{chiral_projector} and the components of torsion and curvature in eq.~\eqref{torsion_comp} we find the following identities
\begin{equation}\begin{aligned}\label{D4_comp}
 \mathcal{D}^2 \mathcal{D}^2 \Phi \rvert & =  \tfrac{16}{3} F \bar{M}  \\
 \bar{\mathcal{D}}^2 \bar{\mathcal{D}}^2 \bar{\Phi} \rvert &= \tfrac{16}{3} \bar{F} M \\
 \mathcal{D}_\alpha \bar{\mathcal{D}}^2 \mathcal{D}^\alpha \Phi \rvert 
 &= - \tfrac{16}{3} MF \\
 \mathcal{D}^2 \bar{\mathcal{D}}^2\bar{\Phi} \rvert &= 16 \Box \bar{A}  + \tfrac{32}{3} i b^m  \partial_m \bar{A} + \tfrac{32}{3} \bar{M} \bar{F} \\
  \mathcal{D}_{\alpha} \bar{\mathcal{D}}_{\dot{\alpha}} \bar{\mathcal{D}}^2 \bar{\Phi} \rvert
  &= \tfrac{8}{3} i M \sigma_{\alpha \dot{\alpha}}^a e_a^m \partial_m \bar{A} \\
  \mathcal{D}_{\alpha} \bar{\mathcal{D}}_{\dot{\alpha}} \mathcal{D}^2 \Phi \rvert 
 &= 8 i \sigma_{\alpha \dot{\alpha}}^a e_a^m (\partial_m F - \tfrac{1}{3} \bar{M} \partial_m A) \ .
\end{aligned}\end{equation}
\subsubsection*{Components of $\mathcal{O}(\mathcal{D}^6)$ acting on $\Phi,\bar{\Phi}$}
We also need certain components with six spinorial super-covariant derivatives, which read
\begin{equation}\begin{aligned}\label{D6_comp}
  \tfrac{1}{64} \mathcal{D}^2 \bar{\mathcal{D}}^2 \bar{\mathcal{D}}^2 \bar{\Phi} \rvert 
 &= \tfrac{1}{3}\bar{F} \left(\tfrac{1}{2}\mathcal{R}- \tfrac{4}{3} \rvert M \rvert^2 -\tfrac{1}{3} b_a b^a + i\mathcal{D}_m b^m  \right) - \tfrac{1}{3} M \Box \bar{A} \\
 &\qquad - \tfrac{2i}{9} M b^m \partial_m \bar{A}  \\
 \tfrac{1}{64} \bar{\mathcal{D}}^2 \mathcal{D}^2 \bar{\mathcal{D}}^2 \bar{\Phi} \rvert 
 &= \tfrac{1}{3} \bar{F} \left(\tfrac{1}{2}\mathcal{R} - 2 \rvert M \rvert^2 - \tfrac{1}{3} b_a b^a + i \mathcal{D}_m b^m \right) - \Box \bar{F} - \tfrac{1}{3} M \Box \bar{A}  \\
  &\qquad + \tfrac{2}{3}  \partial_m M \partial^m \bar{A} + \tfrac{2i}{3} b^m (\partial_m \bar{F} -  M \partial_m \bar{A})
\end{aligned}\end{equation}
\subsubsection*{Components of $R$}
For the convenience of the reader it is useful to display the relevant components of the superfield $R$ here.
\begin{equation}\begin{aligned}\label{eq:comp_R}
 R\rvert &= - \tfrac{1}{6} M \ \\
  \mathcal{D}_a R \rvert &= - \tfrac{1}{6} e_a^m \partial_m M \\
 \mathcal{D}^2 R \rvert &= \tfrac{2}{3}\left(-\tfrac{1}{2}\mathcal{R} + \tfrac{2}{3}\rvert M \rvert^2 + \tfrac{1}{3} b_a b^a - i  \mathcal{D}_m b^m \right) \\
 \bar{\mathcal{D}}^2 \mathcal{D}^2 R \rvert &= - \tfrac{8}{3} \Box M  + \tfrac{16}{9} i b^m \partial_m M - \tfrac{16}{9} M \left(-\tfrac{1}{2}\mathcal{R} + \tfrac{2}{3}\rvert M \rvert^2 + \tfrac{1}{3} b_a b^a - i \mathcal{D}_m b^m \right)
\end{aligned}\end{equation}
\subsubsection*{Components with $\mathcal{D}_a$ acting on $\Phi,\bar{\Phi}$}
In addition, several components involving super-covariant derivatives of $\mathcal{D}_a \bar{\Phi}$ and $\mathcal{D}_a \Phi$ are relevant, in particular for the computation of $\mathcal{O}_{(2 \rvert 3)}$ in tab.~\ref{list_of_four_superspace_ops}. The important identities involving two spinorial covariant derivatives are given by
\begin{equation}\begin{aligned}\label{Da_comp}
  \mathcal{D}^2 \mathcal{D}_a \bar{\Phi} \rvert &= \tfrac{2}{3} \bar{M} e_a^m \partial_m \bar{A} \\
  \bar{\mathcal{D}}^2 \mathcal{D}_a \Phi \rvert &= \tfrac{2}{3} M e_a^m \partial_m A \\
  \mathcal{D}^\alpha \bar{\mathcal{D}}^{\dot{\alpha}} \mathcal{D}^a \Phi \rvert &= \tfrac{i}{3} M F \bar{\sigma}^{a  \dot{\alpha}\alpha } \\
  \mathcal{D}^2 \mathcal{D}_a \Phi \rvert &= -\tfrac{8i}{3}F b_a - \tfrac{2}{3} \bar{M} e_a^m \partial_m A - 4 e_a^m \partial_m F \\
  \mathcal{D}^\alpha \bar{\mathcal{D}}^{\dot{\alpha}} \mathcal{D}^a \bar{\Phi} \rvert &= - 2i \bar{\sigma}_{b}^{ \dot{\alpha}\alpha} ( \tfrac{1}{3} \epsilon_{cd} {}^{ab} b^c e^d_m \partial^m \bar{A} + e^b_m e^a_n \mathcal{D}^m  \partial^n \bar{A} ) - \tfrac{i}{3} \bar{\sigma}^{a  \dot{\alpha} \alpha} \bar{M} \bar{F} \ .
\end{aligned}\end{equation}
The above components were also determined in \cite{Baumann:2011nm}. Note that we find agreement with the results in that reference, apart from the last component identity which differs by a minus sign in the first and last term.

Finally, there are also components with four spinorial covariant derivatives acting on $\mathcal{D}_a \bar{\Phi}$ and $\mathcal{D}_a \Phi$ and they read
\begin{equation}\begin{aligned}\label{D4_Da_comp}
  \tfrac{1}{16} \mathcal{D}^2 \bar{\mathcal{D}}^2 \mathcal{D}_a \Phi\rvert  &= - \tfrac{1}{6} (-\tfrac{1}{2}\mathcal{R}+ \tfrac{5}{6} \lvert M \rvert^2 + \tfrac{1}{3} b_c b^c - i \mathcal{D}_m b^m) \partial_a A \\
  &\quad - \tfrac{i}{9}MFb_a - \tfrac{1}{6} M \partial_a F \\
  \tfrac{1}{16} \mathcal{D}^2 \bar{\mathcal{D}}^2 \mathcal{D}_a \bar{\Phi}\rvert & = e^m_a \Bigl[ \mathcal{D}_m (\Box \bar{A} + \tfrac{2}{3}ib^n \partial_n \bar{A}) - \tfrac{i}{3}  \bar{F} \bar{M} b_m + \tfrac{1}{3}\bar{F} \partial_m \bar{M} \\
 & \quad - (\mathcal{R}_{mn} + \tfrac{2}{9} b_m b_n + \tfrac{2}{3} i \mathcal{D}_n b_m - \tfrac{1}{3} \epsilon_{pmqn} \mathcal{D}^q b^p) \partial^n \bar{A} \\
 & \quad + \tfrac{5}{6} \bar{M} \partial_m \bar{F}  + (\tfrac{1}{12} \mathcal{R} + \tfrac{1}{12} \lvert M \rvert^2 + \tfrac{1}{6} b_c b^c - \tfrac{i}{6} \mathcal{D}_n b^n)  \partial_m \bar{A}\Bigr] \ .
\end{aligned}\end{equation}
Note that the first identity differs from the result in \cite{Baumann:2011nm} by a factor multiplying the first term.
\subsubsection*{Components of $G_a$}
For the computation of higher-components of the superfield $\mathcal{D}_a \Phi$ and its conjugate we need certain components of $G_a$. The relevant identities read
\begin{align} \label{D^2G}
 \mathcal{D}^2 G_a \rvert = \tfrac{2}{3}(\bar{M}b_a - ie_a^m \partial_m \bar{M}) \ , \qquad \qquad \qquad  \qquad \, \\
\begin{aligned}
 \label{DbDG} \bar{\sigma}^{\dot{\beta} \beta}_b \bar{\mathcal{D}}_{\dot{\beta}} \mathcal{D}_\beta G_d \rvert &= \mathcal{R}_{bd} - \eta_{bd}(\tfrac{1}{6} \mathcal{R}+ \tfrac{1}{9} \lvert M \rvert^2 + \tfrac{1}{9} b_a b^a) + \tfrac{2}{9} b_b b_d \\
 & \quad - \tfrac{2i}{3} e_b^m  \mathcal{D}_m b_d + \tfrac{1}{3} \epsilon_{abcd} \, e^c_m \mathcal{D}^m b^a  \ .
 \end{aligned}
\end{align}
These results perfectly agree with \cite{Baumann:2011nm}.
\subsection*{Derivation of Component Identities}
Let us now give some of the derivations of the component identities which listed in the previous section. Here we constrain ourselves to the discussion of the more involved computations. In particular, the identities in eq.~\eqref{D4_comp} are easily obtained by means of eq.~\eqref{chiral_projector} together with the (anti-)commutation relations and the components of the torsion and curvature and, hence, do not discuss them any further here. Instead let us demonstrate the identities in eq.~\eqref{D6_comp}. We rewrite the first component expression using eq.~\eqref{chiral_projector} as follows
\begin{equation}
 \mathcal{D}^2 \bar{\mathcal{D}}^2 \bar{\mathcal{D}}^2 \bar{\Phi} \rvert  = 8 \mathcal{D}^2 R\rvert \, \bar{\mathcal{D}}^2 \bar{\Phi}\rvert \,+\, 8 R\rvert \, \mathcal{D}^2 \bar{\mathcal{D}}^2 \bar{\Phi}\rvert \ .
\end{equation}
Inserting the required components we directly arrive at the result in eq.~\eqref{D6_comp}. The second identity in eq.~\eqref{D6_comp} reads
\begin{equation}
 \bar{\mathcal{D}}^2 \mathcal{D}^2 \bar{\mathcal{D}}^2 \bar{\Phi} \rvert = \bar{\mathcal{D}}^2 \mathcal{D}^2 (\bar{\mathcal{D}}^2 - 8R ) \bar{\Phi} \rvert + 8 \bar{\Phi}\rvert\, \bar{\mathcal{D}}^2 \mathcal{D}^2 R \rvert + 8 \bar{\mathcal{D}}^2 \bar{\Phi} \rvert \, \mathcal{D}^2 R \rvert \ .
\end{equation}
Here, only the first term on the r.h.s. needs further attention, the remaining terms can be read off from eq.~\eqref{eq:comp_R}. Since $(\bar{\mathcal{D}}^2 - 8R ) \bar{\Phi}$ is a covariantly chiral superfield, we need to determine the $\Theta$-expansion of this chiral superfield and then use eq.~\eqref{D4_comp}. Fortunately, this expansion was already given in \cite{Wess:1992cp} and we simply insert the respective components here. Altogether, we then find the result in eq.~\eqref{D6_comp}.

Next, we turn to eq.~\eqref{D4_Da_comp} together with the necessary auxiliary results in eqs.~\eqref{D^2G}, \eqref{DbDG}. Let us mention again that these have already been derived in \cite{Baumann:2011nm}, but as a cross-check we re-derive them here. In particular, since we find disagreement regarding the first identity in eq.~\eqref{D4_Da_comp}, it important to explain why we obtain a different result. This identity can be computed rather straightforwardly by using the algebra of super-covariant derivatives together with the torsion components. More precisely, we find 
\begin{equation}
 \mathcal{D}^2 \bar{\mathcal{D}}^2 \mathcal{D}_a \Phi\rvert = - 4(\mathcal{D}^2 R \rvert \, \mathcal{D}_a \Phi \rvert + R \rvert \,\mathcal{D}^2 \mathcal{D}_a \Phi \rvert)
\end{equation}
After inserting the necessary superfield-component identities we indeed find the displayed result in eq.~\eqref{D4_Da_comp}. 

Next, we turn to the auxiliary results in eqs.~\eqref{D^2G}, \eqref{DbDG}. Firstly, by acting with a super-covariant derivative $\mathcal{D}^\alpha$ on eq.~\eqref{DG_antisymmetric} we find eq.~\eqref{D^2G}. Alternatively, we can act with $\bar{\mathcal{D}}_{\dot{\alpha}}$ on eq.~\eqref{DG_antisymmetric} which yields
\begin{equation}\label{Star1}
 \bar{\mathcal{D}}_{\dot{\alpha}} \mathcal{D}_\alpha G_{\beta \dot{\beta}} - \bar{\mathcal{D}}_{\dot{\alpha}} \mathcal{D}_\beta G_{\alpha \dot{\beta}} = - \tfrac{1}{2} \epsilon_{\alpha \beta} \epsilon_{\dot{\alpha} \dot{\beta}} \bar{\mathcal{D}}^2 \bar{R} \ .
\end{equation}
Analogously, from eq.~\eqref{DbG_antisymmetric} we derive the following identity
\begin{equation}\label{Star2}
   \mathcal{D}_\alpha \bar{\mathcal{D}}_{\dot{\alpha}} G_{\beta \dot{\beta}} -  \mathcal{D}_\alpha \bar{\mathcal{D}}_{\dot{\beta}} G_{\beta \dot{\alpha}} = \tfrac{1}{2} \epsilon_{\alpha \beta} \epsilon_{\dot{\alpha} \dot{\beta}} \mathcal{D}^2 R \ .
\end{equation}
We can decompose $\bar{\sigma}^{\dot{\beta} \beta}_b \bar{\mathcal{D}}_{\dot{\beta}} \mathcal{D}_\beta G_d$ into symmetric and anti-symmetric components as 
\begin{equation}
 \bar{\sigma}^{\dot{\beta} \beta}_b  \bar{\mathcal{D}}_{\dot{\beta}} \mathcal{D}_\beta G_{d} = \tfrac{1}{2}  ( \bar{\sigma}^{\dot{\beta} \beta}_b \bar{\mathcal{D}}_{\dot{\beta}} \mathcal{D}_\beta G_{d} + \bar{\sigma}^{\dot{\delta} \delta }_d \bar{\mathcal{D}}_{\dot{\delta}} \mathcal{D}_\delta G_{b} ) + \tfrac{1}{2} ( \bar{\sigma}^{\dot{\beta} \beta}_b\bar{\mathcal{D}}_{\dot{\beta}} \mathcal{D}_\beta G_{d} -  \bar{\sigma}^{\dot{\delta} \delta }_d\bar{\mathcal{D}}_{\dot{\delta}} \mathcal{D}_\delta G_{b} ) \ .
\end{equation}
The anti-symmetric part can be easily computed by using eqs.~\eqref{Star1}, \eqref{Star2} and the commutation relations.\footnote{It is also necessary to use the $\sigma$-matrix trace identity $Tr(\sigma_a \bar{\sigma}_b \sigma_c \bar{\sigma}_d) = 2(\eta_{ab} \eta_{cd} - \eta_{ac} \eta_{bd} + \eta_{ad} \eta_{bc} - i \epsilon_{abcd})$.} We find
\begin{equation}
 \bar{\sigma}^{\dot{\beta} \beta}_b \bar{\sigma}^{\dot{\delta} \delta }_d( \bar{\mathcal{D}}_{\dot{\beta}} \mathcal{D}_\beta G_{\delta \dot{\delta}} - \bar{\mathcal{D}}_{\dot{\delta}} \mathcal{D}_\delta G_{\beta \dot{\beta}} ) = 4i(\mathcal{D}_d G_b -  \mathcal{D}_b G_d) - 4 \epsilon_{adcb} \mathcal{D}^c G^a \ .
\end{equation}
The symmetric component of $\bar{\sigma}^{\dot{\beta} \beta}_b \bar{\mathcal{D}}_{\dot{\beta}} \mathcal{D}_\beta G_d$ can be derived by computing 
\begin{equation}
 \eta^{ac} R_{abcd} = \tfrac{1}{16} \eta^{ac} \bar{\sigma}^{\dot{\alpha} \alpha}_a \bar{\sigma}^{\dot{\beta} \beta}_b \bar{\sigma}^{\dot{\gamma} \gamma}_c \bar{\sigma}^{ \dot{\delta} \delta}_d R_{\alpha \dot{\alpha} \beta  \dot{\beta} \gamma  \dot{\gamma} \delta \dot{\delta}}
\end{equation}
from the formula displayed in \cite{Wess:1992cp} which is determined by the solution to the Bianchi identities. To simplify the resulting expression we use trace identities for the $\sigma$-matrices and eq.~\eqref{Star1} such that
\begin{equation}\begin{aligned}
 &\frac{1}{2} \left( \bar{\sigma}^{\dot{\beta} \beta}_b \bar{\mathcal{D}}_{\dot{\beta}} \mathcal{D}_\beta G_{d} + \bar{\sigma}^{\dot{\delta} \delta }_d \bar{\mathcal{D}}_{\dot{\delta}} \mathcal{D}_\delta G_{b} \right) \\
 &=  R_{bd} - \eta_{bd} \left[ 12 R \bar{R} + 2G_a G^a - \frac{1}{4} \left(\mathcal{D}^2 R + \bar{\mathcal{D}}^2 \bar{R}\right) \right] + 2 G_b G_d + i(\mathcal{D}_b G_d + \mathcal{D}_d G_b) \ .
\end{aligned}\end{equation}
Altogether, this demonstrates eq.~\eqref{DbDG}. We are now in a position to derive the missing identity in eq.~\eqref{D4_Da_comp}. Making iterative use of the (anti-)commutation relations we find
\begin{equation}\begin{aligned}\label{D2Db2DaP}
 \mathcal{D}^2 \bar{\mathcal{D}}^2 \mathcal{D}_a \bar{\Phi} &= - \mathcal{D}^2 \bar{\mathcal{D}}^{\dot{\alpha}} (T_{\dot{\alpha} a} {}^{\dot{\beta}} \bar{\mathcal{D}}_{\dot{\beta}} \bar{\Phi}) - \mathcal{D}^2(R_{\dot{\alpha}a \dot{\delta}} {}^{\dot{\alpha}} \bar{\mathcal{D}}^{\dot{\delta}} \bar{\Phi}) - \mathcal{D}^2(T_{\dot{\alpha} a} {}^{\gamma}\mathcal{D}_\gamma  \bar{\mathcal{D}}^{\dot{\alpha}} \bar{\Phi}) \\
 &\quad + \mathcal{D}^2(T_{\dot{\alpha} a} {}^{\dot{\gamma}} \bar{\mathcal{D}}_{\dot{\gamma}}  \bar{\mathcal{D}}^{\dot{\alpha}} \bar{\Phi}) + \mathcal{D}^\alpha (- T_{\alpha a} {}^{\gamma} \mathcal{D}_\gamma \bar{\mathcal{D}}^2 \bar{\Phi} + T_{\alpha a} {}^{\dot{\gamma}} \bar{\mathcal{D}}_{\dot{\gamma}} \bar{\mathcal{D}}^2 \bar{\Phi}) \\
 &\quad + R_{\alpha a \delta} {}^{\alpha} \mathcal{D}^\delta  \bar{\mathcal{D}}^2 \bar{\Phi} + T_{\alpha a} {}^\gamma \mathcal{D}_\gamma \mathcal{D}^\alpha \bar{\mathcal{D}}^2 \bar{\Phi}- T_{\alpha a} {}^{\dot{\gamma}}  \bar{\mathcal{D}}_{\dot{\gamma}}\mathcal{D}^\alpha \bar{\mathcal{D}}^2\bar{\Phi} + \mathcal{D}_a \mathcal{D}^2 \bar{\mathcal{D}}^2\bar{\Phi} \ .
\end{aligned}\end{equation}
This precisely coincides with the respective result in \cite{Baumann:2011nm}. To determine the final component form one has to compute the individual terms. The first and second term in eq.~\eqref{D2Db2DaP} require most effort. The remaining terms in eq.~\eqref{D2Db2DaP} are easier to compute and we omit their details here. After some work, using the algebra of super-covariant derivatives and eqs.~\eqref{torsion_comp}, \eqref{curvature_comp}, \eqref{DbG_antisymmetric} we find
\begin{equation}\begin{aligned}
 & - \mathcal{D}^2 \bar{\mathcal{D}}^{\dot{\alpha}} (T_{\dot{\alpha} a} {}^{\dot{\beta}} \bar{\mathcal{D}}_{\dot{\beta}} \bar{\Phi}) \rvert - \mathcal{D}^2(R_{\dot{\alpha}a \dot{\delta}} {}^{\dot{\alpha}} \bar{\mathcal{D}}^{\dot{\delta}} \bar{\Phi})\rvert \\
 & = i \mathcal{D}^2 G_a\rvert \, \bar{\mathcal{D}}^2 \bar{\Phi}\rvert + iG_a \rvert\, \mathcal{D}^2 \bar{\mathcal{D}}^2 \bar{\Phi}\rvert + 64 i \mathcal{D}_b G_a \rvert\,\mathcal{D}^b \bar{\Phi} \rvert - 16 \mathcal{D}^b \bar{\Phi} \rvert\,\bar{\sigma}_b^{\dot{\beta} \beta} \bar{\mathcal{D}}_{\dot{\beta}} \mathcal{D}_\beta G_a \rvert 
\end{aligned}\end{equation}
To obtain the final component expression, it remains to insert the required component formulae which, in particular, encompass eqs.~\eqref{D^2G}, \eqref{DbDG}. Finally, we arrive at the displayed component form in eq.~\eqref{D4_Da_comp} which agrees with the result in \cite{Baumann:2011nm}.
\section{Classification of Four-Superspace Derivative Operators}\label{appendix:integration_by_parts}
In this appendix we classify the possible four superspace-derivative operators and reduce them to a minimal set of relevant operators by using the commutation relations in eq.~\eqref{anticom} as well as integration by parts identities. As mentioned in sec.~\ref{HD_Sugra_overview} we have to include also the gravitational superfields $R, G_a$ and $W_{\alpha \beta \gamma}$ into the analysis. $R$ and $G_a$ count as the equivalent of two and $W_{\alpha \beta \gamma}$ as the equivalent of three spinorial covariant derivatives. Operators which are directly related via the identity $\{ \mathcal{D}_\alpha , \bar{\mathcal{D}}_{\dot{\alpha}} \} \Phi \sim \mathcal{D}_a \Phi$, via eq.~\eqref{DD=D^2} or via the chirality condition in eq.~\eqref{chiral_projector} are easy to identify and, therefore, for the sake of brevity we do not need to distinguish them explicitly here.

We conduct the classification stepwise by listing those operators first which depend only on super-covariant derivatives acting on $\Phi, \bar{\Phi}$ (and, hence, do not involve the gravitational superfield). Any operator of this type can be labeled by the number of chiral or anti-chiral superfields on which covariant derivatives act. This number ranges between four and one. All remaining operators carry an explicit dependence on the superfields $R, \bar{R}$ and $G_a$. The results are displayed in table~\ref{four_der_operators}.
\begin{table}
\onehalfspacing
 \centering
 \begin{tabular}{| l | l | l |}
  \hline
  Label & Form of Operator & Real \\\hline
 $\mathcal{O}_{(4 \rvert 1)}$ & $(\mathcal{D}_\alpha \Phi \mathcal{D}^\alpha \Phi )^2 $ &  \\
 $\mathcal{O}_{(4 \rvert 2)}$ & $\mathcal{D}_\alpha \Phi \mathcal{D}^\alpha \Phi \bar{\mathcal{D}}_{\dot{\alpha}} \bar{\Phi}  \bar{\mathcal{D}}^{\dot{\alpha}} \bar{\Phi} $ & \checkmark \\ \hline 
 $\mathcal{O}_{(3 \rvert 1)}$ & $\mathcal{D}^2 \Phi \mathcal{D}_\alpha \Phi \mathcal{D}^\alpha \Phi $ &\\
 $\mathcal{O}_{(3 \rvert 2)}$ & $\bar{\mathcal{D}}^2 \bar{\Phi} \mathcal{D}_\alpha \Phi \mathcal{D}^\alpha \Phi  $ &\\
 $\mathcal{O}_{(3 \rvert 3)}$ & $\bar{\mathcal{D}}_{\dot{\alpha}} \mathcal{D}_\alpha \Phi  \mathcal{D}^\alpha \Phi \bar{\mathcal{D}}^{\dot{\alpha}} \bar{\Phi}  $ &\\\hline
 $\mathcal{O}_{(2\rvert 1)}$ & $\mathcal{D}^2 \Phi \bar{\mathcal{D}}^2 \bar{\Phi} $ & \checkmark\\
 $\mathcal{O}_{(2\rvert 2)}$ & $(\mathcal{D}^2 \Phi )^2 $ &\\
 $\mathcal{O}_{(2\rvert 3)}$ & $\mathcal{D}_a \Phi \mathcal{D}^a \bar{\Phi} $   & \checkmark \\
 $\mathcal{O}_{(2\rvert 4)}$ & $\mathcal{D}_a \Phi \mathcal{D}^a \Phi  $  &\\
 $\mathcal{O}_{(2\rvert 5)}$ & $\bar{\mathcal{D}}_{\dot{\alpha}} \mathcal{D}^2 \Phi \bar{\mathcal{D}}^{\dot{\alpha}} \bar{\Phi} $ &\\
 {\color{darkblue} $\mathcal{O}_{(2\rvert 6)}$ } & {\color{darkblue} $\mathcal{D}^2 \bar{\mathcal{D}}_{\dot{\alpha}} \bar{\Phi} \bar{\mathcal{D}}^{\dot{\alpha}} \bar{\Phi} $ } &\\
 {\color{darkblue} $\mathcal{O}_{(2\rvert 7)}$ } & {\color{darkblue} $\mathcal{D}^\alpha \bar{\mathcal{D}}_{\dot{\alpha}}\mathcal{D}_\alpha \Phi \bar{\mathcal{D}}^{\dot{\alpha}} \bar{\Phi} $ } &\\ \hline
 {\color{darkblue} $\mathcal{O}_{(1\rvert 1)}$ } & {\color{darkblue} $i \bar{\sigma}^{a  \dot{\alpha}\alpha}  \bar{\mathcal{D}}_{\dot{\alpha}} \mathcal{D}_a \mathcal{D}_\alpha \Phi  $ } &\\
 $\mathcal{O}_{(1\rvert 2)}$ & $\mathcal{D}_a \mathcal{D}^a \Phi  $ &\\
 {\color{darkblue} $\mathcal{O}_{(1\rvert 3)}$ } & {\color{darkblue} $\bar{\mathcal{D}}^2 \mathcal{D}^2 \Phi $ } &\\
 {\color{darkblue} $\mathcal{O}_{(1\rvert 4)}$ } & {\color{darkblue} $\mathcal{D}_\alpha \bar{\mathcal{D}}^2 \mathcal{D}^\alpha \Phi $ } &\\
 {\color{darkblue} $\mathcal{O}_{(1\rvert 5)}$ } & {\color{darkblue} $\bar{\mathcal{D}}_{\dot{\alpha}} \mathcal{D}_\alpha\bar{\mathcal{D}}^{\dot{\alpha}} \mathcal{D}^\alpha \Phi  $ } &\\ \hline
  $\mathcal{O}_{(R \rvert 1)}$ &  $R \mathcal{D}_\alpha \Phi \mathcal{D}^\alpha \Phi $ &\\
  $\mathcal{O}_{(R \rvert 2)}$ &  $R \mathcal{D}^2 \Phi $ &\\
  $\mathcal{O}_{(R \rvert 3)}$ &  $R^2  $& \\
  $\mathcal{O}_{(\bar{R} \rvert 1)}$ &  $\bar{R} \mathcal{D}_\alpha \Phi \mathcal{D}^\alpha \Phi $ &\\
  $\mathcal{O}_{(\bar{R} \rvert 2)}$ &  $\bar{R} \mathcal{D}^2 \Phi $ &\\
  $\mathcal{O}_{(G \rvert 1)}$ &  $G_a \mathcal{D}^a \Phi $ &\\
  $\mathcal{O}_{(G \rvert 2)}$ &  $G_{\alpha \dot{\alpha}} \mathcal{D}^\alpha \Phi \bar{\mathcal{D}}^{\dot{\alpha}} \bar{\Phi}$ & \checkmark \\
  $\mathcal{O}_{(R \rvert \mathcal{D}^2)}$ &  $\mathcal{D}^2 R $ &\\
  $\mathcal{O}_{(R \rvert \mathcal{D})}$ & $\mathcal{D}_\alpha R \mathcal{D}^\alpha \Phi $ &\\
  {\color{darkblue} $\mathcal{O}_{(G \rvert \mathcal{D})}$ } & {\color{darkblue} $\mathcal{D}^a  G_a$ } & {\color{darkblue} \checkmark } \\ \hline
 \end{tabular}
  \caption{List of operators at four superspace-derivative level. Operators whose $\theta=\bar{\theta}=0$-component is not real-valued have to be completed by their conjugate expressions at the level of the action as in eq.~\eqref{LHD}. The individual operators are understood as being multiplied by a superfield $T(\Phi,\bar{\Phi})$ (and $\bar{T}(\Phi,\bar{\Phi})$ for their conjugate parts, in case that the operator is not real). Operators, which are displayed in blue color can be recast in terms of other (black) operators by means of the algebra of covariant derivatives eq.~\eqref{anticom}. Note that we omitted several operators here, for which it can be seen quite easily that they are redundant and, therefore, be recast in terms of other operators.}
  \label{four_der_operators}
\end{table}
Note that we did not include the higher-curvature operators in table~\ref{four_der_operators}. These were already displayed in eq.~\eqref{Sprime1} and briefly discussed in sec.~\ref{sec:higher_curvature_ops}.\footnote{One may wonder why the superfield $W_{\alpha \beta \gamma}$ does not appear in table~\ref{four_der_operators}. The reason for this is that, since $W_{\alpha \beta \gamma}$ has mass dimension three, we have to contract two of its indices to build a scalar operator. However, as $W_{\alpha \beta \gamma}$ is completely symmetric, such operators vanish identically.}

The list of operators in table~\ref{four_der_operators} is highly degenerate and displays several redundant operators. One of the tools that allow us to identify redundant operators is the algebra of super-covariant derivatives given in eq.~\eqref{anticom}. Making repetitive use of eq.~\eqref{anticom} and occasionally also of eq.~\eqref{DG} we deduce the following list of relations among the operators in table~\ref{four_der_operators}
\begin{equation}\begin{aligned}\label{list_redudancies_algebra}
 &\mathcal{O}_{(1\rvert 1)} = - \tfrac{5}{2} \mathcal{O}_{(R\rvert \mathcal{D})} - 10 i \mathcal{O}_{(G \rvert 1)} - \tfrac{1}{2} \mathcal{O}_{(1\rvert 5)} \\
 &\mathcal{O}_{(1\rvert 5)} = 8 \mathcal{O}_{(1\rvert 2)} + 8 \mathcal{O}_{(R\rvert 2)} - 8 \mathcal{O}_{(R\rvert \mathcal{D})} \\
 &\mathcal{O}_{(1\rvert 4)} =  - 8 \mathcal{O}_{(R\rvert 2)} + 8 \mathcal{O}_{(R\rvert \mathcal{D})} \\
 &\mathcal{O}_{(1\rvert 3)} = 16 \mathcal{O}_{(1\rvert 2)} + 16 \mathcal{O}_{(R \rvert 2)} + 32i \mathcal{O}_{(G \rvert 1)} - 8 \mathcal{O}_{(R \rvert \mathcal{D})} \\
 &\mathcal{O}_{(2\rvert 5)} = -8 \mathcal{O}_{(G\rvert 2)} - 2 \mathcal{O}_{(2\rvert 7)} \\
 &\mathcal{O}_{(2\rvert 6)} = - 8 \mathcal{O}_{(R\rvert 1)} \\
 &\mathcal{O}_{(R\rvert \mathcal{D}^2)} - \text{h.c.} = 4 i \mathcal{O}_{(G \rvert \mathcal{D})}  \ .
\end{aligned}\end{equation}
In total this reduces the list in table~\ref{four_der_operators} by seven operators. In the following we make the particular choice to delete the operators $\{\mathcal{O}_{(1\rvert 5)}$, $\mathcal{O}_{(1\rvert 4)}$, $\mathcal{O}_{(1\rvert 3)}$, $\mathcal{O}_{(1\rvert 1)}$, $\mathcal{O}_{(2\rvert 7)}$, $\mathcal{O}_{(2\rvert 6)}$, $\mathcal{O}_{(G \rvert \mathcal{D})} \}$ from the list of relevant operators. To indicate this we marked these operators in blue color in table~\ref{four_der_operators}.

To further reduce the number of independent operators we now apply a second tool, namely integration by parts identities in curved superspace. These identities can be stated as the fact that the following integrals are equivalent to superspace-surface terms \cite{Binetruy:2000zx}
\begin{equation}
 \int \d^8 z  E \mathcal{D}_\alpha V^\alpha \ , \qquad \int \d^8 z  E  \bar{\mathcal{D}}^{\dot{\alpha}} V_{\dot{\alpha}} \ , \qquad \int\d^8 z  E  \mathcal{D}_a V^a \ ,
\end{equation}
where $V^A$ is an arbitrary covariant superfield. As a first example we now consider 
\begin{equation}
 V^\alpha = S(\Phi,\bar{\Phi}) \mathcal{D}^\alpha \Phi \ ,
\end{equation}
where $S$ is an arbitrary scalar superfield and depends on the superfields $\Phi, \bar{\Phi}$. For the above $V^\alpha$ we find that up to boundary terms 
\begin{equation}\label{int_by_parts}
 0 = \int   \d^8 z  E  \mathcal{D}_\alpha ( S(\Phi,\bar{\Phi}) \mathcal{D}^\alpha \Phi)  = \int  \d^8 z  E  (-  S(\Phi, \bar{\Phi}) \mathcal{D}^2 \Phi + \frac{\partial S}{\partial \Phi} \mathcal{D}_\alpha \Phi \mathcal{D}^\alpha \Phi)  \ ,
\end{equation}
which together with the analogous result for $V_{\dot{\alpha}} = \bar{S}(\Phi,\bar{\Phi}) \bar{\mathcal{D}}_{\dot{\alpha}} \bar{\Phi}$ shows that the operators $\mathcal{O}_{(1)}$ and $\mathcal{O}_{(2)}$ in eq.~\eqref{two_derivative_operators} are equivalent. 

Next we turn to integration by parts identities at the four-superspace derivative level. Different identities arise from choosing distinct superfields $V^a, V^\alpha, V_{\dot{\alpha}}$. However, it suffices to take into account only those identities which arise from superfields with an undotted spinor index $V^\alpha$, since for any superfield $V_{\dot{\alpha}}$ there exists a conjugate superfield $V^\alpha$ whose integration by parts identity expresses the conjugate of the integration by parts identity for $V_{\dot{\alpha}}$. Moreover, superfields $V^a$ fulfill the relation
\begin{equation}
 \mathcal{D}_a V^a = - \tfrac{i}{4} \{ \mathcal{D}_\alpha, \bar{\mathcal{D}}_{\dot{\alpha}} \} V^{\alpha \dot{\alpha}} = - \tfrac{i}{4} (\mathcal{D}_\alpha \tilde{V}^\alpha - \bar{\mathcal{D}}^{\dot{\alpha}} \tilde{V}_{\dot{\alpha}}) \ ,
\end{equation}
where $\tilde{V}^\alpha = \bar{\mathcal{D}}_{\dot{\alpha}} V^{\alpha \dot{\alpha}}$ and $ \tilde{V}_{\dot{\alpha}} = -\mathcal{D}^{\alpha} V_{\alpha \dot{\alpha}}$ and, hence, yield integration by parts identities that can be rewritten in terms of the spinorial superfields $\tilde{V}^\alpha, \tilde{V}_{\dot{\alpha}}$. Our task is, therefore, to classify all possible higher-derivative spinorial superfields $V^\alpha$, such that the collective mass-dimension of the objects $(\mathcal{D}_A, R,\bar{R}, G_a, W_{\alpha \beta \gamma},\bar{W}_{\dot{\alpha} \dot{\beta} \dot{\gamma}})$ appearing in $V^\alpha$ is given by $\Lambda^{3/2}$ (i.e.~superfields of order $\Delta=3$). Again care must be taken, since some of these superfields are related via the algebra of covariant derivatives in eq.~\eqref{anticom}. To give some examples, consider $V^\alpha = \bar{\mathcal{D}}^{\dot{\alpha}} G_{\dot{\alpha}}^\alpha$ which is equivalent to $\mathcal{D}^\alpha R$ via eq.~\eqref{DG}. Further examples are given by
\begin{equation}\begin{aligned}
 V^\alpha = \bar{\mathcal{D}}^2 \mathcal{D}^\alpha \Phi = 8 R \mathcal{D}^\alpha \Phi \ , \qquad V^\alpha = \mathcal{D}^\alpha \bar{\mathcal{D}}^2 \bar{\Phi} =  - 8G_{\dot{\alpha}}^\alpha \bar{\mathcal{D}}^{\dot{\alpha}} \bar{\Phi} + 2 \bar{\mathcal{D}}^{\dot{\alpha}}\mathcal{D}^\alpha \bar{\mathcal{D}}_{\dot{\alpha}} \bar{\Phi} \ .
\end{aligned}\end{equation}
\begin{table}[!htb]
\onehalfspacing
 \centering
 \begin{tabular}{| l | l |}\hline
  $ \qquad \qquad V^\alpha$ & \text{Equivalence of operators} \\ \hline 
  $ T(\Phi,\bar{\Phi}) \mathcal{D}_\beta \Phi \mathcal{D}^\beta \Phi \mathcal{D}^\alpha \Phi $ &  $\mathcal{O}_{(4 \lvert 1)} \simeq 0$ \\
  $ T(\Phi,\bar{\Phi}) \bar{\mathcal{D}}^{\dot{\alpha}} \bar{\Phi} \bar{\mathcal{D}}_{\dot{\alpha}} \bar{\Phi} \mathcal{D}^\alpha \Phi $ & $ \mathcal{O}_{(3 \lvert 2)}\simeq \mathcal{O}_{(4 \lvert 2)} + \mathcal{O}_{(3 \lvert 3)}$  \\
  $ T(\Phi,\bar{\Phi}) \bar{\mathcal{D}}^2 \bar{\Phi} \mathcal{D}^\alpha \Phi $ & $ \mathcal{O}_{(2 \lvert 5)} \simeq \mathcal{O}_{(2 \lvert 1)} + \mathcal{O}_{(3 \lvert 2)}$  \\
  $ T(\Phi,\bar{\Phi}) \bar{\mathcal{D}}^{\dot{\alpha}} \bar{\Phi} \bar{\mathcal{D}}_{\dot{\alpha}} \mathcal{D}^\alpha \Phi $ &  $ \mathcal{O}_{(2 \lvert 7)} \simeq \mathcal{O}_{(3 \lvert 3)} + \mathcal{O}_{(2 \lvert 3)}$ \\
  $ T(\Phi,\bar{\Phi}) \bar{\mathcal{D}}^{\dot{\alpha}} \bar{\Phi}  \mathcal{D}^\alpha \bar{\mathcal{D}}_{\dot{\alpha}} \bar{\Phi} $ &  $ \mathcal{O}_{(3 \lvert 3)} \simeq \mathcal{O}_{(2 \lvert 4)} + \mathcal{O}_{(2 \lvert 6)}$ \\
  $ T(\Phi,\bar{\Phi}) \mathcal{D}^2 \Phi \mathcal{D}^\alpha \Phi $ & $ \mathcal{O}_{(\bar{R} \lvert 1)} \simeq \mathcal{O}_{(3 \lvert 1)} + \mathcal{O}_{(2 \lvert 2)}$  \\
  $ T(\Phi,\bar{\Phi}) \bar{\mathcal{D}}_{\dot{\alpha}}  \mathcal{D}^\alpha \bar{\mathcal{D}}^{\dot{\alpha}}  \bar{\Phi} $ &  $\mathcal{O}_{(1\rvert 5)} \simeq \mathcal{O}_{(2\rvert 7)}$ \\
  $ T(\Phi,\bar{\Phi}) R \mathcal{D}^\alpha \Phi $ & $\mathcal{O}_{(R \lvert \mathcal{D})} \simeq \mathcal{O}_{(R \lvert 2)} + \mathcal{O}_{(R \lvert 1)}$ \\
  $ T(\Phi,\bar{\Phi}) \bar{R} \mathcal{D}^\alpha \Phi $ &  $\mathcal{O}_{(\bar{R} \lvert 1)} \simeq \mathcal{O}_{(\bar{R} \lvert 2)}$ \\ 
  $ T(\Phi,\bar{\Phi})  \mathcal{D}^\alpha R $ & $\mathcal{O}_{(R\lvert \mathcal{D}^2)} \simeq \mathcal{O}_{(R\lvert \mathcal{D})} $  \\ 
  $ T(\Phi,\bar{\Phi})  G_{\dot{\alpha}}^\alpha \bar{\mathcal{D}}^{\dot{\alpha}} \bar{\Phi} $ & $\mathcal{O}_{(G \lvert 1)} \simeq \mathcal{O}_{(G \lvert 2)} + \mathcal{O}_{(R \lvert \mathcal{D})}$ \\ \hline
  \end{tabular}
  \caption{Table of independent integration by parts identities. Here the symbol $\simeq$ is understood as merely indicating that the particular operator can be recast into a combination of other operators (up to some numerical coefficients). 
  Here we display only those identities which give rise to a set of linearly independent constraints.}
  \label{integration_by_parts}
\end{table}
We collect the remaining, independent integration by parts identities in table~\ref{integration_by_parts}. It can be explicitly checked that the resulting system of constraints is non-degenerate by computing the rank of the corresponding matrix and, hence, that each identity is independent from the others. In total we find 11 constraints which together with the redundancy coming from the use of the commutation relations reduces the number of relevant operators down to a set of 9. According to the identities in table~\ref{integration_by_parts}, there is some freedom in choosing this set of operators.
\footnote{For instance, we find that the following operators are equivalent to each other $\mathcal{O}_{(3 \rvert 2)} \longleftrightarrow \mathcal{O}_{(3 \rvert 3)}$ and $\mathcal{O}_{(2 \rvert 3)} \longleftrightarrow \mathcal{O}_{(G \rvert 1)} \longleftrightarrow \mathcal{O}_{(G \rvert 2)}$. In fact the last equivalence was identified as a generalized K\"ahler transformation in \cite{Baumann:2011nm} and used to simplify the component Lagrangian.} Here we make the following choice of a basis of independent operators
\begin{equation}\label{eq:app_complete_list_4der_ops}
 \{ \mathcal{O}_{(4 \rvert 2)},\mathcal{O}_{(3 \rvert 1)},\mathcal{O}_{(3 \rvert 3)},\mathcal{O}_{(2 \rvert 1)},\mathcal{O}_{(2 \rvert 2)},\mathcal{O}_{(2 \rvert 3)},\mathcal{O}_{(R \rvert 1)},\mathcal{O}_{(R \rvert 2)},\mathcal{O}_{(R \rvert 3)} \} \ .
\end{equation}
Note that six of these operators induce four-derivative terms for the chiral scalar off-shell. These are precisely the operators which involve an equal number of spinorial and anti-spinorial covariant derivatives and they read \footnote{In the counting of spinorial super-covariant derivatives we consider $\bar{R} \sim \mathcal{D}^2$ and $R \sim \bar{\mathcal{D}}^2$.}
\begin{equation}\label{eq:app_list_4der_ops}
  \{ \mathcal{O}_{(4 \rvert 2)},\mathcal{O}_{(3 \rvert 3)},\mathcal{O}_{(2 \rvert 1)},\mathcal{O}_{(2 \rvert 3)},\mathcal{O}_{(R \rvert 1)},\mathcal{O}_{(R \rvert 2)} \} \ .
\end{equation}
The minimal list of operators in eq.~\eqref{eq:app_complete_list_4der_ops} is the result of the classification of the $\Delta=4$ operators. One may now proceed to compute the component versions of these operators. In sec.~\ref{sec:four_derivative_operators} we display the component forms of the operators in eq.~\eqref{eq:app_list_4der_ops}.
\section{Curvature Constraints from Killing Spinors}\label{app:Killing_spinor}
As we stated in sec.~\ref{sec:sugra_vacua} the preservation of supersymmetry in vacua of $\mathcal{N}=1$ supergravity requires that
\begin{equation}\label{Killing_spinor_int}
 \langle \mathcal{R} \rangle = \tfrac{4}{3} \langle \lvert M \rvert^2 \rangle \ ,
\end{equation}
which also follows from the integrability condition for the Killing spinor and is associated with the vanishing of the gravitino variation in the vacuum. 
For the ordinary two-derivative supergravity this equation is satisfied at the level of the Einstein equations after evaluating the action at the supersymmetric condition $\langle F^i \rangle = 0$ and after integrating out the auxiliary field $M$. In this appendix we investigate the situation for the general higher-derivative supergravity. In principle, we need to compute the component form of the general higher-derivative action which is displayed in eq.~\eqref{LHD}. However, it is not necessary to use the most general higher-derivative theory, but it suffices to analyse a simpler Lagrangian. Since we are interested in the Lagrangian at the supersymmetric points, we want to evaluate the action at $\langle F^i \rangle = 0$. Higher-derivative operators for the chiral superfields, such as operators involving $\mathcal{D}_A \Phi^i$ and $\mathcal{D}_A \bar{\Phi}^{\bar{\jmath}}$, neither contribute to the scalar potential nor to the gravitational part of the action after evaluating at $\langle F^i \rangle = 0$. Therefore, without loss of generality it is sufficient to ignore these operators here. What remains are higher-order operators involving only the gravitational superfields.\footnote{The couplings of these operators are allowed to depend on the chiral multiplets.} We are interested on examining whether the general form of the scalar potential, which is now corrected by higher-powers of $M$ and $\bar{M}$, still allows the condition in eq.~\eqref{Killing_spinor_int} to be satisfied. To this end it is convenient to discuss the following Lagrangian
\begin{equation}\label{Uapp}
 \mathcal{L} = \int \mathrm{d}^4 \theta E \left(-3 U(\Phi,\bar{\Phi},R,\bar{R}) + \frac{W(\Phi)}{2R} + \frac{\bar{W}(\bar{\Phi})}{2\bar{R}} \right) \ .
\end{equation}
Note that there is an infinite tower of purely gravitational higher-order operators which we excluded here. The excluded operators involve super-covariant derivatives acting on $R,\bar{R}$ and the superfields $G_a$ or $W_{\alpha \beta \gamma}$. In particular, these operators can also contribute higher monomials in $M$ and $\bar{M}$ to the scalar potential. However, any contribution of that type can equivalently be generated via operators of the form $R^n \bar{R}^m$ and, therefore, be rewritten via operators involving only $R, \bar{R}$. These considerations are also in agreement with the conjectured action in eq.~\eqref{conjectured_L}. Therefore, by choosing the Lagrangian in eq.~\eqref{Uapp} we only constrain the allowed higher-curvature terms, but not the form of the scalar potential. More specifically, eq.~\eqref{Uapp} implies off-shell only a $\mathcal{R} + \mathcal{R}^2$ gravity. Naturally, the form of the higher-curvature terms has an effect on the vacuum-structure of the theory \cite{Hindawi:1995cu}. For instance, this can be seen when we rewrite the $f(\mathcal{R})$-degree of freedom in terms of a scalar field, which may induce supersymmetry breaking \cite{CECOTTI198786, Hindawi:1995qa}. However, while in higher-curvature theories additional non-supersymmetric vacua may exist, they still admit supersymmetric vacua where eq.~\eqref{Killing_spinor_int} holds. To conclude, from now on we consider eq.~\eqref{Uapp} and return to the discussion of the higher-curvature terms at the end of this appendix. 

The component version of eq.~\eqref{Uapp} can easily be computed following the steps in sec.~\ref{sugraaction}. Evaluating the result at the vacuum conditions 
\begin{equation}\label{FdAbla0}
 \langle F^i \rangle = \langle \partial_a A^i \rangle = \langle \partial_a \bar{A}^{\bar{\jmath}} \rangle = \langle \partial_a M \rangle = \langle b_a \rangle = \dots = 0 \ ,
\end{equation}
we arrive at the component Lagrangian
\begin{equation}\begin{aligned}\label{app:L}
 \mathcal{L}/e &= - \tfrac{1}{2} \Omega \mathcal{R} - \tfrac{3}{4} U_{M\bar{M}} \mathcal{R}^2 - V_J \\
 \Omega &= U + M U_M + \bar{M} U_{\bar{M}} - 4 \lvert M \rvert^2 U_{M \bar{M}} \\
 V_J &= W\bar{M} + \bar{W} M + \tfrac{1}{3} \lvert M \rvert^2 \Xi \\
 \Xi &=  U -2 M U_M -2 \bar{M} U_{\bar{M}} + 4 \lvert M \rvert^2 U_{M \bar{M}}
\end{aligned}\end{equation}
For clarity we set the kinetic terms for $M$ to zero, since in an EFT we expect that we must integrate $M$ out and, hence, contributions to $\mathcal{L}$ involving $\partial_a M$ generate kinetic terms for the chiral scalars.\footnote{It is also possible that higher-derivatives terms for the spacetime-metric are introduced this way. However, for the maximally symmetric solutions to the Einstein equations these terms are irrelevant as well.} Furthermore, we do not explicitly indicate anymore that all quantities are understood as being evaluated in the supersymmetric vacuum. After performing the Weyl transformation to the Einstein frame we arrive at the following action
\begin{equation}\begin{aligned}\label{app:L_E}
 \mathcal{L}/e = - \tfrac{1}{2} \mathcal{R}_E - \tfrac{3}{4} U_{M\bar{M}} \mathcal{R}_E^2 - V_E \ , \qquad  V_E = \frac{V_J}{\Omega^2} \ ,
\end{aligned}\end{equation}
where $\mathcal{R}_E$ denotes the Einstein frame scalar curvature. Let us first have a look at the respective Einstein equations for eq.~\eqref{app:L_E}. In the vacuum $\langle V_E \rangle$ sets the value of the cosmological constant via 
\begin{equation}
 \Lambda = - V_E \ .
\end{equation}
Since we have evaluated the Lagrangian at eq.~\eqref{FdAbla0} the coupling $U_{M\bar{M}}$ does not carry any dependence on the space-time metric and, hence, the Einstein equations read
\begin{equation}\label{Einstein_eq}
 9 \, U_{M\bar{M}} \Box \mathcal{R}_E +  \mathcal{R}_E  = - 4 \Lambda \ ,
\end{equation}
where we used the general Einstein equations for $\mathcal{R}^2$-gravity as displayed in \cite{Stelle:1976gc, Stelle:1977ry}. Since we are looking for maximally symmetric backgrounds we set $\Box \mathcal{R}_E = 0$ and, hence, we obtain
\begin{equation}\label{sol_EH}
 \mathcal{R}_E  = - 4 \Lambda = 4 V_E\ ,
\end{equation}
which coincides with the solution of the ordinary Einstein-equations. At the level of eq.~\eqref{sol_EH} the curvature constraint in eq.~\eqref{Killing_spinor_int} can thereby be recast as
\begin{equation}\label{integrability_cond}
 V_J = - \frac{1}{3} \Omega \lvert M \rvert^2 \ .
\end{equation}
It remains to check whether this condition is indeed fulfilled after replacing $M$ in $V_J$ via the solution to its respective equations of motion. The equations of motion for $\bar{M}$ read
\begin{equation}
 W + \frac{1}{3} M \Xi + \frac{1}{3} \lvert M \rvert^2 \frac{\partial \Xi}{\partial \bar{M}} - 2 \frac{V_J}{\Omega} \frac{\partial \Omega}{\partial \bar{M}} + \frac{3}{4} \Omega^2  U_{M \bar{M} \bar{M}} \mathcal{R}_E^2 = 0 \ .
\end{equation}
Let us instead analyse a real-valued version of this equation. For instance, we may investigate
\begin{equation}
 \bar{M} \frac{\partial \mathcal{L}}{\partial \bar{M}} + M \frac{\partial \mathcal{L}}{\partial M} = 0 \ .
\end{equation}
Eliminating the terms involving the superpotential via eq.~\eqref{app:L} and replacing $\mathcal{R}_E$ via eq.~\eqref{sol_EH} we find that the above expression reduces to
\begin{equation}\begin{aligned}
 &\frac{1}{3} \, \lvert M \rvert^2 \left( \Xi + \bar{M} \frac{\partial \Xi}{\partial \bar{M}} + M \frac{\partial \Xi}{\partial M} \right) + \frac{V_J}{\Omega} \left(\Omega - 2\bar{M} \frac{\partial \Omega}{\partial \bar{M}} - 2M \frac{\partial \Omega}{\partial M} \right) \\
 &+ 12 \left(\frac{V_J}{\Omega}\right)^2 \left(M U_{MM\bar{M}} + \bar{M} U_{M \bar{M} \bar{M}} \right) = 0
\end{aligned}\end{equation}
We read this equation as a quadratic equation determining $V_J/\Omega$. Inserting the expressions for $\Xi, \Omega$ in eq.~\eqref{app:L} we find that the above simplifies to
\begin{equation}\label{app:quadratic_eq}
 \mathcal{E}\left( \frac{1}{3} \, \lvert M \rvert^2 + \frac{V_J}{\Omega} \right) +  \left(M U_{MM\bar{M}} + \text{h.c.} \right) \left(\frac{4}{3} \, \lvert M \rvert^4 + 8 \lvert M \rvert^2 \frac{V_J}{\Omega} + 12 \frac{V_J^2}{\Omega^2} \right) = 0
\end{equation}
where $\mathcal{E}$ is an expression in $M, \bar{M}, U$ and derivatives thereof. It is now easily seen that one of the two solutions of eq.~\eqref{app:quadratic_eq} is precisely given by eq.~\eqref{integrability_cond}. This concludes the demonstration that the curvature constraint as displayed in eq.~\eqref{Killing_spinor_int} is still automatically satisfied for eq.~\eqref{Uapp}.

There exists a second solution of eq.~\eqref{app:quadratic_eq} which leads to a violation of eq.~\eqref{integrability_cond} and, hence, corresponds to a non-supersymmetric vacuum. We already mentioned at the beginning of this appendix that higher-curvature terms may lead to supersymmetry breaking. Here we indeed find a second non-supersymmetric vacuum induced by the $\mathcal{R}^2$-term. If we had allowed for additional higher-curvature terms in $\mathcal{L}$ then further vacua might have appeared. However, the supersymmetric vacuum where eq.~\eqref{integrability_cond} is satisfied should still be allowed.
\bibliographystyle{JHEP}
\bibliography{C_2016_Final_v2}

\end{document}